\def\mum {\hbox{$\mu$m}}

\def\cii{\hbox{{\rm [C {\scriptsize II}]}}}
\def\ci{\hbox{{\rm [C {\scriptsize I}]}}}
\def\nii{\hbox{{\rm [N {\scriptsize II}]}}}
\def\sii{\hbox{{\rm [S {\scriptsize II}]}}}
\def\oiii{\hbox{{\rm [O {\scriptsize III}]}}}

\documentclass{aa} 
\usepackage{natbib}
\usepackage{multirow}
\usepackage[flushleft]{threeparttable}
\usepackage{float}
\usepackage{placeins}
\usepackage{graphicx}
\usepackage{gensymb}
\usepackage{txfonts}
\usepackage{subfigure}
\usepackage{fixltx2e}
\usepackage[colorinlistoftodos]{todonotes}
\usepackage[normalem]{ulem}

\begin{document}

   \title{Unveiling the remarkable photodissociation region of M8}

   \subtitle{}

   \author{M. Tiwari\inst{1}\fnmsep\thanks{Member of the International Max Planck Research School (IMPRS) for Astronomy and Astrophysics at the Universities of Bonn and Cologne.}\fnmsep, K. M. Menten\inst{1}, F. Wyrowski\inst{1}, J.P. P\'{e}rez-Beaupuits\inst{1,2}, H. Wiesemeyer\inst{1}, R. G\"{u}sten \inst{1}, B. Klein \inst{1}, C. Henkel \inst{1,3} }

   \institute{Max-Planck-Institut f{\" u}r Radioastronomie,
              Auf dem H\"{u}gel, 53121 Bonn, Germany \\
              \email{mtiwari@mpifr-bonn.mpg.de}
              \and European Southern Observatory, Alonso de C\'{o}rdova 3107,Vitacura Casilla 7630355, Santiago, Chile
              \and Astron. Dept., King Abdulaziz University, P.O. Box 80203, Jeddah 21589, Saudi Arabia}
              %\email{mtiwari@mpifr.mpg.de}{European Southern Observatory, Santiago, }}

   \date{Received September ....; accepted ...}

% \abstract{}{}{}{}{} 
% 5 {} token are mandatory
 \abstract
   {}
   %{%\todo[inline]{JPE: (2017-10-23) changes in the text are indicated with bold face}
{Messier 8 is one of the brightest HII regions in the sky. We collected an extensive dataset comprising multiple submillimeter spectral lines from neutral and ionized carbon and from CO. Based on it, we aim to understand M8's morphology and that of its associated photo dissociation region and to carry out a quantitative analysis of the regions' physical conditions such as kinetic temperatures and volume densities.}
   {Using the Stratospheric Observatory For Infrared Astronomy (SOFIA), the Atacama Pathfinder Experiment (APEX) 12 m and the Institut de Radioastronomie Millim\'{e}trique (IRAM) 30 m telescopes, we have performed a comprehensive imaging survey of the emission from the fine structure lines of \cii\ and \ci\ and multiple rotational transitions of carbon monoxide (CO) isotopologues within 1.3 $\times$ 1.3 pc around the dominant Herschel 36 (Her 36) system composed of at least three massive stars. To further explore the morphology of the region we compare archival infrared, optical and radio images of the nebula with our newly obtained fine structure line and CO data, in particular with the velocity information they provide. We perform a quantitative analysis, using both LTE and non-LTE methods to determine the abundances of some of the observed species as well as kinetic temperatures and volume densities.} 
   %e used GREAT, FLASH\textsuperscript{+}, CHAMP\textsuperscript{+} and EMIR receivers to get maps of \textsuperscript{12}C\textsuperscript{+}; \textsuperscript{12}C; high, mid and low J transitions of isotopologues of CO of this region.  }
   {Bright CO, \cii\ and \ci\ emission has been found towards the HII region and the photodissociation region (PDR) in M8. Our analysis places the bulk of the molecular material in the background of the nebulosity illuminated by the bright stellar systems Her 36 and 9 Sagitarii. Since the emission from all observed atomic and molecular tracers peaks at or close to the position of Her 36, we conclude that the star is still physically close to its natal dense cloud core and heats it. A veil of warm gas moves away from Her 36 toward the Sun and its associated dust contributes to the foreground extinction in the region. One of the most prominent star forming regions in M8, the Hourglass Nebula, is particularly bright due to cracks in this veil close to Her 36. By using radiative transfer modeling of different transitions of CO isotopologues, we obtain H\textsubscript{2} densities ranging from $\sim$ 10\textsuperscript{4} -- 10\textsuperscript{6} cm\textsuperscript{-3} and kinetic temperatures of 100 -- 150 K in the bright PDR caused by Her 36.}
  {}

\authorrunning{M. Tiwari et al.}
  \maketitle
%
%________________________________________________________________
\section{Introduction}

The influence of bright stars on their surrounding interstellar medium is immense. Their strong ultraviolet and far-ultraviolet fields give rise to bright HII regions and photodissociation regions (PDRs). These are the best grounds to study the effect of UV and far-UV photons on the heating and chemistry of the interstellar medium (ISM). The fine structure lines of C$^+$ and O, observable at far-IR wavelengths, are mainly responsible for the cooling in these regions \citep{1985ApJ...291..722T}, which allow us to deduce the amount and sometimes also the source of heating. The fine structure line of C$^{+}$ at 158~$\mu$m is one of the brightest lines in PDRs and traces the transition from H$^{+}$ to H and H$_{2}$ as C has an ionization potential of 11.3~eV (e.g., \citet{2017A&A...606A..29P}). In PDRs, a C$^{+}$ layer extends to a depth of $A$\textsubscript{v} $\sim$ 2 -- 4, after which C$^{+}$ recombines to C probing the interface to CO \citep{1999RvMP...71..173H}. Deeper into the associated molecular clouds, cooling is dominated by the transitions of CO, observable at (sub)millimeter and far infrared wavelengths. Modeling the relative intensity distributions of multiple lines from various molecular and atomic species allows us to derive the physical conditions in PDRs. 

Messier 8 (M8) is located in the Sagittarius-Carina arm, near our line of sight towards the Galactic Center. It is located at a distance $\sim$1.25~kpc (1$\arcmin$ corresponds to 0.36~pc) from the Sun (\citet{2004ApJ...608..781D}; \citet{2006MNRAS.366..739A}) with an error of $\sim$~0.1~kpc \citep{2008hsf2.book..533T} and is about 34 $\times$ 12~pc in diameter. M8 is associated with the open young stellar cluster NGC6530, the HII region NGC6523/33 and large quantities of molecular gas \citep{2008hsf2.book..533T}.

The open cluster NGC 6530 (centered at R.A. 18\textsuperscript{h}04\textsuperscript{m}24\textsuperscript{s}; Dec. $-$24$\degree$21$\arcmin$12$\arcsec$(J2000)), is a relatively young cluster
\citep[formed about 2 -- 4~Myrs ago,][]{2007AJ....134.1368C} and contains several bright O-type stars. The brightest among them is Her 36 \citep{1961PASP...73..206W} at R.A. 18\textsuperscript{h}03\textsuperscript{m}40\textsuperscript{s}.3; Dec. $-$24$\degree$22$\arcmin$43$\arcsec$(J2000). It is resolved into three main components: a close massive binary consisting of an O9 V and a B0.5 V star and a more distant companion O7.5 V star \citep{2010ApJ...710L..30A,2014A&A...572L...1S}. Her 36 is responsible for ionizing the gas in the western half of the HII region of NGC~6523 including the bright Hourglass Nebula \citep{1961PASP...73..206W, 1976ApJ...203..159L, 1986AJ.....91..870W}. 
%This cluster lies in front of the molecular cloud as the far-IR luminosity of the M8 region is significantly lower than the integrated light of several OB stars combined in the cluster \citep{1973ApJ...182..497J}. 
\citet{1976ApJ...203..159L} compared the optical and the millimeter-wave observations of the M8 region and suggested the molecular cloud is located behind the HII region of the nebula, similar to the Orion-KL nebula. An ultra compact HII region, G5.97--1.17 is also very close to Her 36 at (R.A. 18\textsuperscript{h}03\textsuperscript{m}40.5\textsuperscript{s}; Dec. $-$24$\degree$22$\arcmin$44.3$\arcsec$(J2000)) \citep{2014ApJ...797...60M}.\

\begin{figure*}[htp]
  \centering
  \subfigure{\includegraphics[height=80mm]{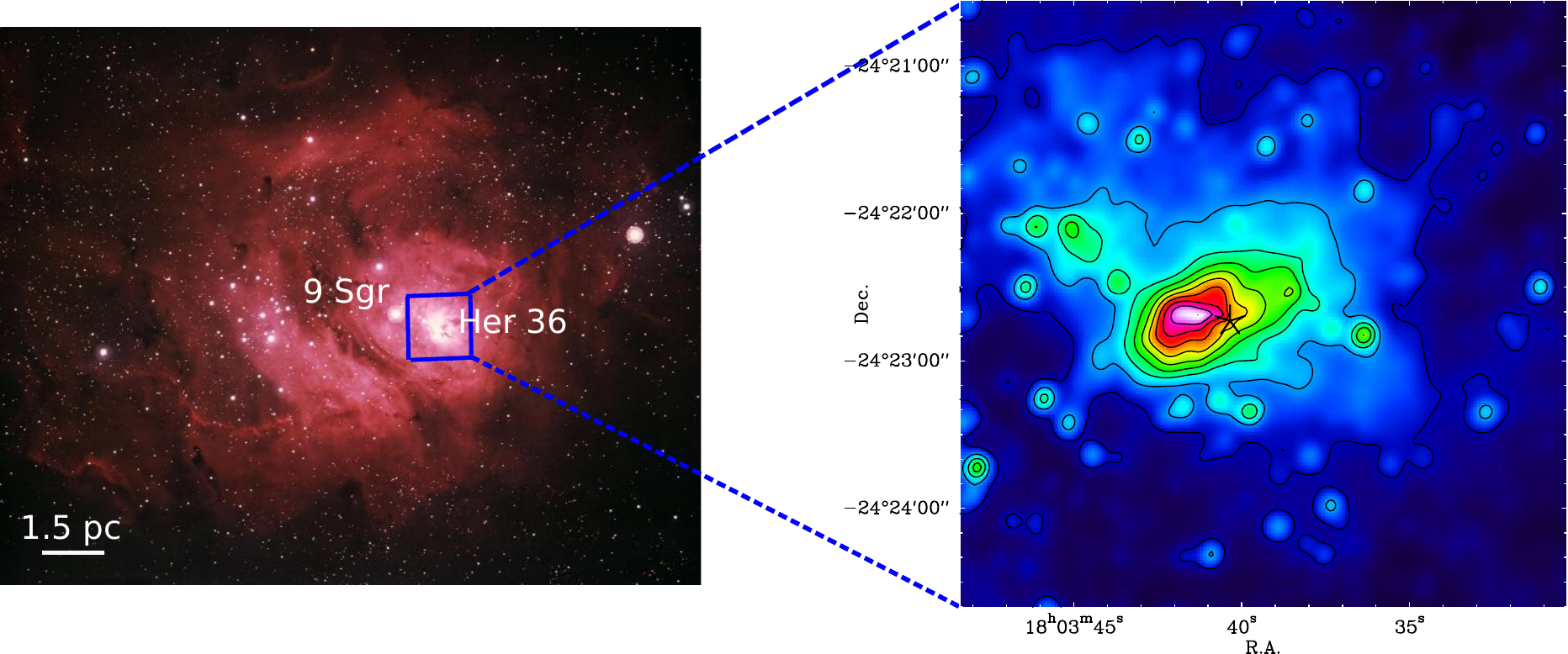}}
 
  \caption{The left panel shows the Lagoon Nebula, Messier object 8 (M8) or NGC~6523, in the constellation of Sagittarius, as seen by the Kitt Peak 4-meter Mayall telescope in 1973. North is up, east to left. Credits to "National Optical Astronomy Observatory/Association of Universities for Research in Astronomy/National Science Foundation" Copyright WIYN Consortium, Inc., all rights reserved. The right panel shows the WISE 3.4 $\mu$m image of the M8 region around Her 36 (marked with a star) investigated in this paper. The contour levels are 10$\%$ to 100$\%$ in steps of 10\% of the peak emission $\sim$ 4000 data number. The 1.5 pc indicated in the lower left correspond to 4$\arcmin$.}
\end{figure*}

A multiband near-IR image of Her 36 and its surroundings presented in \citet{2006ApJ...649..299G} shows the IR source Her 36 SE lying 0.25$\arcsec$ SE of Her 36 and being completely obscured. It is inferred to be an early-type B star with a visual extinction $A_{\rm V} > 60$~mag that is deeply embedded in dense, warm dust and is powering the ultra compact HII region G5.97--1.17. The morphology of H\textsubscript{2} and CO $J=3 \to 2$ emission around Her 36 \citep{1997A&A...323..529W, 2002PASA...19..260B} is in accordance with the HST jet-like feature detections extending 0.5$\arcsec$ south-east of Her 36 \citep{1995ApJ...445L.153S} that suggest there might be a molecular outflow in the core of M8 \citep{2002PASA...19..260B}. X-ray emission from Her 36 and diffuse X-ray emission from the Hourglass region, the brightest part of the optically visible nebula located $\sim$ 15$\arcsec$ east from Her 36 \citep{2002A&A...395..499R}, suggest the presence of a bubble of hot gas of size 0.4 pc that is produced by the interaction of the stellar wind of Her 36 with the denser part of the molecular cloud in the background. Anomalously broad diffuse interstellar bands (DIBs) at 5780.5, 5797.1, 6196.0 and 6613.6 {\AA} along with CH\textsuperscript{+} and CH are found in absorption along the line of sight to Her 36 \citep{2013ApJ...773...41D}. CH\textsuperscript{+} and CH are radiatively excited by strong far-IR emission from the adjacent IR source Her 36 SE \citep{2006ApJ...649..299G} and the broadening of DIBs is attributed to radiative pumping of closely spaced high-$J$ rotational levels of small polar carrier molecules \citep{2013ApJ...773...41D, 2014ApJ...793...68O, 2014IAUS..297...89Y}. We performed (sub)millimeter observations related to these species that will be discussed in a future paper.

 The eastern half of the HII region is illuminated by the 9 Sgr stellar system as shown in Fig.~1. 9 Sgr is a well known binary with an orbit of $\sim$ 9 yr duration, consisting of an O3.5 V primary and an O5-5.5 V secondary \citep{2012A&A...542A..95R}. South-east of the cluster core (NGC 6530), another cluster, M8E, though optically invisible, is associated with two massive star forming regions \citep{2008hsf2.book..533T}. A superposition of four HII regions seems to be responsible for the ionization of the gas in M8: the Hourglass nebula illuminated by Her 36, the core of NGC 6523 illuminated by Her 36, the remaining parts of NGC 6523 and NGC 6533 illuminated by 9 Sgr (O4V) \citep{2008hsf2.book..533T} and M8E illuminated by HD 165052 \citep{1982ApJ...263..130L,1986AJ.....91..870W}.\

Though M8 has been studied extensively in the X-ray, optical and IR regimes \citep{1995ApJ...445L.153S, 2004ApJ...608..781D, 2006MNRAS.366..739A, 2006ApJ...649..299G,2017A&A...604A.135D}, only few studies have been performed at millimeter and submillimeter wavelengths. \citet{1997A&A...323..529W} reported the discovery of the second strongest source of mm and submm wavelength CO line emission in our Galaxy towards Her 36 in M8 \citep{1997A&A...323..529W}. \citet{1976ApJ...203..159L} compared optical and millimeter-wave observations to sketch the morphology of M8 where the core surrounding Her 36, the hourglass nebula with its structure and the eastern part of M8 are described. \citet{2002ApJ...580..285T} presented submillimeter- and millimeter-wave maps of the $J=2\to 1$ and $J=3\to 2$ transitions of \textsuperscript{12}CO tracing the molecular gas and dust around Her 36.\ 
%the edge of the HII region in M8.\
   
    We report a comprehensive survey of the 1.5 $\times$ 1.5 pc (4$\arcmin$ $\times$ 4$\arcmin$) region around Her 36 (as shown by the blue square in Fig.~1) at far-IR, millimeter- and submillimeter wavelengths to probe the physical conditions and to image the morphology of this exceptional photodissociation region. We present for the first time extended maps of this region in the 158 $\mu$m fine structure line of C\textsuperscript{+}, high-$J$ transitions of \textsuperscript{12}CO emission, observed with the GREAT\footnote{GREAT is a development by the MPI f{\"u}r Radioastronomie and KOSMA/Universit{\"a}t zu K{\"o}ln, in cooperation with the MPI f{\"u}r Sonnensystemforschung and the DLR Institut f{\"u}r Planetenforschung.} receiver on board of the SOFIA observatory, the mid-$J$ transitions of \textsuperscript{12}CO and \textsuperscript{13}CO using the PI230, FLASH\textsuperscript{+} and CHAMP\textsuperscript{+} receivers of the APEX\footnote{This publication is based on data acquired with the Atacama Pathfinder EXperiment (APEX). APEX is a collaboration between the Max-Planck-Institut f{\"u}r Radioastronomie, the European Southern Observatory, and the Onsala Space Observatory.} telescope, and low-$J$ transitions of \textsuperscript{12}CO and \textsuperscript{13}CO using the EMIR receiver of the IRAM\footnote{Based on observations carried out with the IRAM 30~m telescope. IRAM is supported by INSU/CNRS (France), the MPG (Germany), and IGN (Spain).} 30 m telescope. 

 \begin{figure*}[htp]
  \centering
% \subfigure{\includegraphics[width=80mm]{m8-CIIn.pdf}}\quad
 % \subfigure{\includegraphics[width=80mm]{m8-12CO11-10n.pdf}}
 % \subfigure{\includegraphics[width=80mm]{m8-CO13-12n.pdf}}\quad
 % \subfigure{\includegraphics[width=80mm]{m8-12CO16-15n.pdf}}
 
 \subfigure{\includegraphics[width=0.48\textwidth]{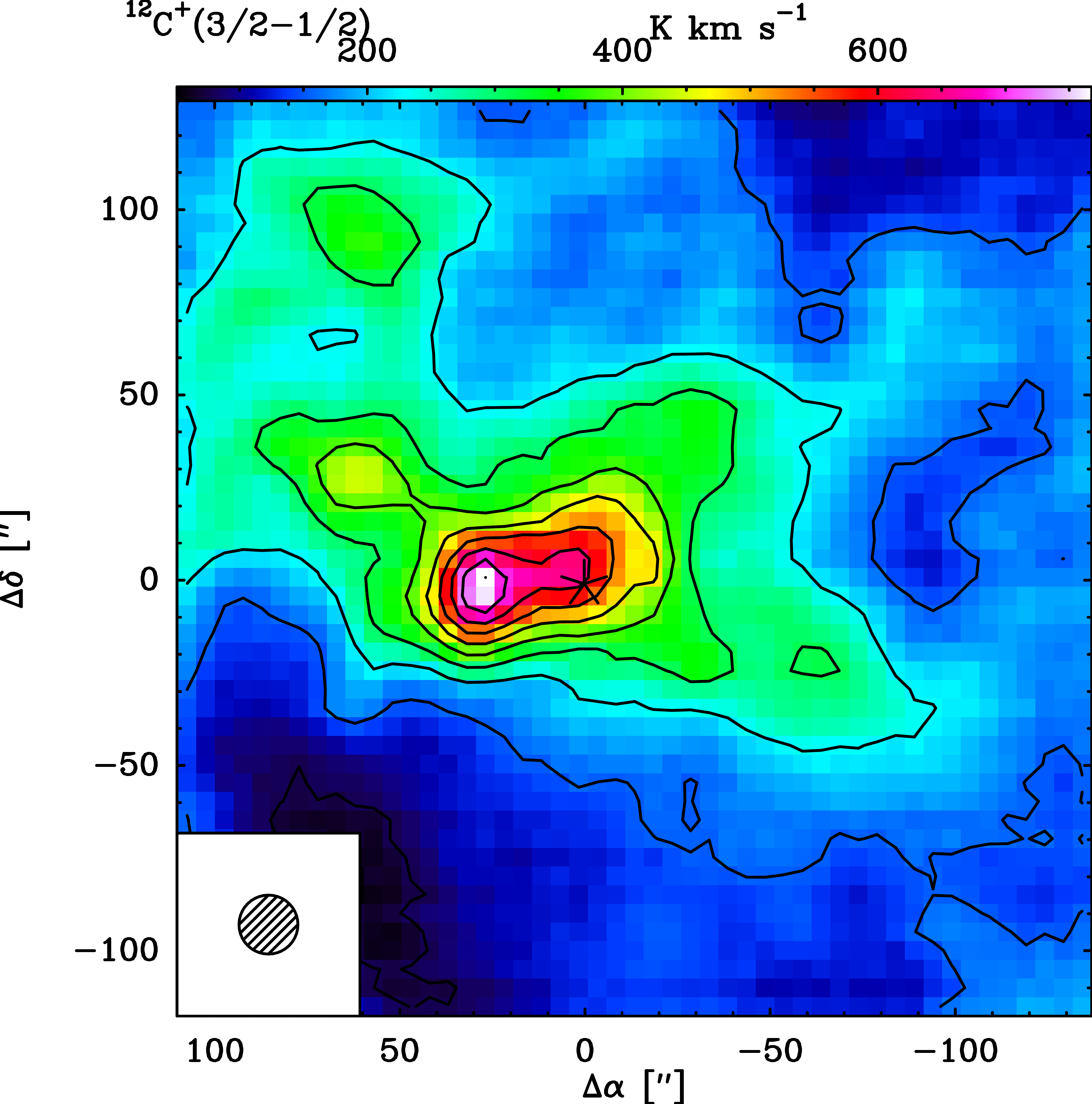}}\quad
  \subfigure{\includegraphics[width=0.48\textwidth]{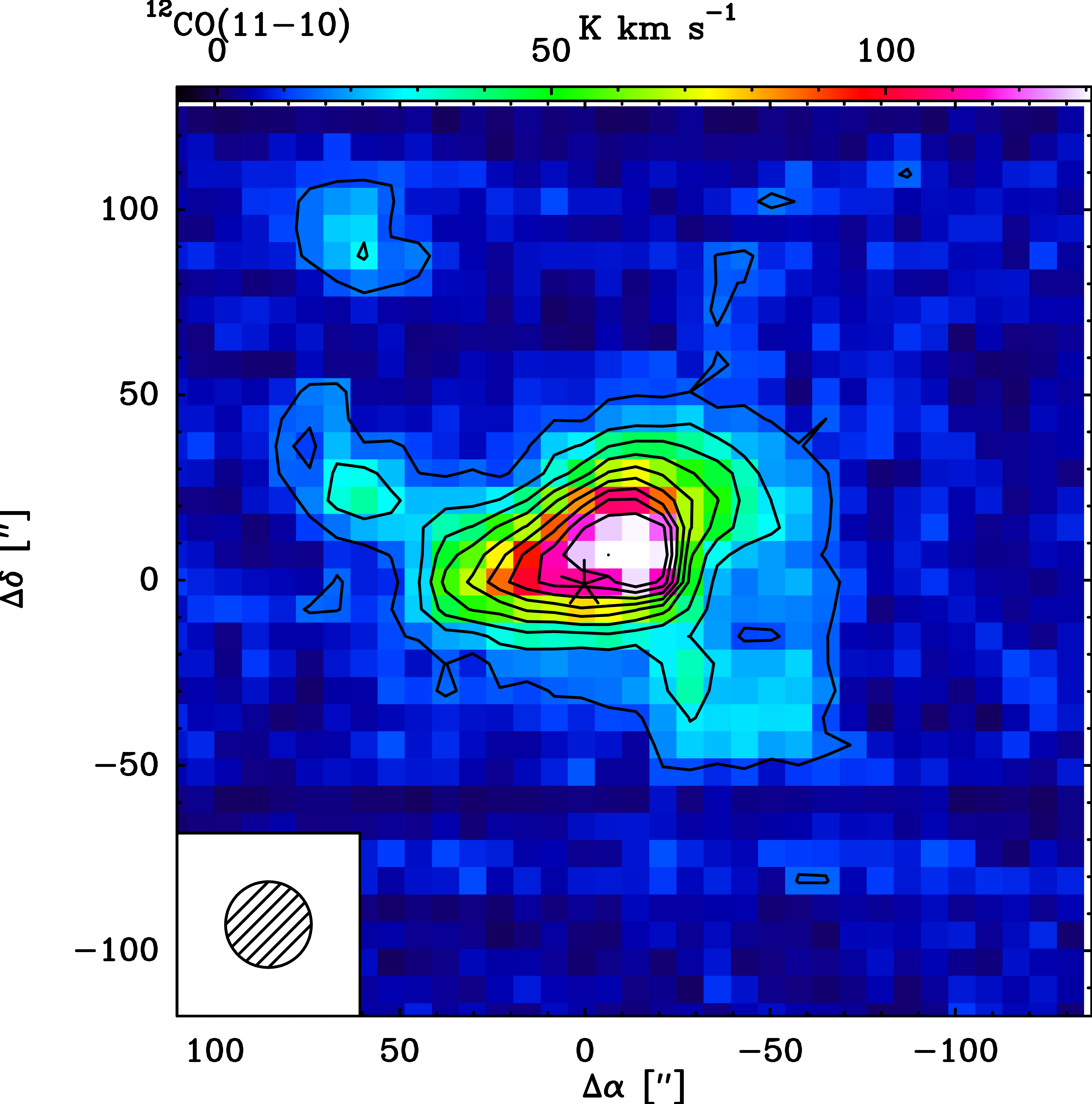}}
  \subfigure{\includegraphics[width=0.48\textwidth]{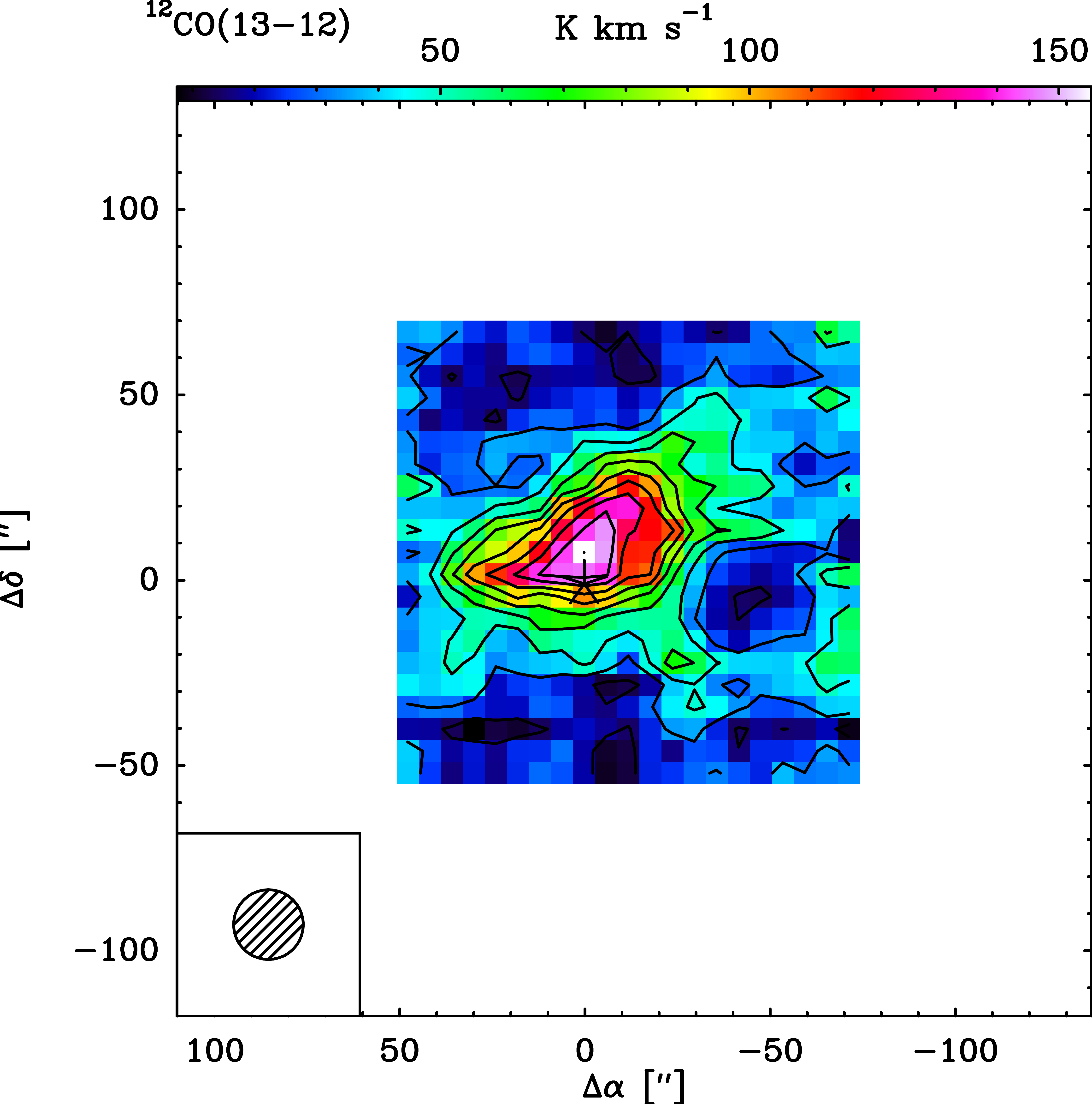}}\quad
  \subfigure{\includegraphics[width=0.48\textwidth]{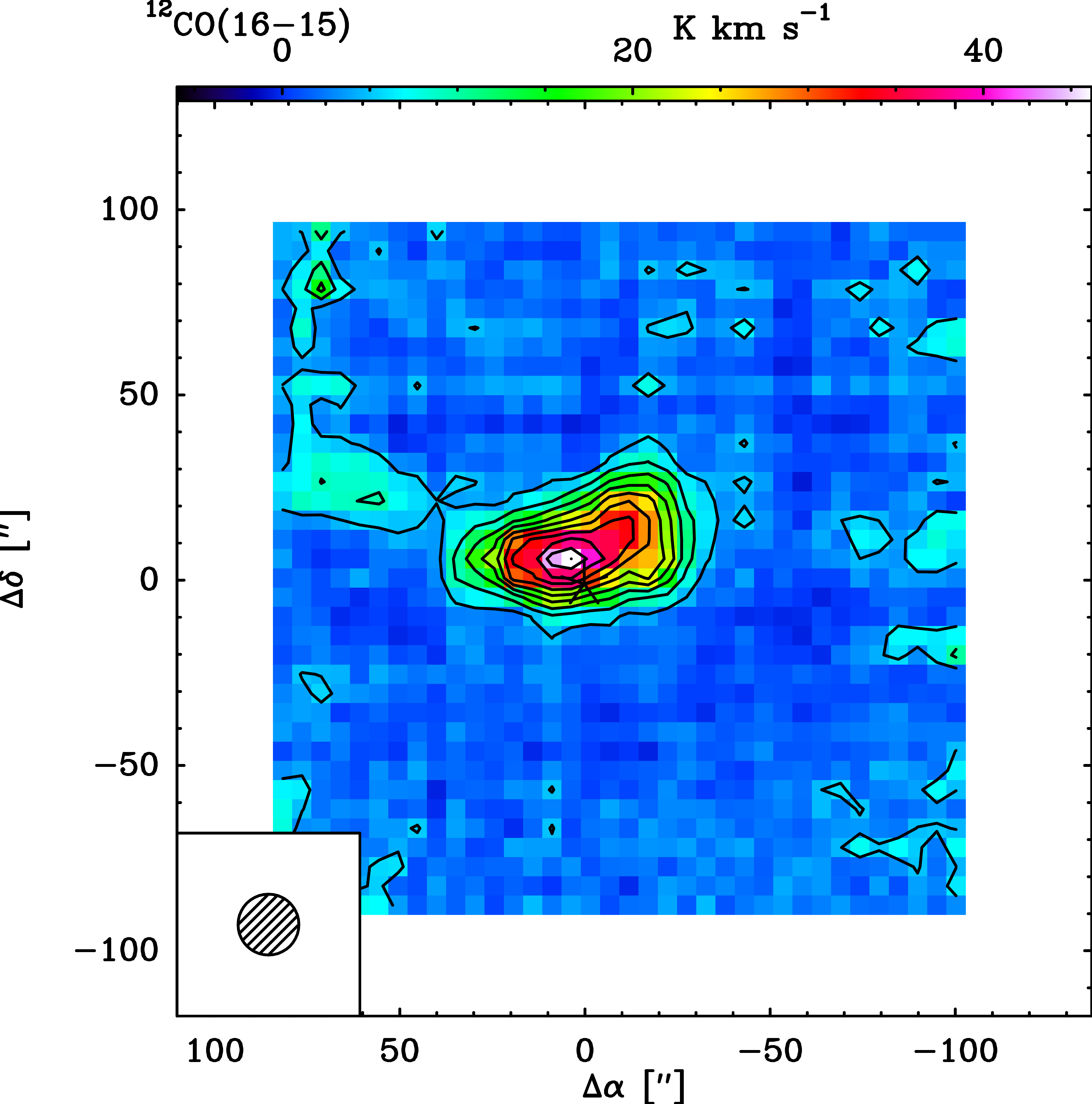}}
 
  \caption{Colour maps of the integrated intensity of the \cii~158~\mum and $J$ = 11 $\to$ 10, $J$ = 13 $\to$ 12 and $J$ = 16 $\to$ 15 transitions of \textsuperscript{12}CO toward Her 36 which is the central position ($\Delta\alpha$ = 0, $\Delta\delta$ = 0) at R.A.(J2000) = 18\textsuperscript{h}03\textsuperscript{m}40.3\textsuperscript{s} and Dec.(J2000) = $-$24\degree22$\arcmin$43$\arcsec$, marked with an asterisk. The contour levels are 10\% ($>$~3 $\times$ rms, given in Table 1) to 100\% in steps of 10\% of the corresponding peak emission given in Table 1. All maps are plotted using original beam sizes shown in the bottom left of each map.}
\end{figure*}

\section{Observations}
\subsection{The SOFIA/GREAT data}

The high-$J$ CO and \cii\ 158~$\mu$m observations summarized in
Table 1 were conducted with the L1 channel of the German Receiver for Astronomy at Terahertz frequencies (GREAT, \citet{2012A&A...542L...1H})
and the upGREAT LFA arrays \citep{2016A&A...595A..34R} on board of the
Stratospheric Observatory for Infrared Astronomy (SOFIA, \citet{2012ApJ...749L..17Y}). The data was acquired during the observatory's flight \#297 on 2016 May 14, at 14.2~km altitude and under a median water vapor column of 11~$\mu$m. upGREAT was employed with 14 pixels (seven pixels for each polarization, with a hexagonal layout). The spectral analysis was performed by means of Fast Fourier Transform spectrometers \citep{2012A&A...542L...3K}, in a mode providing 4.0~GHz bandwidth with $2^{14}$ spectral channels. 

In the first setup we simultaneously mapped the $^{12}\mathrm{CO}$ $J = 11 \rightarrow 10$ transition at 1267.014.486 GHz and the \cii\ $^{2}P_{3/2} \rightarrow \,^{2}P_{1/2}$ fine structure line at $1900.537$~GHz.  In a second setup the $^{12}\mathrm{CO}$ $J = 13 \rightarrow 12$ and $J = 16 \rightarrow 15$ transitions at 1496.923~GHz and 1841.345~GHz were mapped, respectively. Typical single-sideband system temperatures ranged between 1600 and 1800~K for the lower frequency L1 channel and between 2080 and 2260~K for the higher frequency upGREAT array, with atmospheric transmissions of 0.90 to 0.94 and 0.85 to 0.88, respectively. All maps shown in Fig.~2 were observed in on-the-fly (OTF) total power mode with a sampling of $6''$ centered on R.A. $18^{\rm h}03^{\rm m}40\fs 33$, Dec. $-24^\circ 22'42\farcs 7$ (J2000), the Her 36 location.
We integrated 0.4~s per record for the CO $(11 \to 10)/$[CII]
setup, and 0.8~s for the CO $(13 \to 12)/(16 \to 15)$ setup. The originally chosen reference position at $(\Delta\alpha,\Delta\delta) = (+500'',-500'')$ (relative to the map center) was found to be contaminated in both transitions of the first setup, and was therefore changed in favor of a second, clean reference at offset $(+30',-30')$, while the pointing accuracy of $<$~3$\arcsec$ was maintained.

The raw data was calibrated with the program kalibrate \citep{2012A&A...542L...4G}, which is part of the KOSMA software package. The resulting antenna temperatures $T_{\rm A}^*$ were converted to main beam brightness temperatures $T_{\rm mb} = T_{\rm A}^* \cdot F_{\rm eff}/B_{\rm eff}$ using a forward efficiency $F_{\rm eff} = 0.97$ and beam efficiencies of 0.66 for the L1 channel of GREAT, and 0.58 to 0.68 for the upGREAT pixels, with a median value of 0.65.
In order to optimize the signal-to-noise ratio per channel, the spectra were smoothed to a resolution of 2.44~MHz, corresponding to 0.4~km s$^{-1}$ for upGREAT and 0.6~km s$^{-1}$ for the L1 channel. We subtracted spectral baselines of first order and subsequently produced data cubes with beam/3 sampling. For the beam sizes, see Table~1.

 \begin{figure*}[htp]
  \centering
 
  %\subfigure{\includegraphics[width=58mm]{m8-12CO21n.pdf}}\quad
  %\subfigure{\includegraphics[width=58mm]{m8-13CO21n.pdf}}\quad
  %\subfigure{\includegraphics[width=58mm]{m8-12C18O21n1.pdf}}
  %\subfigure{\includegraphics[width=58mm]{m8-12CO32n.pdf}}\quad
  %\subfigure{\includegraphics[width=58mm]{m8-13CO43n.pdf}}\quad
  %\subfigure{\includegraphics[width=58mm]{m8-12C10n2.pdf}}
 %\subfigure{\includegraphics[width=58mm]{m8-12CO65n.pdf}}\quad
 %\subfigure{\includegraphics[width=58mm]{m8-13CO65n.pdf}}\quad
  %\subfigure{\includegraphics[width=58mm]{m8-12CO76n.pdf}}

   \subfigure{\includegraphics[width= 0.3\textwidth]{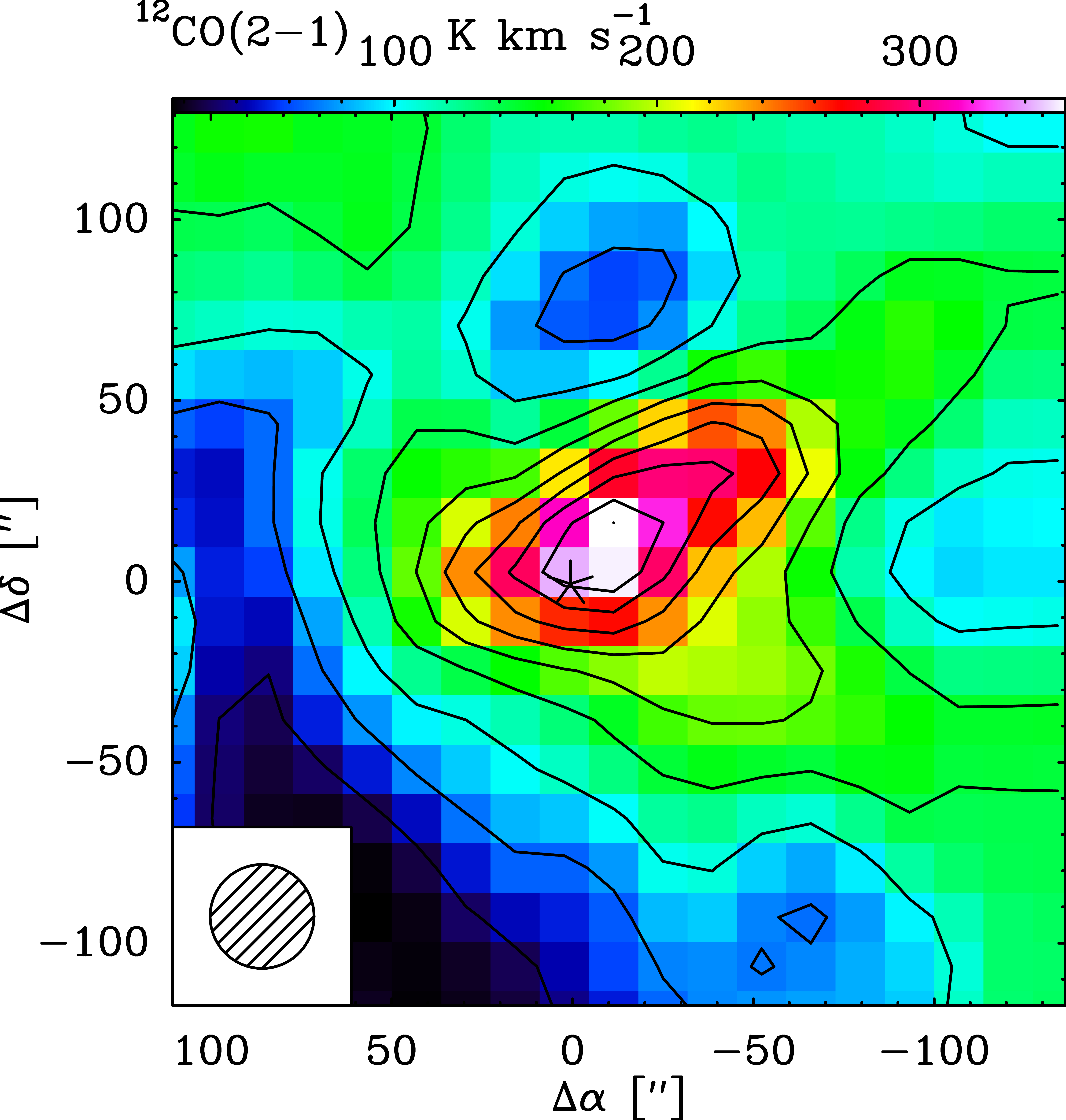}}\quad
  \subfigure{\includegraphics[width=0.3\textwidth]{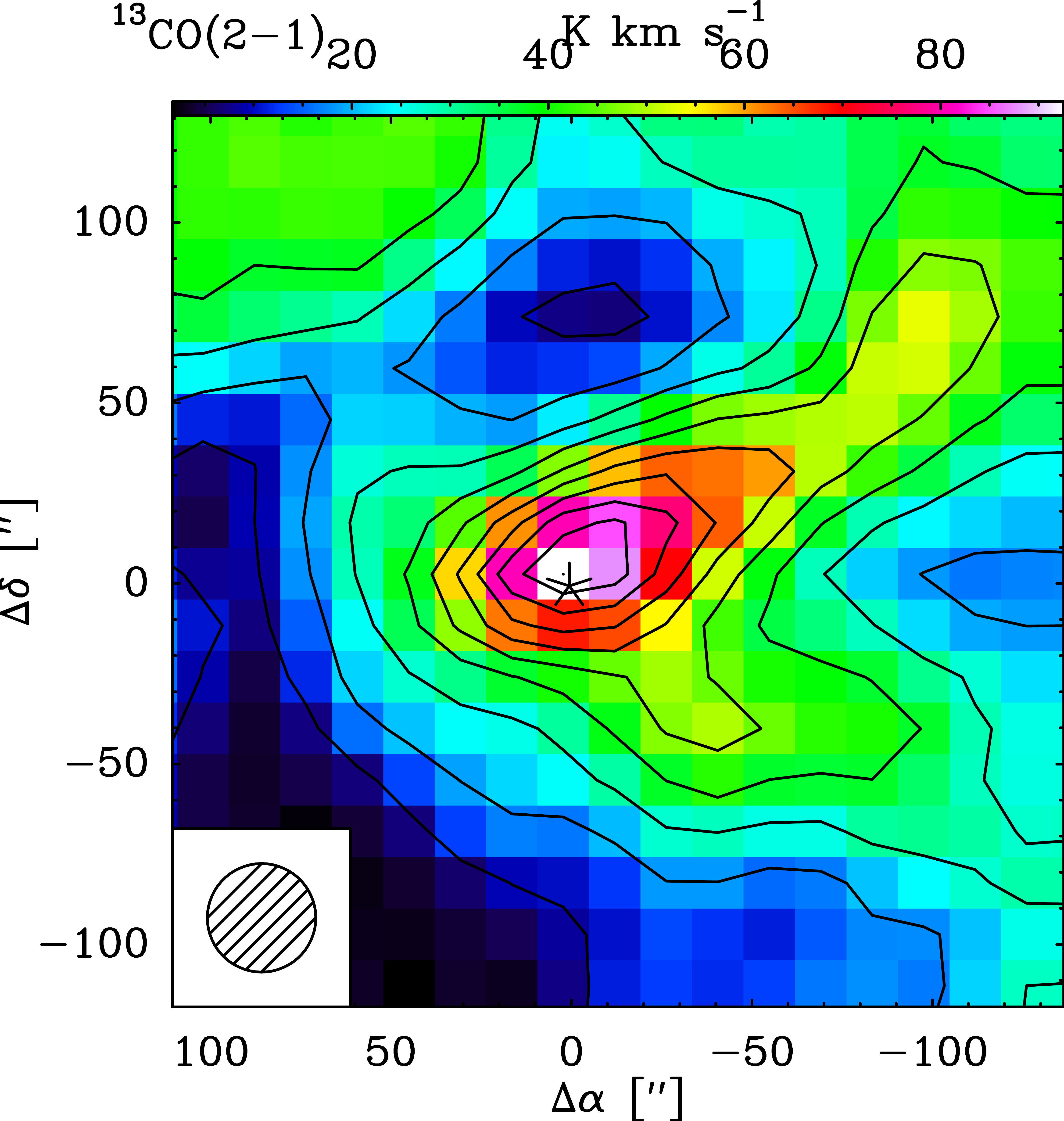}}\quad
  \subfigure{\includegraphics[width=0.3\textwidth]{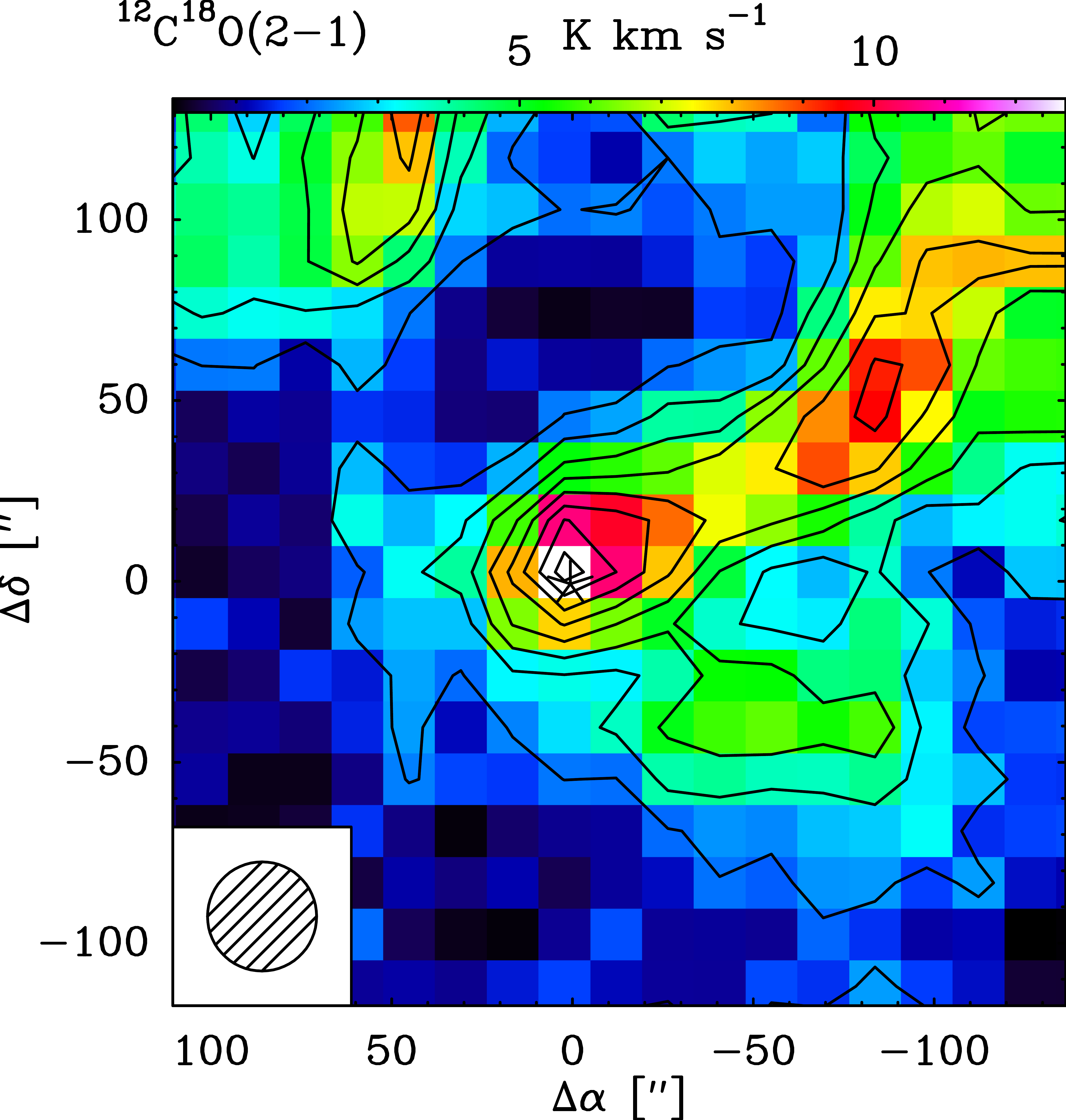}}
  \subfigure{\includegraphics[width=0.3\textwidth]{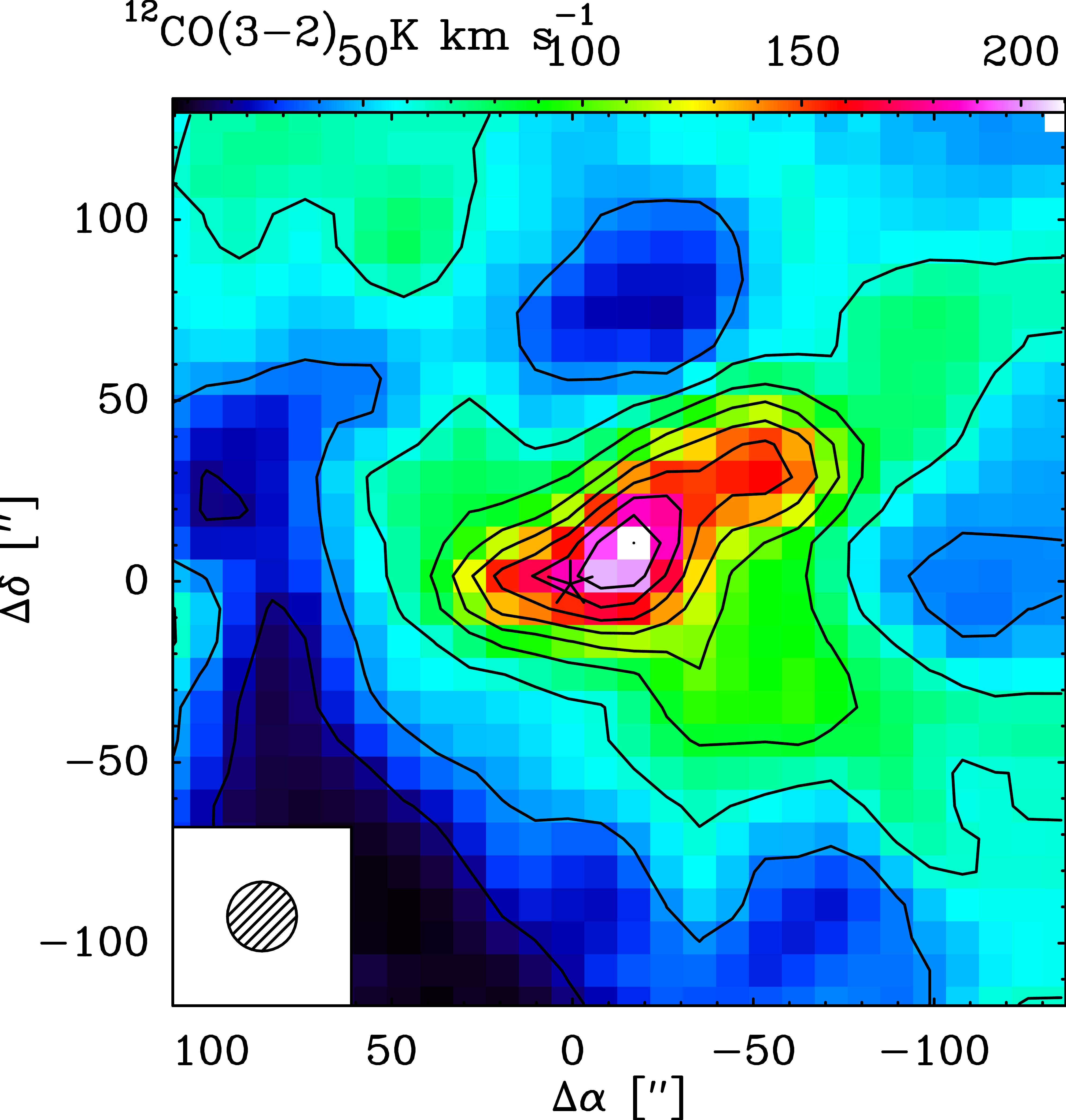}}\quad
  \subfigure{\includegraphics[width=0.3\textwidth]{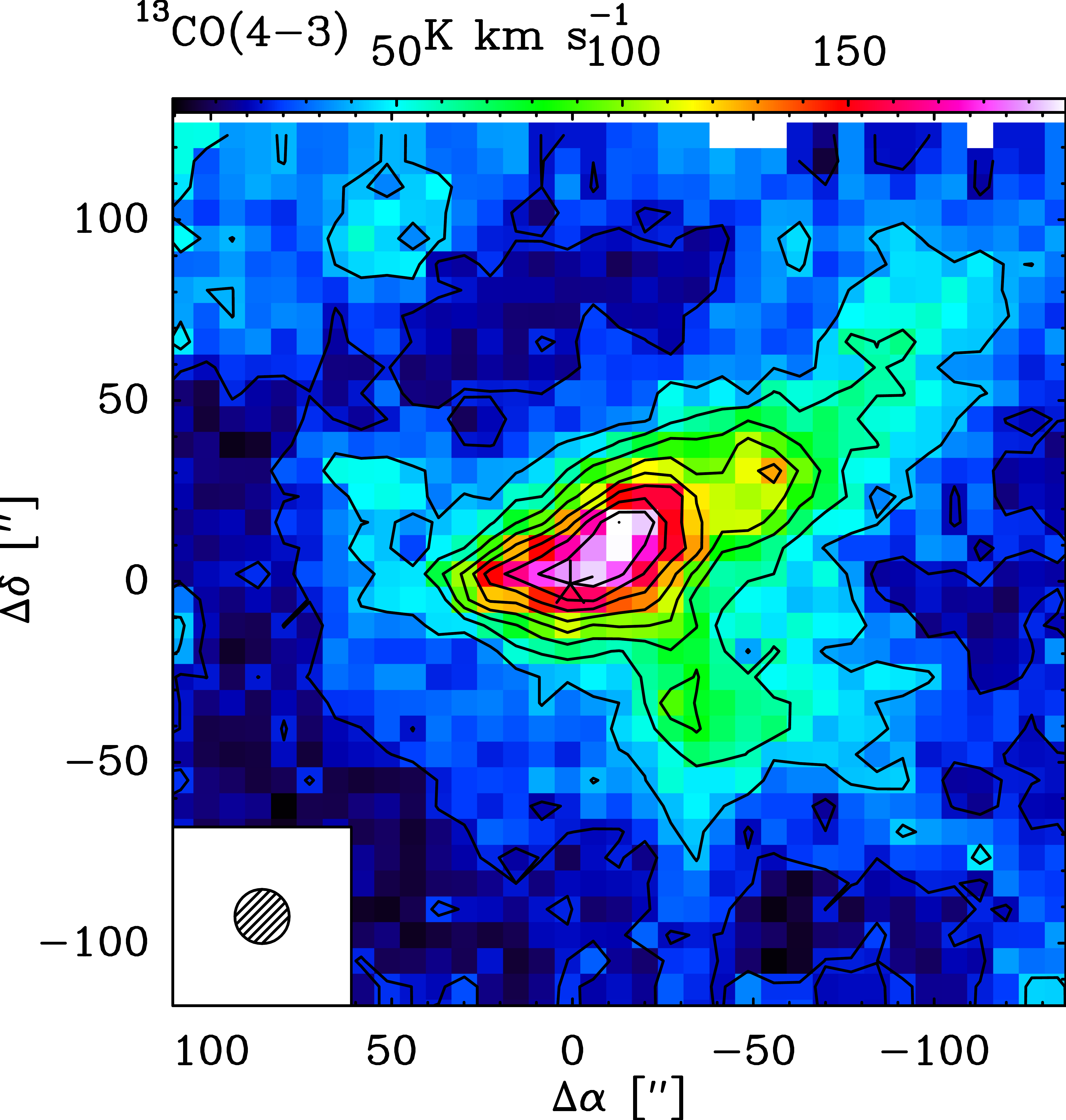}}\quad
  \subfigure{\includegraphics[width=0.3\textwidth]{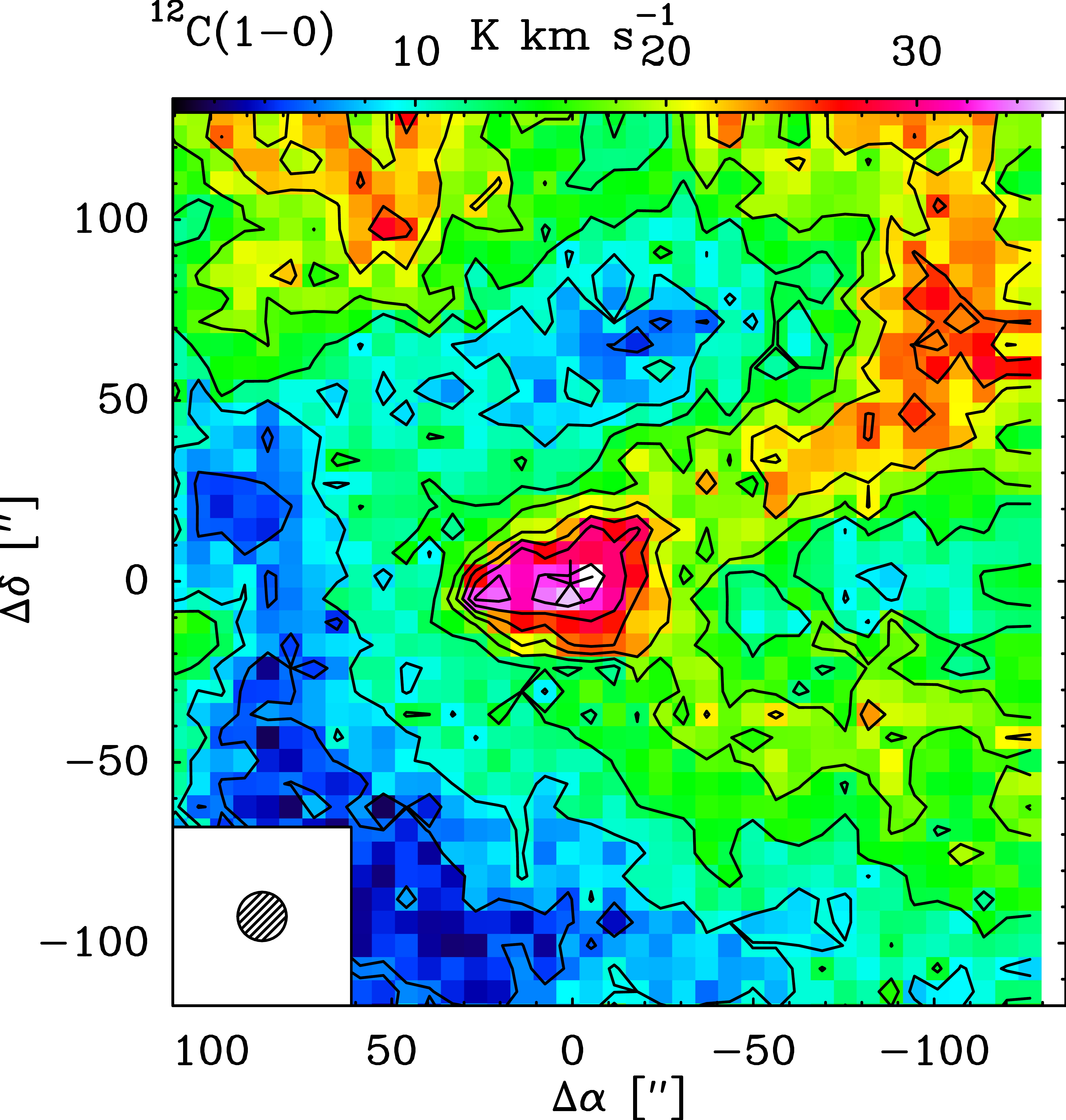}}
 \subfigure{\includegraphics[width=0.3\textwidth]{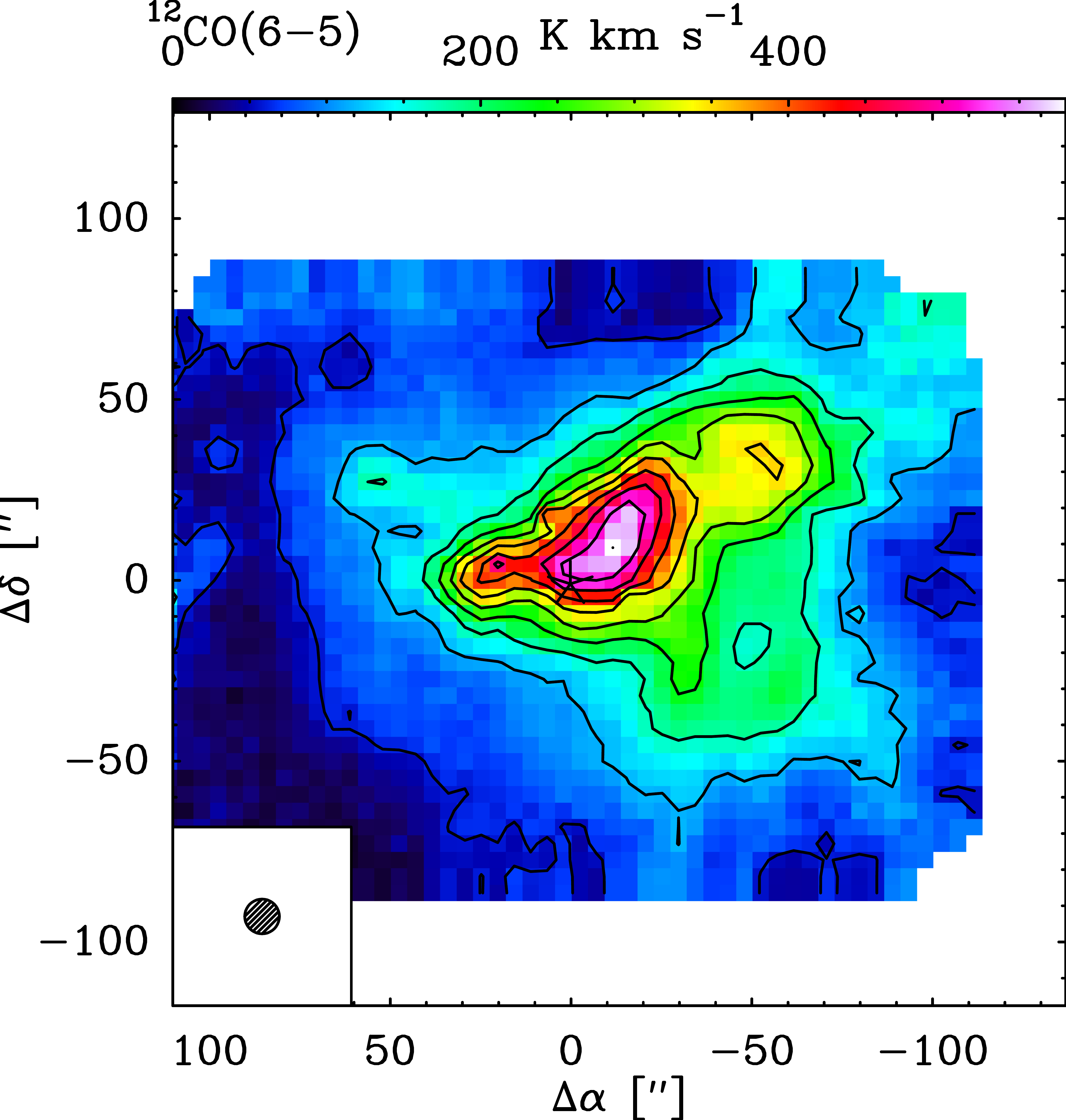}}\quad
 \subfigure{\includegraphics[width=0.3\textwidth]{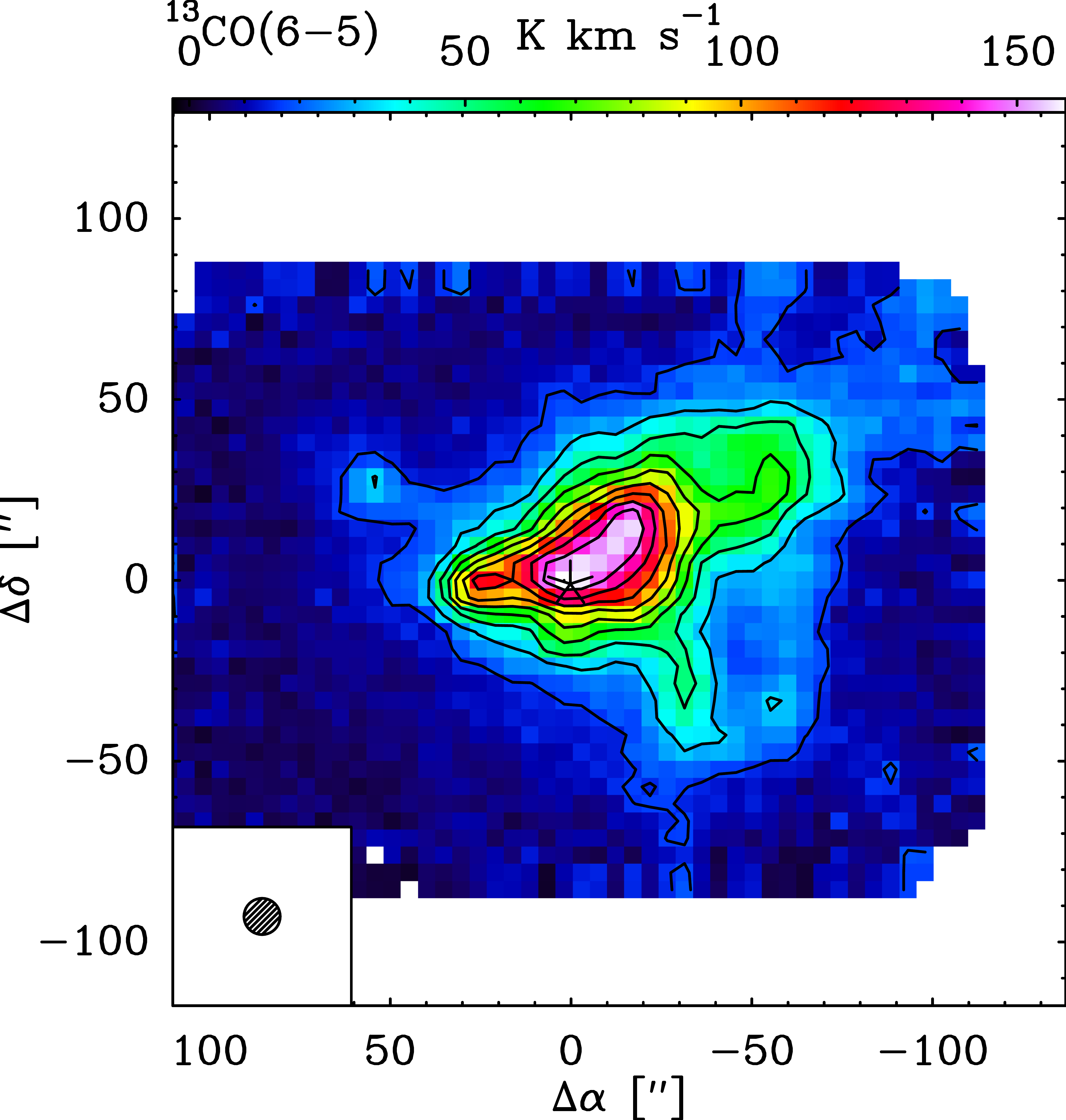}}\quad
  \subfigure{\includegraphics[width=0.3\textwidth]{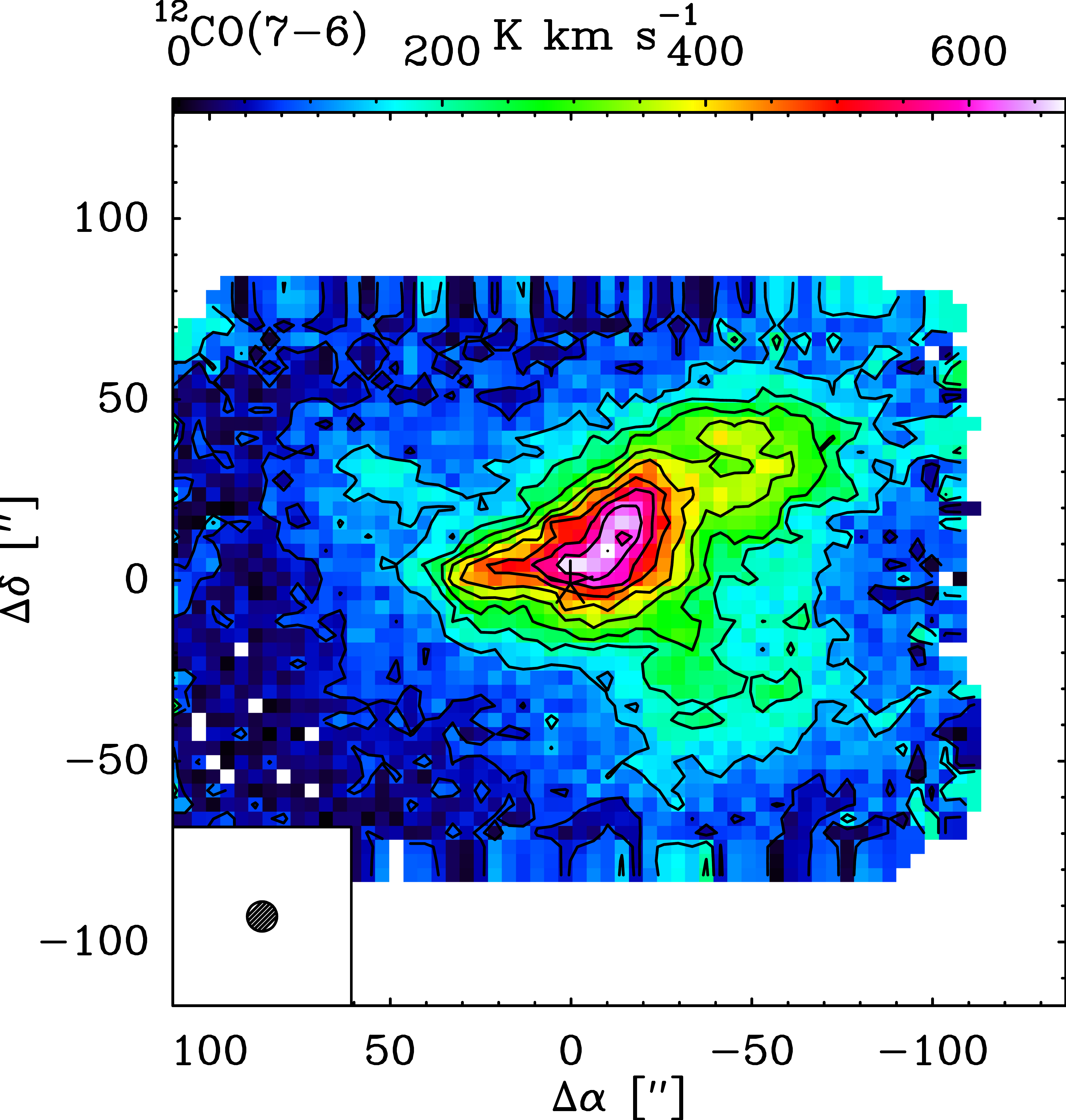}}
 
  \caption{Colour maps of the integrated intensity of the $J=2\to 1$, $J=3\to 2$, $J=6\to 5$ and $J=7\to 6$ transitions of \textsuperscript{12}CO, the $J=2\to 1$, $J=4\to 3$ and $J=6\to 5$ transitions of \textsuperscript{13}CO, the $J=2\to 1$ line of C\textsuperscript{18}O and \ci\ $1\to 0$ toward Her 36 which corresponds to the central position ($\Delta\alpha$ = 0, $\Delta\delta$ = 0) at R.A.(J2000) = 18\textsuperscript{h}03\textsuperscript{m}40.3\textsuperscript{s} and Dec.(J2000) = $-$24$\degree$22$\arcmin$43$\arcsec$, marked with an asterisk. The contour levels of $^{12}$C$^{18}$O and \ci\ are 3 $\times$ rms in steps of 2 $\times$ rms while of other molecules are from 10\% ($>$ 3 $\times$ rms, given in Table 1) to 100\% in steps of 10\% of the corresponding peak emission given in Table 1. All maps are plotted using original beam sizes shown in the bottom left of each map.}
  
\end{figure*}

 \subsection{The APEX data}
Observations of low- and mid-$J$ \textsuperscript{12}CO and \textsuperscript{13}CO transitions were performed with the Atacama Pathfinder Experiment (APEX) 12m submillimeter telescope \citep{2006SPIE.6267E..14G} during 2015 June - August and 2016 July, September and October. As shown in Table~1, we used the following receivers: PI230 to map the low-$J$ $^{12}$CO and $^{13}$CO transitions, FLASH\textsuperscript{+} in the 345 and 460 GHz bands to map the mid-$J$ CO transitions and CHAMP\textsuperscript{+} in low and high frequency sub arrays to map the higher frequency mid-$J$ CO transitions.\
       
We used the PI230 receiver to map C\textsuperscript{18}O $J$ = 2 $\to$ 1 at 219.560 GHz , \textsuperscript{13}CO $J$ = 2 $\to$ 1 at 220.398 GHz and \textsuperscript{12}CO $J$ = 2 $\to$ 1 at 230.538 GHz. FLASH\textsuperscript{+} was used in the 345 GHz band to map the $^{12}$CO $J$ = 3 $\to$ 2 transition at 345.795 GHz. FLASH\textsuperscript{+} was also used in the 460 GHz band to map the \textsuperscript{13}CO $J$ = 4 $\to$ 3 transition at 440.765 GHz, C\textsuperscript{18}O $J$ = 4 $\to$ 3 at 439.088 GHz and \ci\ $^3P_1 \rightarrow$ $^3P_0$ fine structure line at 492.160 GHz. The CHAMP\textsuperscript{+} receiver was used to map \textsuperscript{12}CO $J$ = 6 $\to$ 5 at 691.473 GHz and \textsuperscript{13}CO $J$ = 6 $\to$ 5 at 661.067 GHz in the low frequency sub array complemented by \textsuperscript{12}CO $J$ = 7 $\to$ 6 at 806.651 GHz and \textsuperscript{13}CO $J$ = 8 $\to$ 7 at 881.272 GHz in the high frequency sub array.\ 
%and in the low frequency sub array complemented by in the low frequency sub array complemented by \textsuperscript{13}CO $J$ = 6 $\to$ 5 at 661.067 GHz and \textsuperscript{13}CO $J$ = 8 $\to$ 7 at 881.272 GHz in the high frequency sub array.\

  \begin{figure*}[htp]
  \centering
  \subfigure{\includegraphics[width=58mm]{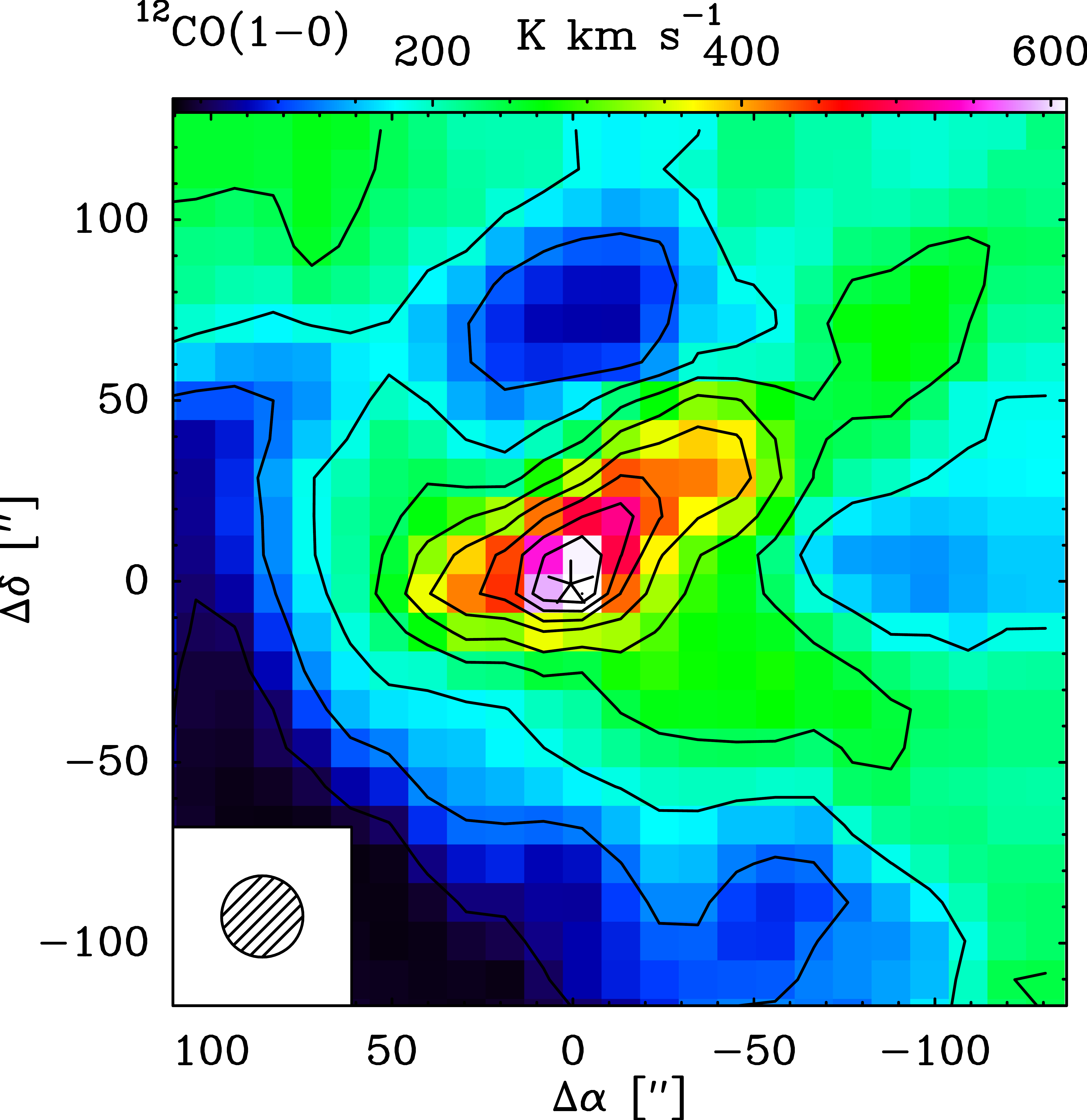}}\quad
  \subfigure{\includegraphics[width=58mm]{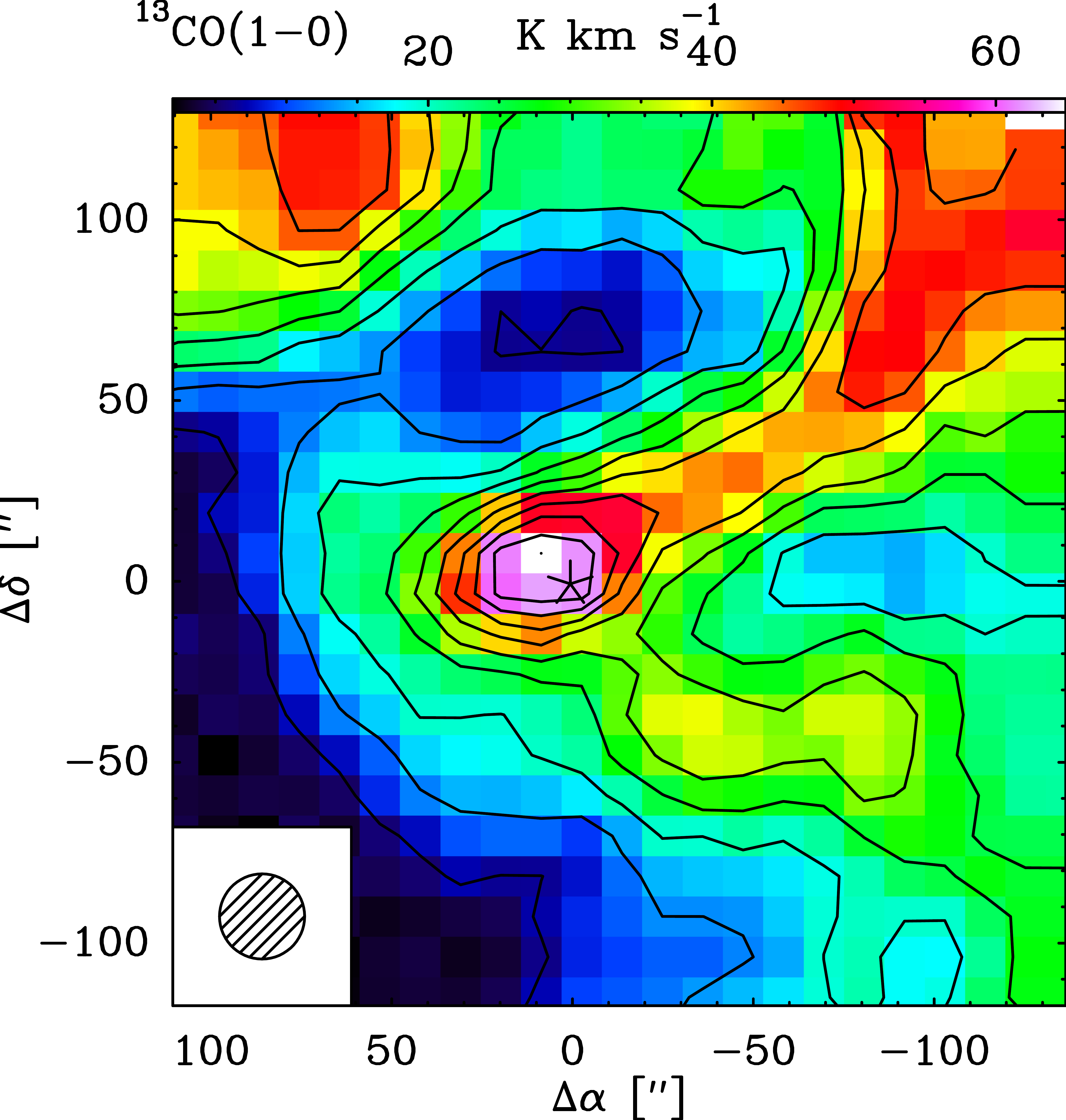}}\quad
  \subfigure{\includegraphics[width=58mm]{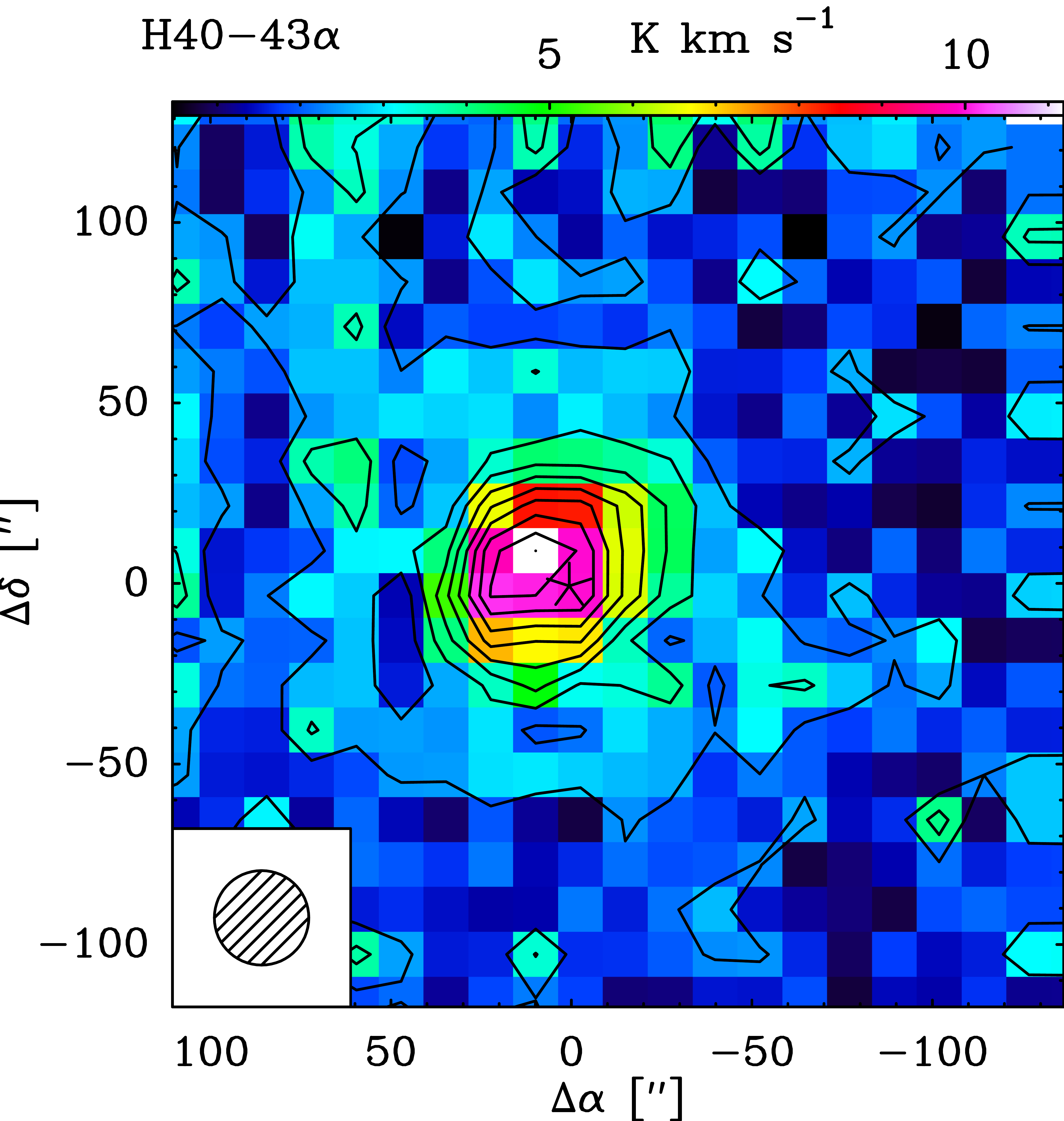}}

  \caption{Colour maps of the integrated intensity (left to right) of the $J=1\to 0$ transition of \textsuperscript{12}CO, \textsuperscript{13}CO and average of H$\alpha$40, 41,42 and 43 lines toward Her 36 which is the central position ($\Delta\alpha$ = 0, $\Delta\delta$ = 0) at R.A.(J2000) = 18\textsuperscript{h}03\textsuperscript{m}40.3\textsuperscript{s} and Dec.(J2000) = $-$24$\degree$22$\arcmin$43$\arcsec$, marked with an asterisk. The contour levels are 10\% ($>$ 3 $\times$ rms, given in Table 1) to 100\% in steps of 10\% of the corresponding peak emission given in Table 1. All maps are plotted using original beam sizes shown in the bottom left of each map.}

\end{figure*}

\begin{table*}[ht]
\centering
\caption{Line parameters of observed transitions.}

\begin{tabular}{c c c c c c c}
 \hline\hline
 \noalign{\smallskip}
 Transition & Frequency (GHz) & $\eta$\textsubscript{mb} & $\theta$\textsubscript{mb}($\arcsec$)& Peak line flux (K km s$^{-1}$) & rms (K km s$^{-1}$) & Telescope\\
 \hline
   \noalign{\smallskip}
    \multicolumn{6}{c}{\textsuperscript{12}CO}\\
 \hline
 \noalign{\smallskip}
 $J$ = 1 $\to$ 0 & 115.271 & 0.73 & 22.5 & 610.2 & 1.1 & IRAM 30m/EMIR\\
 $J$ = 2 $\to$ 1 & 230.538 & 0.65 & 28.7 & 355.5 & 0.4 &APEX/PI230\\ 
 $J$ = 3 $\to$ 2 & 345.796 & 0.73 & 19.2 & 210.2 & 0.8 & APEX/FLASH\textsuperscript{+}\\
 $J$ = 6 $\to$ 5 & 691.473 & 0.43 & 9.6 & 580.1 & 4.0 & APEX/CHAMP\textsuperscript{+}\\
 $J$ = 7 $\to$ 6 & 806.652 & 0.34 & 8.2 & 673.4 & 19.0 & APEX/CHAMP\textsuperscript{+}\\
 $J$ = 11 $\to$ 10 & 1267.014 & 0.68 & 22.9 & 130.9 & 3.5 & SOFIA/GREAT\\
 $J$ = 13 $\to$ 12 & 1496.923 & 0.68 & 19.1 & 155.3 & 2.4 & SOFIA/GREAT\\
 $J$ = 16 $\to$ 15 & 1841.346 & 0.70 & 14.8 & 46.2 & 1.8 & SOFIA/GREAT\\  
 \hline
 \noalign{\smallskip}
 \multicolumn{6}{c}{\textsuperscript{13}CO}\\
 \hline
 \noalign{\smallskip}
 $J$ = 1 $\to$ 0 & 110.201 & 0.73 & 23.5 & 64.9 & 0.4 & IRAM 30m/EMIR\\
 $J$ = 2 $\to$ 1 & 220.399 & 0.65 & 30.1 & 92.6 & 0.6 & APEX/PI230\\
 $J$ = 4 $\to$ 3 & 440.765 & 0.59 & 15.0 & 198.1 & 2.5 &APEX/FLASH\textsuperscript{+}\\
 $J$ = 6 $\to$ 5 & 661.067 & 0.45 & 10.0 & 158.8 & 3.2 & APEX/CHAMP\textsuperscript{+}\\
 %J = 8 - 7 & 881.272 & 0.34 & 7.52 & & APEX/CHAMP\textsuperscript{+}\\
 \hline
 \noalign{\smallskip}
 \multicolumn{6}{c}{C\textsuperscript{18}O}\\
 \hline
 \noalign{\smallskip}
 $J$ = 2 $\to$ 1 & 219.561 & 0.65 & 30.2 &  12.7 & 0.6 & APEX/PI230\\
 \hline
 \noalign{\smallskip}
  \multicolumn{6}{c}{\textsuperscript{12}C}\\
  \hline
   \noalign{\smallskip}
 $^{3}P_{1}$ $\to$ $^{3}P_{0}$ & 492.160 & 0.59 & 13.5 & 34.0 & 1.8 & APEX/FLASH\textsuperscript{+}\\ 
 \hline
 \noalign{\smallskip}
  \multicolumn{6}{c}{\textsuperscript{12}C$^{+}$}\\
  \hline
   \noalign{\smallskip}
   $^{2}P_{3/2}$ $\to$ $^{2}P_{1/2}$ & 1900.53 & 0.70 & 14.8 & 728.5 & 2.1 & SOFIA/GREAT\\ 
  \hline
 \noalign{\smallskip}
  \multicolumn{6}{c}{H$\alpha$}\\
  \hline
   \noalign{\smallskip}
  H40 -- 43$\alpha$ & 80--90 & 0.73 & $\approx26$ & 11.2 & 0.2 & IRAM 30m/EMIR\\ 
  \hline\hline
 \noalign{\smallskip}
 
  \end{tabular}
 \label{table:table}
\end{table*}
  
   All maps shown in Fig.~3 were observed in on-the-fly (OTF) total power mode centered on R.A. 18\textsuperscript{h}03\textsuperscript{m}40\textsuperscript{s}.3; Dec. $-$24$\degree$22'43$\arcsec$(J2000), which corresponds to the position of Her 36. The maps obtained from the observations done in July, September and October 2016 have a size of 240$\arcsec$ x 240$\arcsec$. Maps that were obtained from the observations done in 2015 are comparatively smaller in size. We integrated 0.7 s per dump for all maps and enough coverages were performed to reach the rms noise levels as mentioned in Table~1. The offset position relative to the centre at (30$\arcmin$ , $-$30$\arcmin$) was chosen as reference, similar to the SOFIA observations. The pointing accuracy ($<$ 3$\arcsec$) was maintained by pointing at bright sources such as RAFGL5254 and R Dor every 1 -- 1.5 hrs. A forward efficiency $F_{\rm eff}$ = 0.95 was used for all receivers, and the beam coupling efficiencies $B_{\rm eff}$ = 0.62, 0.69, 0.63, 0.43 and 0.32 were used for the PI230, FLASH\textsuperscript{+}340, FLASH\textsuperscript{+}460, CHAMP\textsuperscript{+}660 and CHAMP\textsuperscript{+}810 receivers, respectively.\
 
 \subsection{The IRAM 30 m data}
  
  \begin{figure*}[htp]
  \centering
  \subfigure{\includegraphics[width=58mm]{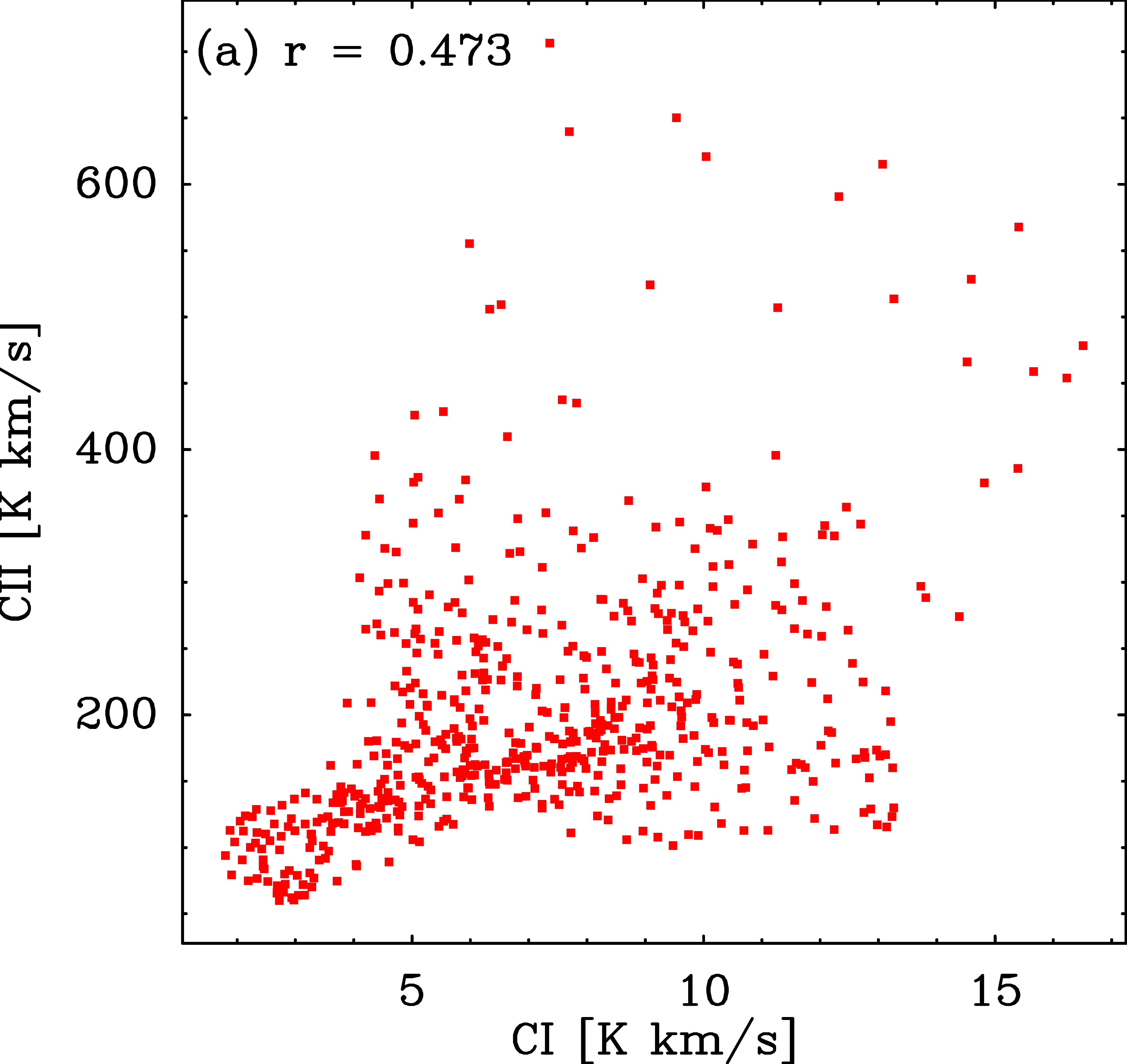}}\quad
  \subfigure{\includegraphics[width=58mm]{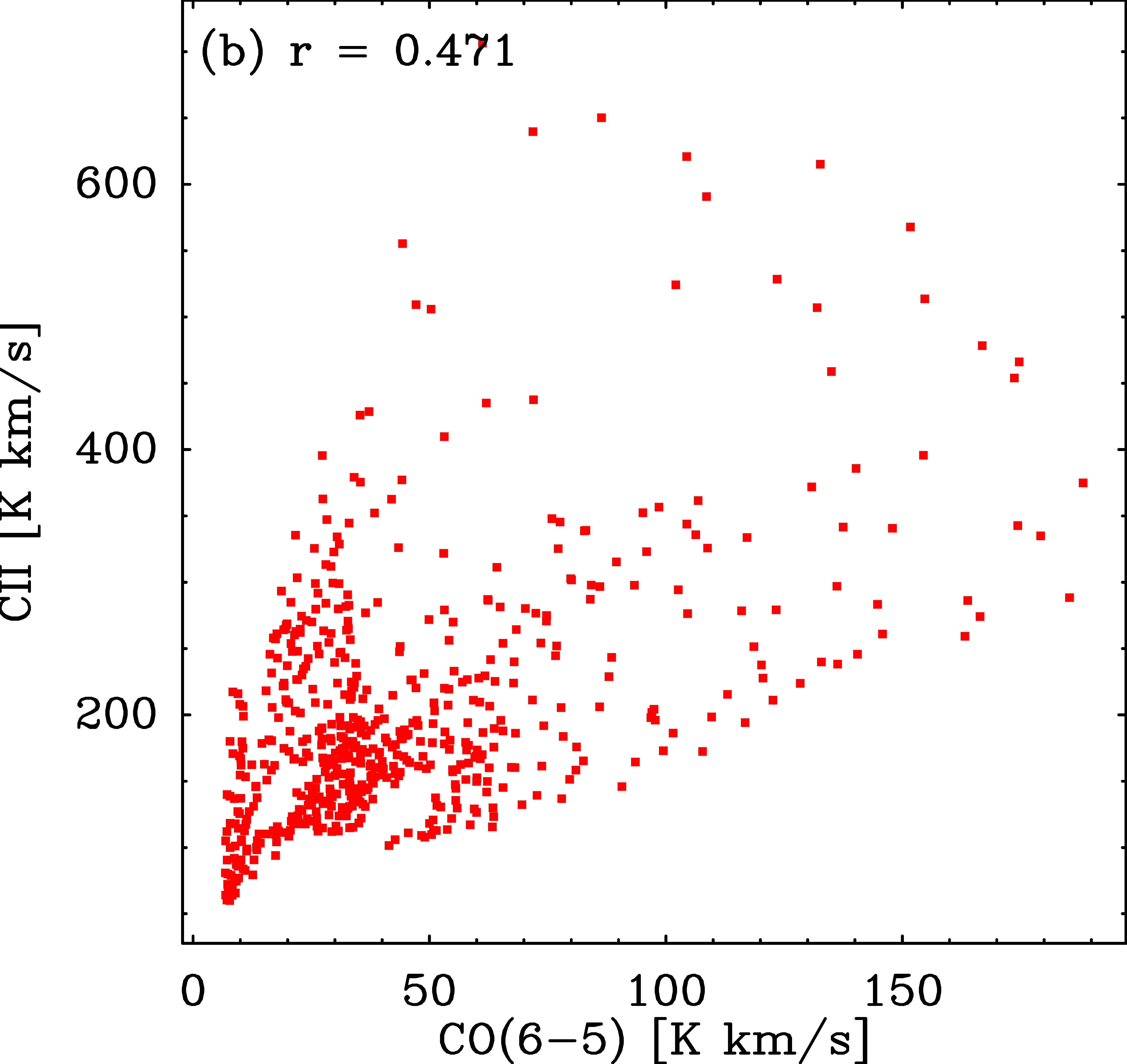}}\quad
  \subfigure{\includegraphics[width=58mm]{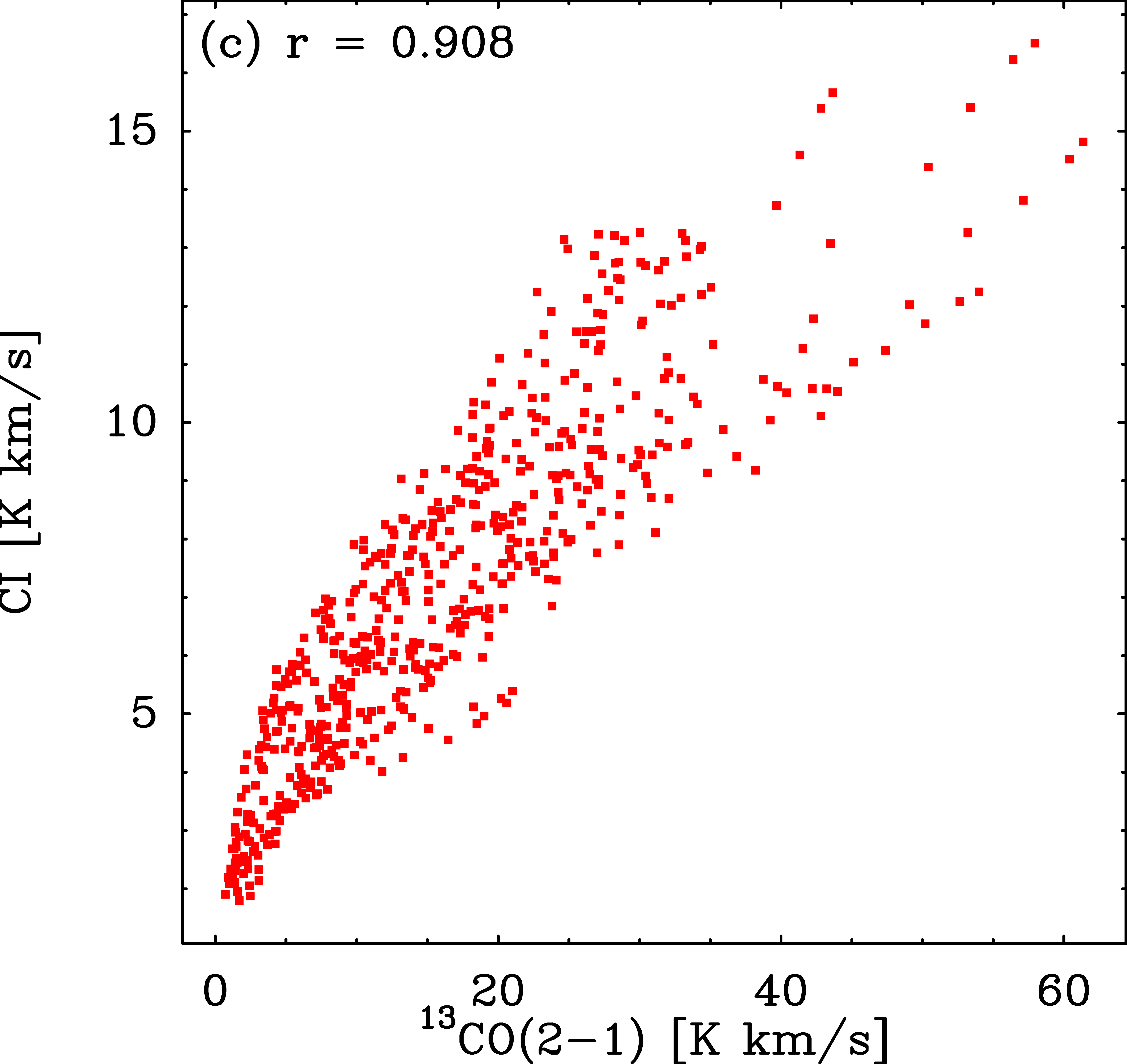}}

   \caption{Scatter plots and correlation coefficients $r$ between the velocity integrated intensity of: (a) \cii\ and \ci\ (b) \cii\ and the J = $6\to$ 5 transition of \textsuperscript{12}CO and (c) \ci\ and the J = $2\to 1$ transition of \textsuperscript{13}CO. All data points were extracted from velocity integrated intensity maps convolved to the same beam size of 31$\arcsec$.}
 \label{correlation}
\end{figure*}
  
Observations of low-$J$ \textsuperscript{12}CO, \textsuperscript{13}CO and hydrogen recombination line observations were performed with the IRAM 30 m telescope in August 2016. We observed the whole 3 mm range using the EMIR receivers \citep{2012A&A...538A..89C}. We simultaneously mapped a region of 240$\arcsec$ x 240$\arcsec$, which is similar to the size of most other maps previously observed with SOFIA and APEX, in the \textsuperscript{12}CO and \textsuperscript{13}CO $J=1\to 0$ transitions (Table 1) and the hydrogen recombination lines H40$\alpha$ to H43$\alpha$ at 99.023 GHz, 92.034 GHz, 85.688 GHz and 79.912 GHz, respectively. Molecular high density tracers, also detected in our wide spectral band observation, will be analyzed in a subsequent paper.

    All maps shown in Fig.~4 were observed in OTF total power mode centered on Her 36. Each subscan lasted 25 s and the integration time on the off-source reference position was 5 s. The offset position relative to the centre at (30$\arcmin$, $-$30$\arcmin$) was similar to that used for the SOFIA and APEX mapping and the pointing accuracy ($<$~3$\arcsec$) was maintained by pointing at the bright calibrator 1757-240 every 1 -- 1.5 hrs. A forward efficiency $F_{\rm eff}$ = 0.95 and a beam coupling efficiency $B_{\rm eff}$ = 0.69 were adopted for the EMIR receivers. These values were taken from the latest (2015) commissioning report\footnote{www.iram.es/IRAMES/mainWIKI/IRAM30mEfficiencies}.\

All data reduction was done using the CLASS and MIRA programs that are a part of the GILDAS\footnote{www.iram.fr/IRAMFR/GILDAS/} software package and all the observations are summarized in Table~1.

\section{Results}

\begin{table}[ht]
\centering
\caption{$^{12}$CO, $^{13}$CO, \ci\ and \cii\ line parameters.}
\begin{threeparttable}
\begin{tabular}{c c c c }
 \hline\hline
 \noalign{\smallskip}
 Offset ($\arcsec$)\tnote{a} & $V$ (km s$^{-1}$) & $\Delta V$ (km s$^{-1}$) & $T_{\rm peak}$\tnote{b} (K)\\
 \hline
   \noalign{\smallskip}
    \multicolumn{4}{c}{\textsuperscript{12}CO $J$ = 6 $\to$ 5}\\
 \hline
 \noalign{\smallskip}
 ($-$40, 35) & 10.33 (0.06) & 3.62 (0.16) & 77.24 \\
 \multirow{2}{*}{($-$13, 8)} & 5.92 (0.10) &3.30 (0.26) & 58.41 \\ &10.43 (0.04) & 2.92 (0.10) & 134.43  \\ 
 \multirow{2}{*}{(0, 0)} &6.39 (0.03)  & 2.42 (0.07) & 45.67 \\ &10.42 (0.02) & 3.58 (0.04) & 97.94\\
 (30, $-$2)& 10.28 (0.09) &4.73 (0.20)  &62.60 \\
 (60, 27) & 11.49 (0.07) &2.51 (0.17)  & 53.77  \\
 \hline
 \noalign{\smallskip}
 \multicolumn{4}{c}{\textsuperscript{13}CO $J$ = 1 $\to$ 0}\\
 \hline
 \noalign{\smallskip}
 ($-$40, 35) & 8.47 (0.04) & 3.00 (0.11) & 8.26  \\
 ($-$13, 8) & 8.76 (0.03) & 2.58 (0.08) & 12.61\\
 (0, 0) & 8.91 (0.01) & 2.68 (0.03) & 15.16\\
 (30, $-$2) & 9.37 (0.03) & 2.48 (0.09) & 11.15\\
 (60, 27) & 10.48 (0.02) & 1.45 (0.05) & 8.46 \\
 \hline
 \noalign{\smallskip}
 \multicolumn{4}{c}{\textsuperscript{12}C $^{3}P_{1}$ $\to$ $^{3}P_{0}$}\\
 \hline
 \noalign{\smallskip}
  ($-$40, 35) & 9.40 (0.63) & 5.91 (2.16) & 2.84 \\
 ($-$13, 8) & 9.47 (0.25) & 3.63 (0.63) & 5.62 \\ 
 (0, 0) & 9.89 (0.18) & 3.86 (0.47) & 7.65 \\
 (30, $-$2)& 10.05 (0.30) & 4.01 (0.64) & 5.45 \\
 (60, 27) & 11.80 (0.09) &0.68 (0.35)  & 5.85 \\
 \hline
 \noalign{\smallskip}
  \multicolumn{4}{c}{\textsuperscript{12}C\textsuperscript{+} $^{2}P_{3/2}$ $\to$ $^{2}P_{1/2}$ }\\
  \hline
   \noalign{\smallskip}
  \multirow{2}{*}{($-$40, 35)} & 5.40 (0.14) & 2.84 (0.30) & 33.45\\ &9.76 (0.11) &3.80 (0.31) & 45.05\\
 \multirow{2}{*}{($-$13, 8)} & 4.92 (0.10) &3.18 (0.25)  &43.04\\ &9.71 (0.04) &2.89 (0.10) & 110.50\\ 
 \multirow{2}{*}{(0, 0)} &5.12 (0.23)  &3.31 (0.57)  &24.97\\ &9.94 (0.06) &4.11 (0.14) &114.49\\
 \multirow{2}{*}{(30, $-$2)}&3.88 (0.16)  &4.01 (0.36)  &37.66\\ &9.72 (0.07) &4.86 (0.16) &100.10\\
 (60, 27) &10.22 (0.04)  &3.34 (0.09)  &112.83  \\ 
  \hline\hline
 \noalign{\smallskip}
 
  \end{tabular}
 \label{table:table}
 \begin{tablenotes}
 \item[a] The reference position is that of Her 36.
  \item[b] In units of main beam brightness temperature.
  \end{tablenotes}
\end{threeparttable}
\end{table}

\subsection{Peak intensities of the molecular line emission}

The maxima of the distributions of the velocity integrated intensities of the emission in the \ci\ and \cii\ lines and various transitions of $^{12}$CO, $^{13}$CO and C$^{18}$O are presented in Table~1. Fig.~2 shows velocity integrated intensity maps of the \textsuperscript{12}CO $J=11 \to 10$, $13 \to 12$ and $16 \to 15$ transitions. The emission in all the lines has a similar spatial distribution and peaks are found at about the same offset position ($\Delta \alpha = 5.0\arcsec$, $\Delta \delta = 5.0\arcsec$) north-west of Her 36.\ 

%\todo[inline]{FWY: the order of the maps in Fig. 3 is strange. First show the three 2-1 maps, then 3-2, 4-3, ci and in the last row 6-5 and 7-6.}
Figs.~3 and 4 show velocity integrated intensity maps of low and mid-$J$ transitions of \textsuperscript{12}CO, \textsuperscript{13}CO and C\textsuperscript{18}O, i.e. the $J$ = 1 $\to$ 0, 2 $\to$ 1, 3 $\to$ 2, 6 $\to$ 5, 7 $\to$ 6 transitions of \textsuperscript{12}CO, the $J$ = 1 $\to$ 0, 2 $\to$ 1, 4 $\to$ 3 and 6 $\to$ 5 transitions of \textsuperscript{13}CO and the $J$ = 1 $\to$ 0 transition of C\textsuperscript{18}O. The intensities of the low-$J$  transitions peak close to Her 36 ($\Delta \alpha = 0.0\arcsec$, $\Delta \delta = 0.0\arcsec$), for $^{12}$CO mid-J transitions, the peaks shift towards the north-west of Her 36 with offsets of ($\Delta \alpha = -13.0\arcsec$, $\Delta \delta = 8.0\arcsec$). It seems like there is a systematic shift in the peak emission of CO transitions with low-$J$ peaking near Her 36, mid-$J$ peaking towards the north-west, while high-$J$ lines peak again closer to Her 36. Nevertheless, all maps show at least a small offset towards the north-west and the emission from CO transitions becomes more and more compact with increasing $J$.\

Figs.~2 and 3 show velocity integrated intensity maps of the \cii\ $\textsuperscript{2}P$\textsubscript{3/2} $\to$ $\textsuperscript{2}P$\textsubscript{1/2} and \ci\ \textbf{$\textsuperscript{3}P$\textsubscript{1} $\to$ $\textsuperscript{3}P$\textsubscript{0}} transitions. \ci\ peaks at Her 36 and is very bright towards the north-west of it. \cii\ peaks at an offset of ($\Delta \alpha = 30.6\arcsec$, $\Delta \delta = -1.6\arcsec$), which is towards the east of Her 36 and the emission extends even further. This extended emission comes from the part of the HII region that is illuminated by the stellar system 9 Sgr \citep{2008hsf2.book..533T}.\

Fig.~4 shows a velocity integrated intensity map of an average of the H40$\alpha$ to H43$\alpha$ hydrogen recombination lines. We have taken the average in order to obtain a better signal to noise ratio. The distribution or the radio recombination line emission agrees well with the 5~GHz continuum VLA interferometric map in Fig.~4 of \citet{1986AJ.....91..870W} and the peak of the H$\alpha$ lines is at ($\Delta \alpha = 10.0\arcsec$, $\Delta \delta = 9.0\arcsec$), close to the center of the Hourglass nebula which indicates the presence of hot ionized gas in the east of Her 36.

\begin{figure*}[htp]
  \centering
  \subfigure{\includegraphics[width=150mm]{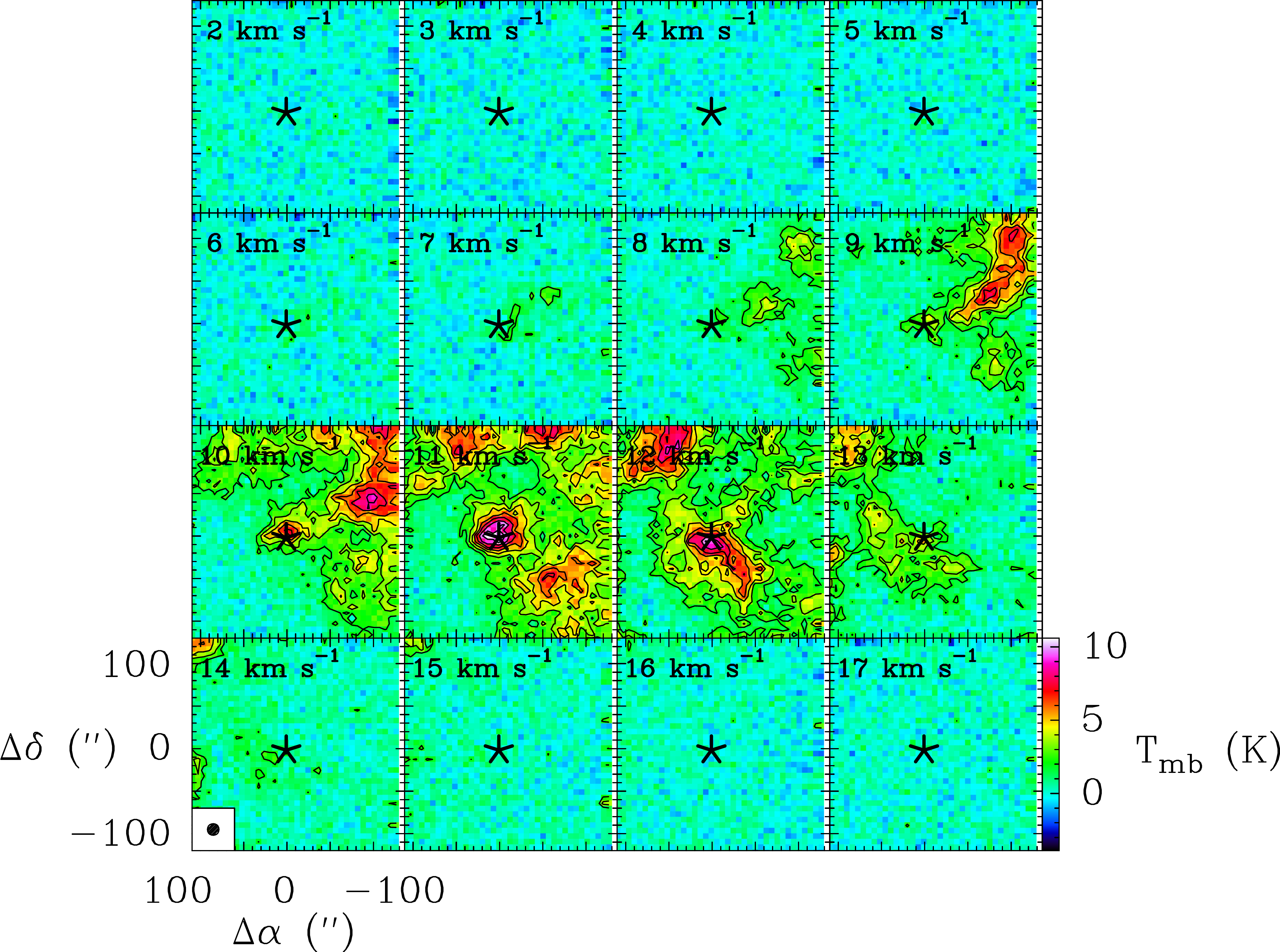}}
  \subfigure{\includegraphics[width=150mm]{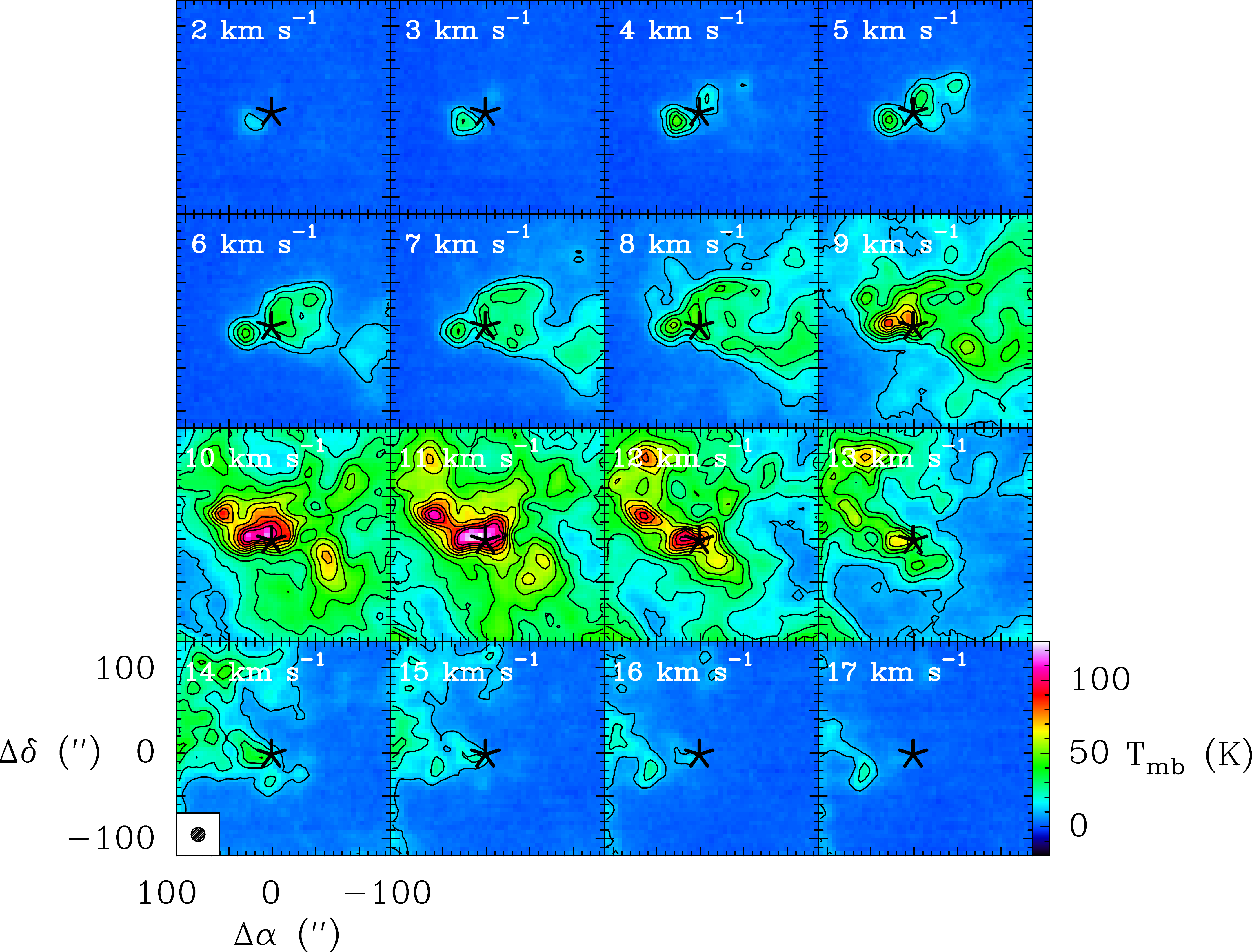}}

  \caption{Velocity channel maps of the $^{3}P_{1}$ $\to$ $^{3}P_{0}$ transition of \ci\ (upper 16 panels) and the $^{2}P_{3/2}$ $\to$ $^{2}P_{1/2}$ transition of \cii\ (lower 16 panels) in a range of 2 -- 17 km s$^{-1}$ with a channel width of 1 km s$^{-1}$ toward Her 36 which is the central position ($\Delta\alpha$ = 0, $\Delta\delta$ = 0) at R.A.(J2000) = 18\textsuperscript{h}03\textsuperscript{m}40.3\textsuperscript{s} and Dec.(J2000) = $-$24$\degree$22$\arcmin$43$\arcsec$, marked with a black asterisk. The contour levels of \ci\ are 3 $\times$ rms in steps of 2 $\times$ rms while those of \cii\ are from 10$\%$ ($>$ 3 $\times$ rms) to 100$\%$ in steps of 10$\%$ of the corresponding peak emission. All maps are plotted using original beam sizes shown in the bottom left of both panels.}
 
\end{figure*}

\begin{figure*}[htp]
  \centering
 % \subfigure{\includegraphics[width=80mm]{m8-agaln.pdf}}\quad
 % \subfigure{\includegraphics[width=80mm]{m8-vlan.pdf}}
 % \subfigure{\includegraphics[width=80mm]{m8-wise1n.pdf}}\quad
%\subfigure{\includegraphics[width=80mm]{m8-wise2n.pdf}}
 
 \subfigure{\includegraphics[width=0.48\textwidth]{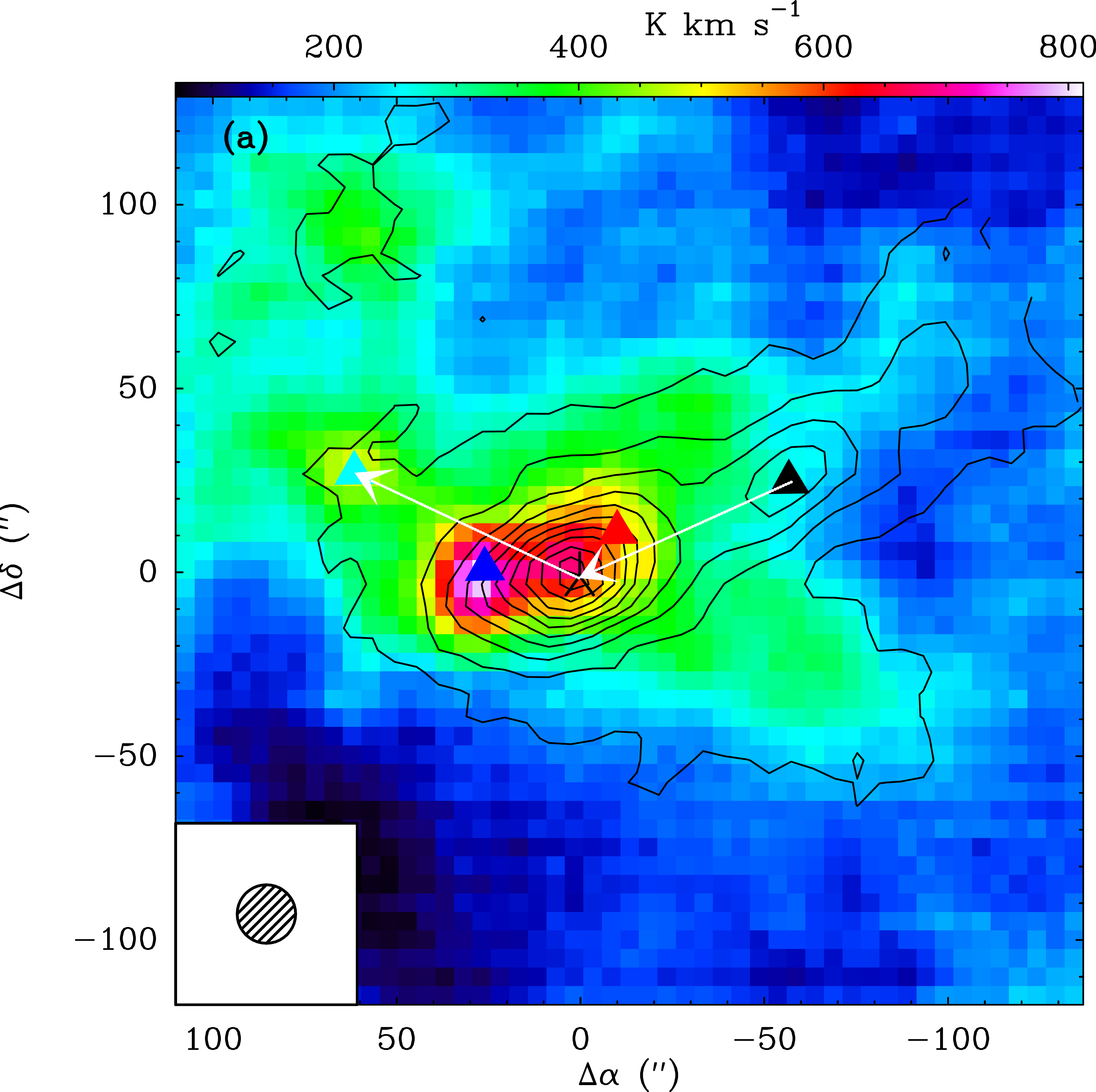}}\quad
  \subfigure{\includegraphics[width=0.48\textwidth]{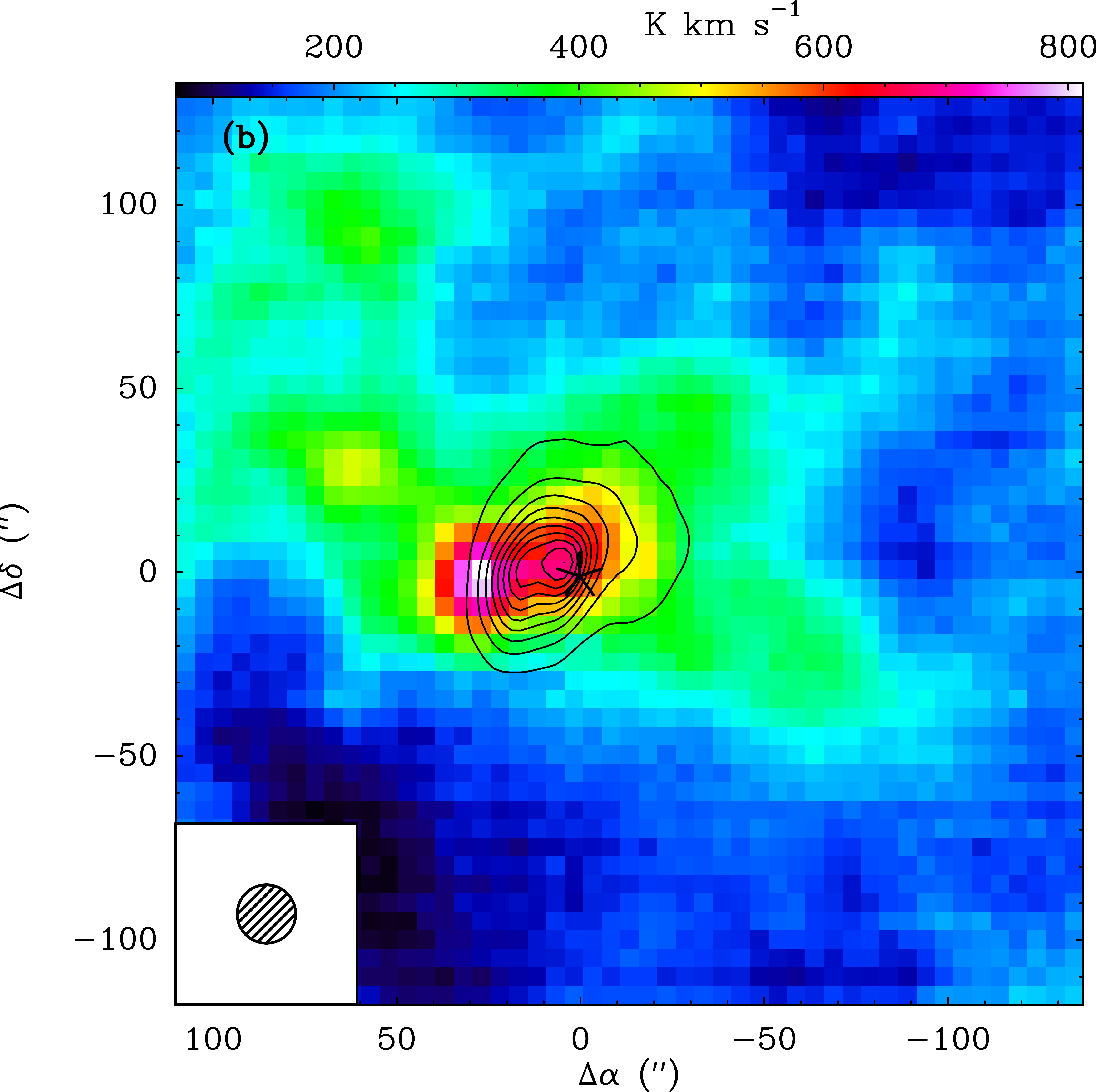}}
  \subfigure{\includegraphics[width=0.48\textwidth]{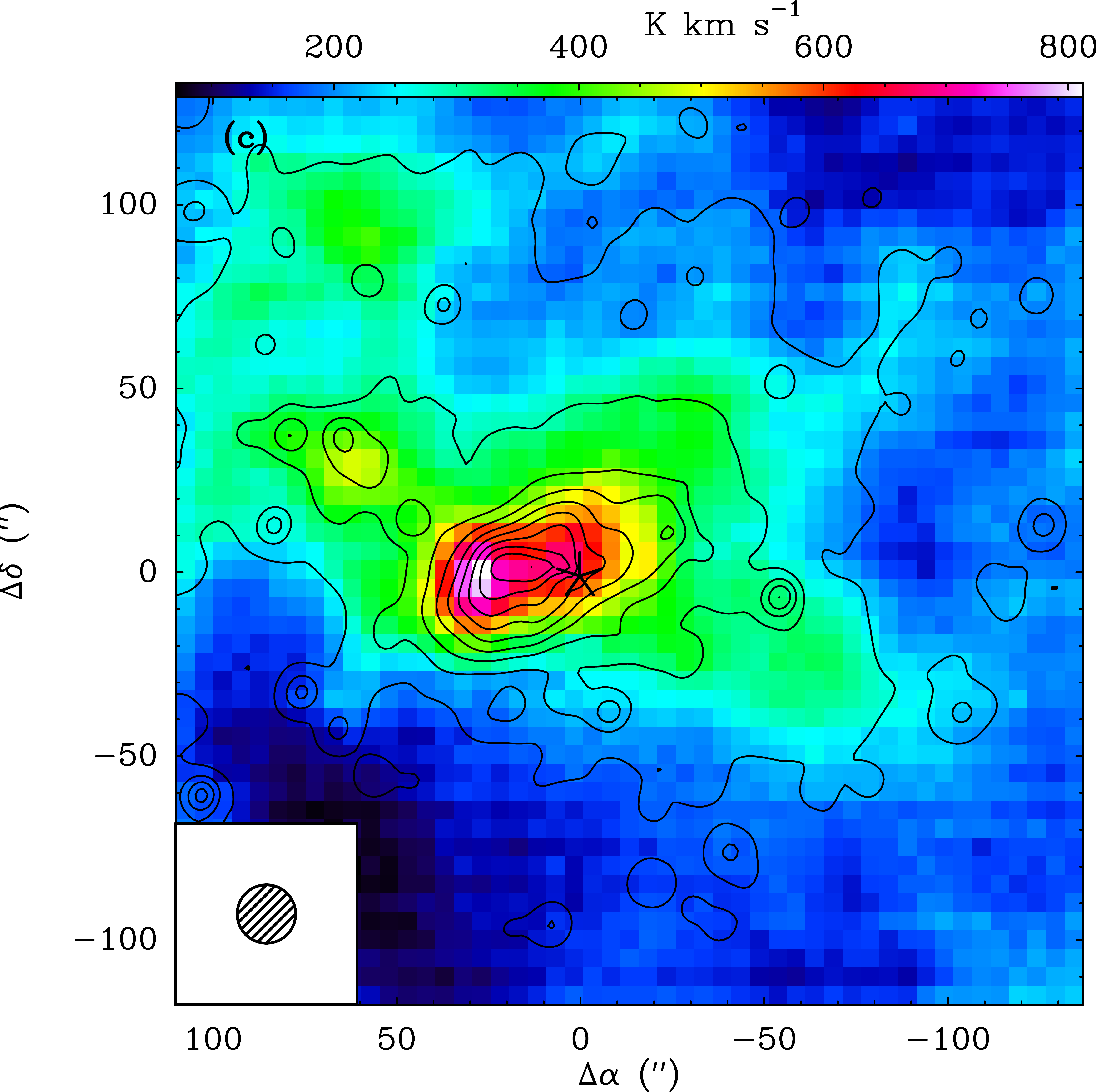}}\quad
\subfigure{\includegraphics[width=0.48\textwidth]{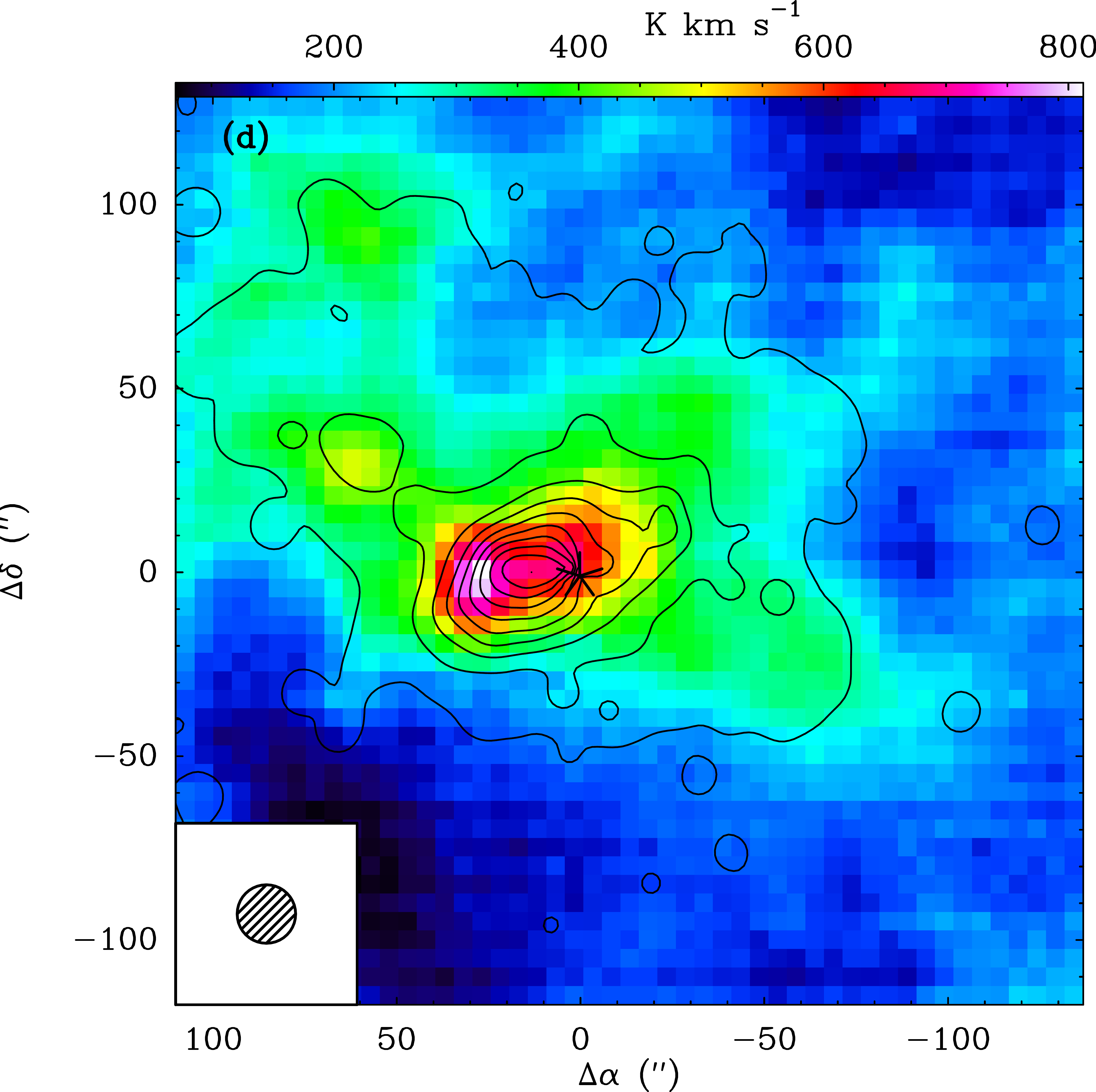}}
 
  \caption{A \cii\ velocity integrated intensity map overlaid with contours of: (a) ATLASGAL 870 $\mu$m continuum emission. Here, important offset positions discussed in the text are marked with triangles of different colors: black representing the secondary ATLASGAL peak deep inside the molecular cloud at ($\Delta \alpha = $-$53.0\arcsec$, $\Delta \delta = 23.0\arcsec$), red representing the emission peak of mid-$J$ transitions of \textsuperscript{12}CO and \textsuperscript{13}CO at ($\Delta \alpha = $-$13.0\arcsec$, $\Delta \delta = 8.0\arcsec$), blue representing the peak of the \cii\ 158 $\mu$m emission at ($\Delta \alpha = 30.0\arcsec$, $\Delta \delta = $-$2.0\arcsec$) and light blue representing the clump to the east of Her 36 observed in the channel maps of \cii\ at ($\Delta \alpha = 60.0\arcsec$, $\Delta \delta = 27.0\arcsec$). The white arrows point along the molecular cloud in the west to the HII regions in the east of Her 36; (b) VLA 1.3 cm free-free continuum emission; (c) WISE 3.4~$\mu$m and (d) WISE 4.6~$\mu$m mid-infrared continuum. Her 36 is the central position marked with an asterisk. The contour levels are 5\% to 95\% in steps of 10\% of the peak emission for (a) and 10\% to 100\% in steps of 10\% of the peak emission for (b), (c) and (d).}
\label{fig:ancillary_data} 
\end{figure*}

\subsubsection{Correlation between \textsuperscript{12}CO, \textsuperscript{13}CO, \ci\ and \cii\ }

As can be seen from the velocity integrated intensity maps, emission from \cii\ is spread out the most as compared to \ci, \textsuperscript{12}CO and \textsuperscript{13}CO. In order to visualize the correlation between these species, scatter plots of \cii\ vs. \ci\ , \cii\ vs. \textsuperscript{12}CO $J$ = 6 $\to$ 5 and \ci\ vs. \textsuperscript{13}CO $J$ = 2 $\to$ 1 are shown in Fig.~5. CO 6 $\to$ 5 is chosen owing to its association to the warm PDR  due to its higher upper level energy compared to low-$J$ CO transitions, while we chose $^{13}$CO 2 $\to$ 1 in particular, as its critical density is comparable to that of \ci\/. \cii\ is correlated the least with \textsuperscript{12}CO $J$ = 6 $\to$ 5. The Pearson correlation coefficient is $r$ = 0.471. Two branches appear to bud out in the upper left and in the lower right of this correlation. The upper left, where the \cii\ emission intensifies for a slowly strengthening \textsuperscript{12}CO $J$ = 6 $\to$ 5 emission, corresponds to the north-east of Her 36 where \cii\ is more extended. The lower right, where \textsuperscript{12}CO $J$ = 6 $\to$ 5 emission intensifies at a faster rate than \cii\/, corresponds to the south-west of Her 36, where \textsuperscript{12}CO $J$ = 6 $\to$ 5 is much more prominent. The correlation of \cii\ with \ci\ has a correlation coefficient of $r$ = 0.473 and again shows two different branches corresponding to different regions. The upper left, where \cii\ emission gets brighter for an almost constant \ci\ emission, corresponds to the north-east of Her 36. The branch in the lower right, similar to the situation shown in Fig.~5 (b), corresponds to the south-west of Her 36. 
%and they have a correlation coefficient of $r$ = 0.473 with a false alarm probability $p$ $\lll$ 0.0001. Two branches appear to bud out in the upper left and in the lower right of this correlation. The upper left, where the \cii\ emission intensifies for a steady \ci\ emission, corresponds to the north - east of Her 36 where \cii\ is more extended. The lower right, where \ci\ emission intensifies for an almost steady \cii\ emission, corresponds to the north-west of Her 36 where \ci\ is much more prominent.  
%The correlation of \cii\ with \textsuperscript{12}CO $J$ = 6 $\to$ 5 has a correlation coefficient of $r$ = 0.471 with a false alarm probability $p$ $\lll$ 0.0001 and again shows two different branches corresponding to different regions. In the left, where the \cii\ emission gets brighter for a slowly increasing \textsuperscript{12}CO $J$ = 6 $\to$ 5, corresponds to the north - east of Her 36 and in the lower right similar to (a), this branch corresponds to the north-west of Her 36. 
In contrast to these correlations, \ci\ is well correlated with \textsuperscript{13}CO $J$ = 2 $\to$ 1 with $r$ = 0.908. This resembles the case M17~SW, for which a correlation coefficient of \ci\ with \textsuperscript{13}CO $J$ = 2 $\to$ 1 was reported to be 0.942 \citep{2015A&A...575A...9P}.

\begin{figure*}[htp]
  \centering
  \subfigure{\includegraphics[width=90mm]{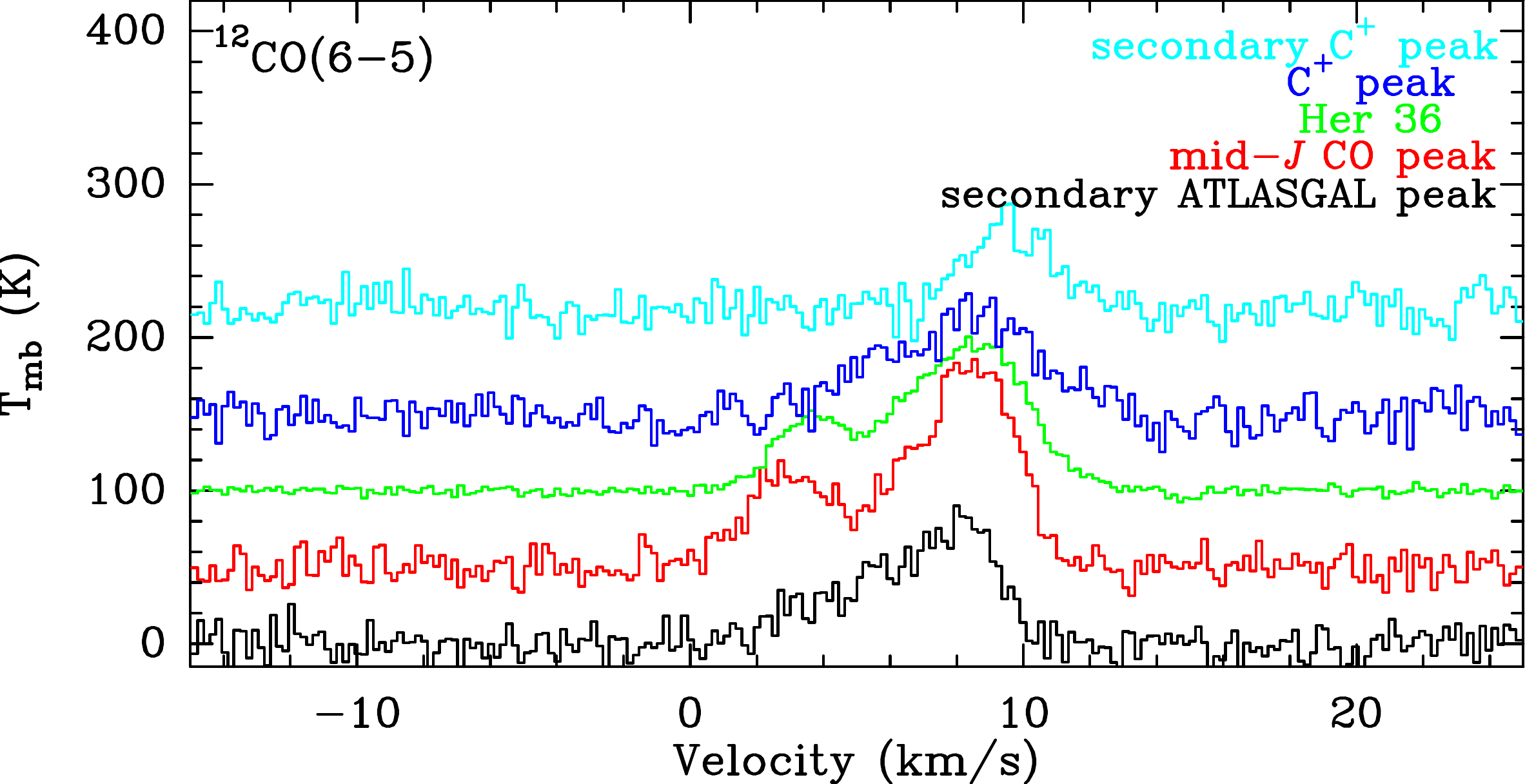}}\quad
  \subfigure{\includegraphics[width=90mm]{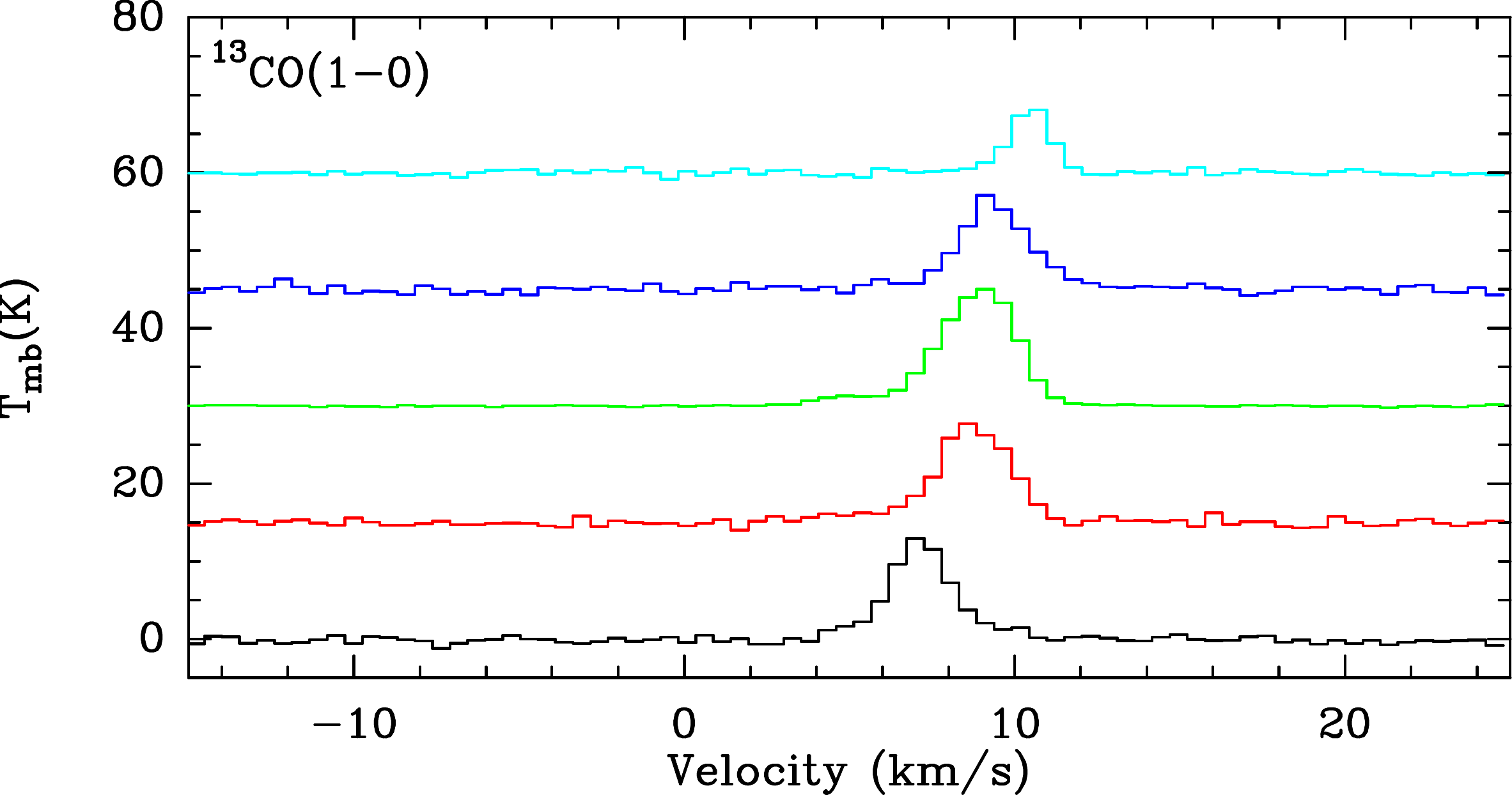}}
  \subfigure{\includegraphics[width=90mm]{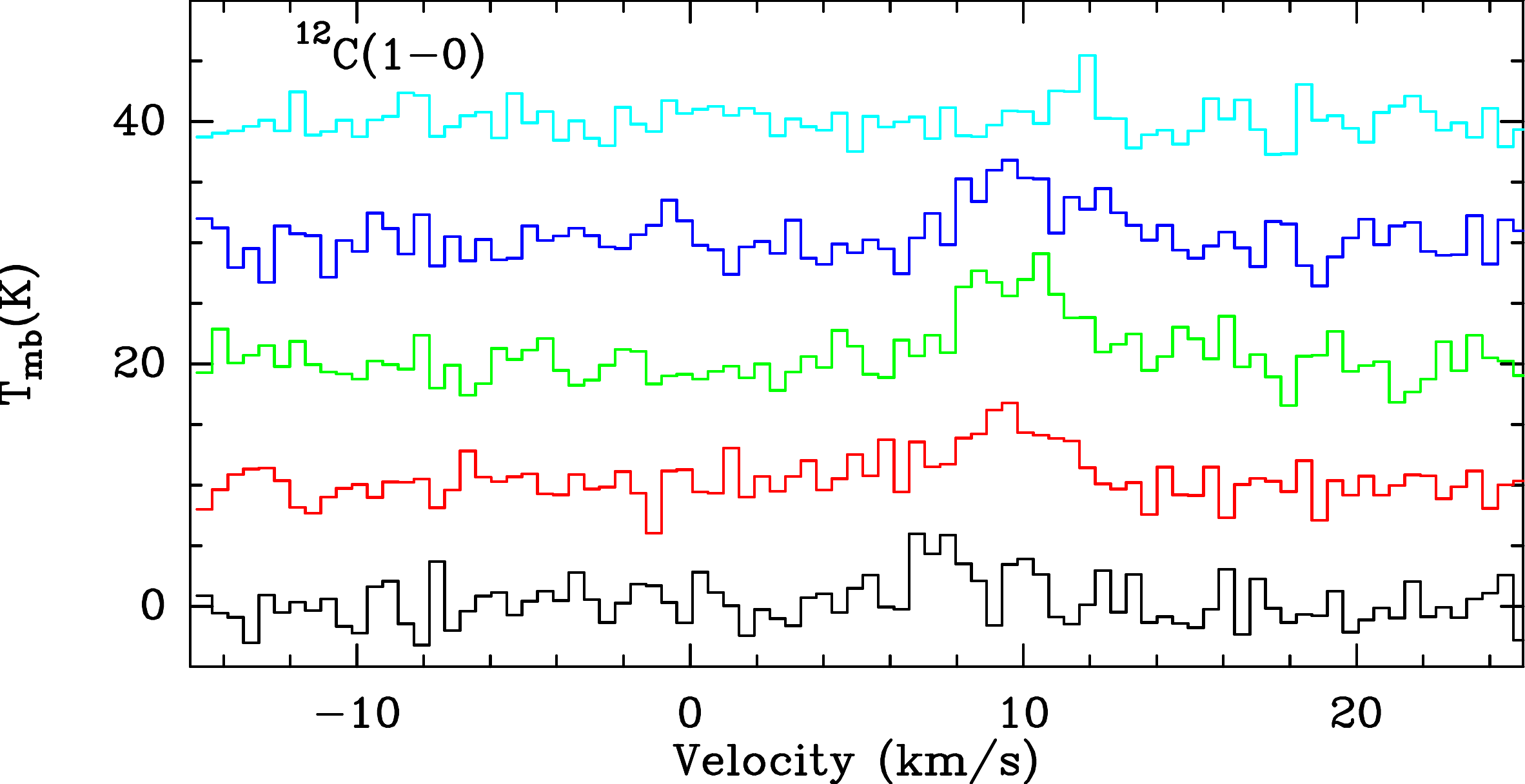}}\quad
  \subfigure{\includegraphics[width=90mm]{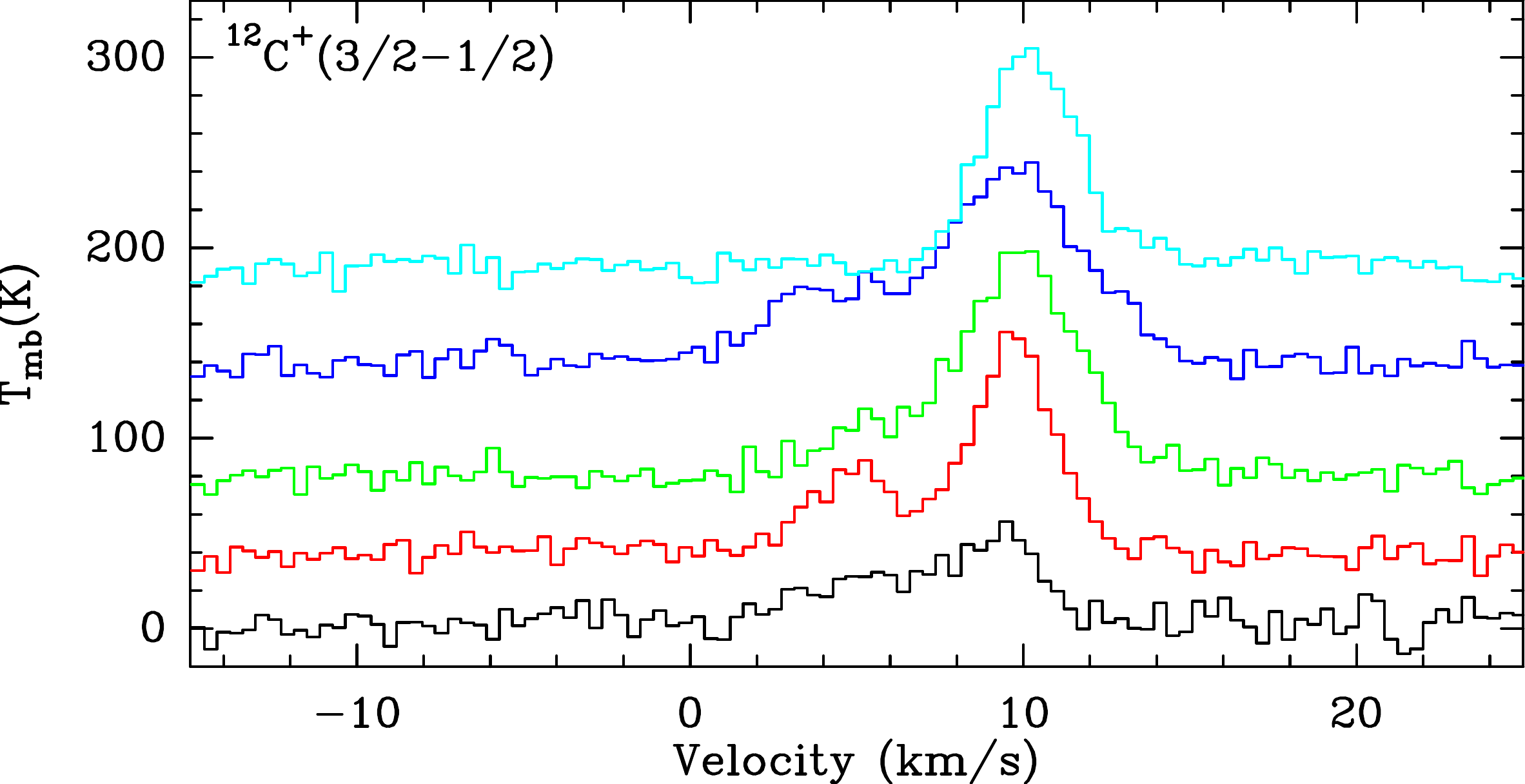}}

  \caption{Line profiles at different offsets ($\arcsec$) relative to Her 36 are shown in different colours at five positions mentioned in the upper left plot. For the detailed positions, see Sect.~3.4. Upper panels: J = 6 $\to$ 5 \textsuperscript{12}CO and J = 1 $\to$ 0 \textsuperscript{13}CO spectra, lower panels: $^{3}P_{1}$ $\to$ $^{3}P_{0}$ and $^{2}P_{3/2}$ $\to$ $^{2}P_{1/2}$ transitions of \ci\ and \cii\~. All spectra were extracted from their original beam sizes as mentioned in Table 1.}
\end{figure*}

\subsection{Channel maps}
In order to investigate the differences in the distribution of ionized and atomic carbon, channel maps of the $^{2}P_{3/2} \to ^{2}P_{1/2}$ transition of \cii\ are compared to those of the \ci\ $^{3}P_{1} \to ^{3}P_{0}$ transition. In Fig.~6, it can be seen that the emission from \cii\ (lower panels) is more spread out as compared to that from \ci\ (upper panels). In the velocity range from 2~km s$^{-1}$ to 6~km s$^{-1}$ and 15~km s$^{-1}$ to 17~km s$^{-1}$ there is no emission from \ci\ while there is emission from \cii\ close to Her 36 and towards the east of it, respectively. In the range from 7 to 9~km s$^{-1}$ both \cii\ and \ci\ emission are found towards the west. These structures extend further towards the north-east for higher velocities in the range of 12~km s$^{-1}$ to 15~km s$^{-1}$. This is quite similar to the case of M17 SW, where the \cii\ channel map shows a strong spatial association with \ci\ and CO channel maps only at intermediate 10~km s$^{-1}$ to 24~km s$^{-1}$ velocities. While at lower ($<$10~km s$^{-1}$) and higher ($>$24~km s$^{-1}$) velocity channels, \cii\ emission is mostly not associated with the other tracers of dense and diffuse gas \citep{2015A&A...575A...9P}. Notably, our \cii\ channel maps show a clumpy structure at an offset of ($\Delta \alpha = 60.0\arcsec$, $\Delta \delta = 27.0\arcsec$) which is missing in the \ci\ maps, complementing the argument that the east of Her 36 is comprised of hot gas and strong UV fields capable of ionizing carbon, i.e., it is part of an HII region. This is consistent with the H$\alpha$ and 5~GHz continuum VLA interferometric maps presented by \citet{1986AJ.....91..870W} in their figs.~1 and 4, which also have their peak intensities east from Her 36.\\
   
 \begin{figure}[htp]
    \centering
    \subfigure{\includegraphics[width=90mm]{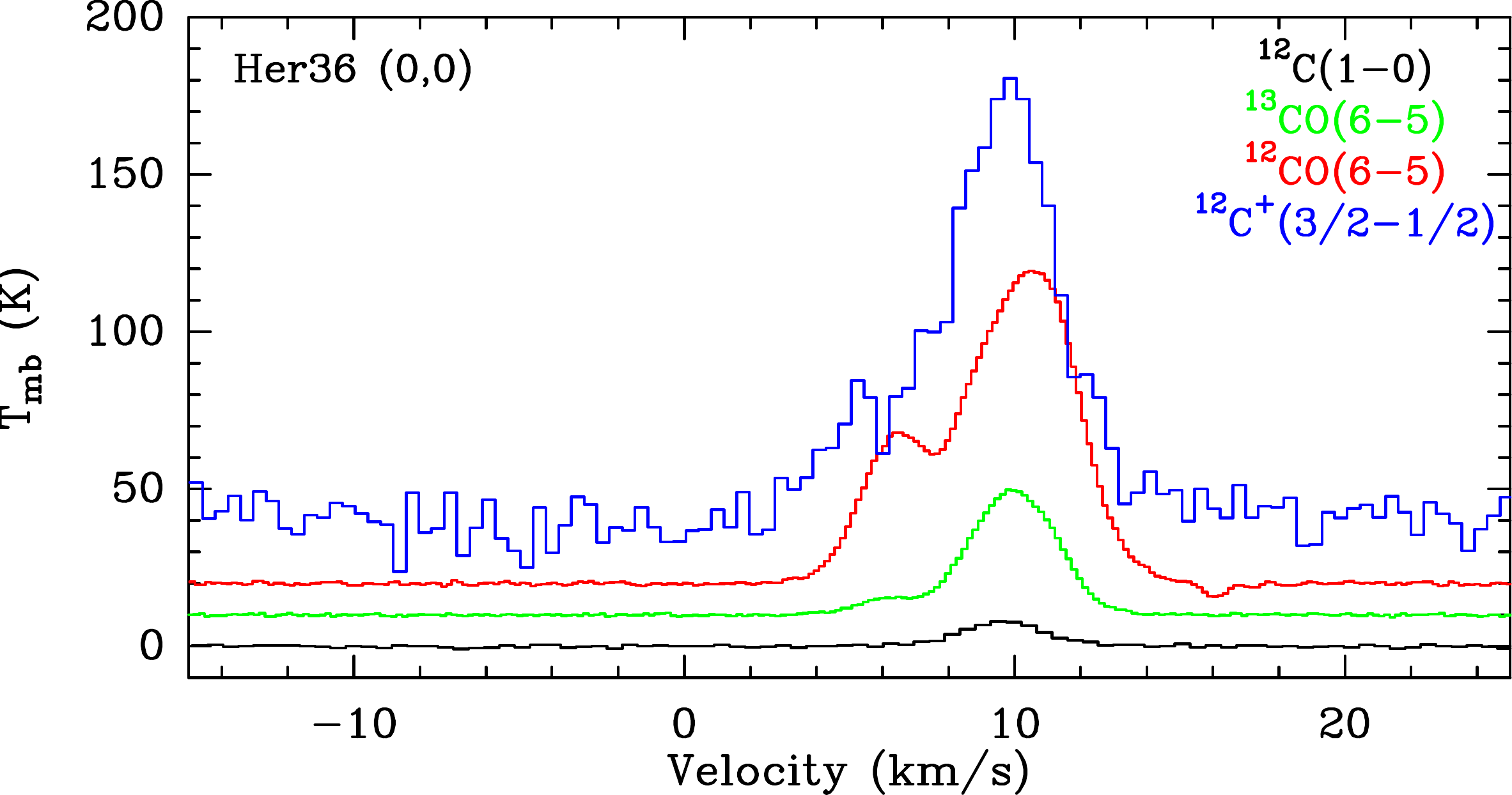}}
    \caption{Line profiles towards Her 36 for \ci\ $^{3}P_{1}$ $\to$ $^{3}P_{0}$, $^{13}$CO $J$ = 6$\to 5$, $^{12}$CO $J$ = 6$\to$ 5 and \cii\ $^{2}P_{3/2}$ $\to$ $^{2}P_{1/2}$. All spectra are extracted from maps which were convolved to the same beam size of 15$\arcsec$.}
    \label{fig:my_label}
\end{figure}

\begin{figure*}[htp]
  \centering
 
  \subfigure{\includegraphics[width=80mm]{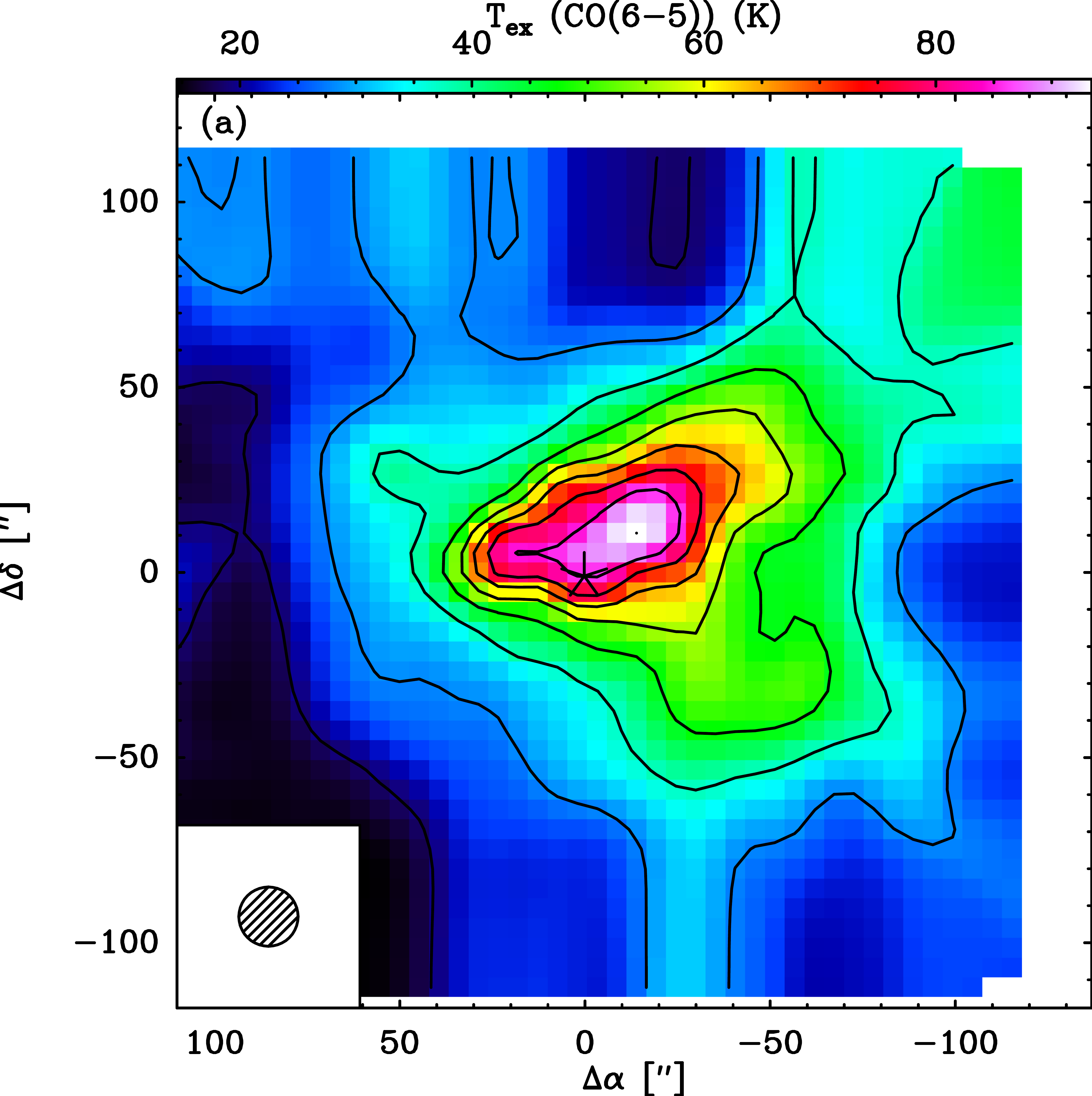}}\quad
 \subfigure{\includegraphics[width=80mm]{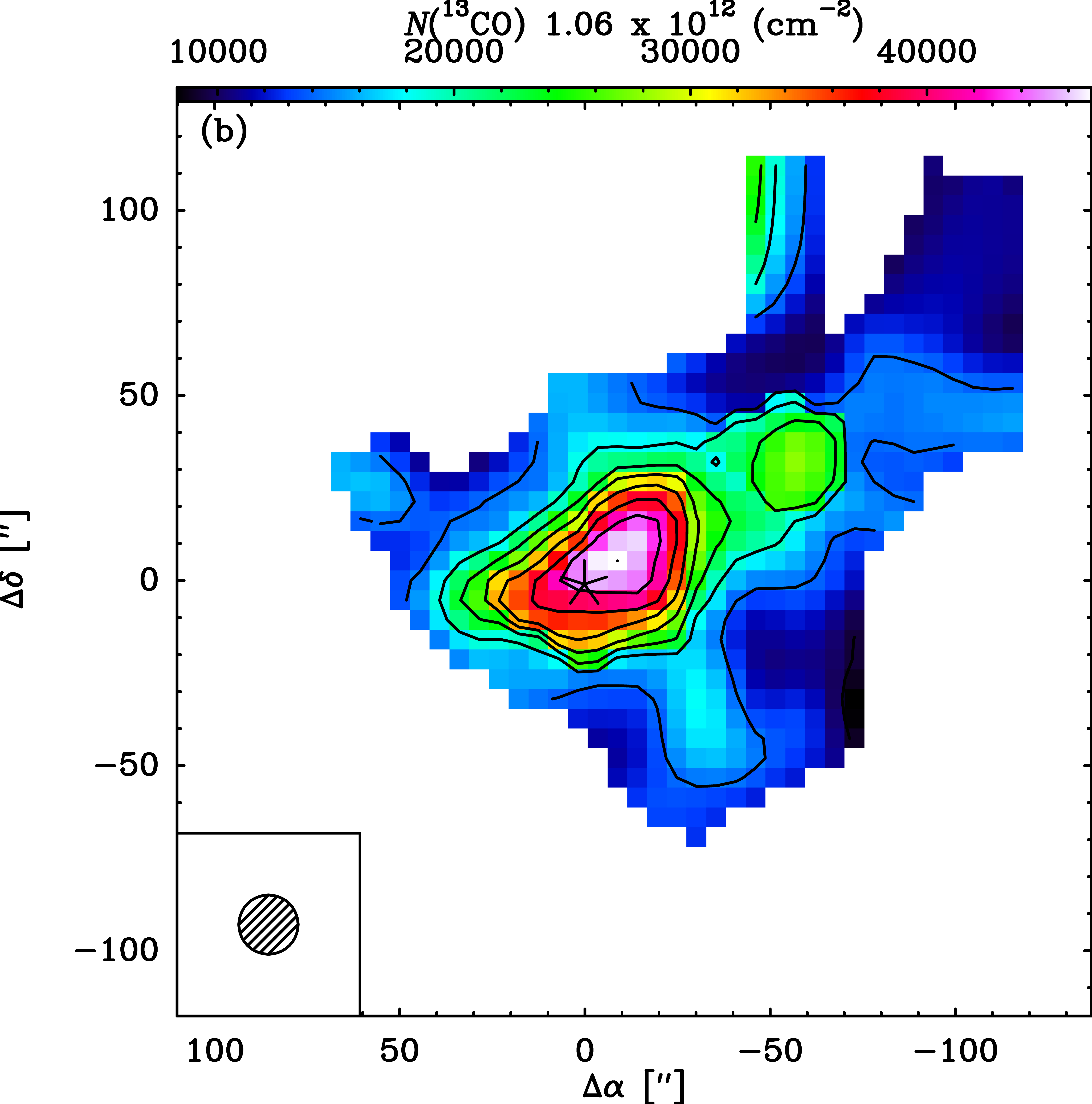}}
  \caption{For the $J$ = 6 $\to$ 5 transition (a) shows the excitation temperature of $^{12}$CO which is assumed to be equal to that of $^{13}$CO and (b) shows the column density of $^{13}$CO. The asterisk represents Her 36 which is the central position ($\Delta\alpha$ = 0, $\Delta\delta$ = 0) at R.A.(J2000) = 18\textsuperscript{h}03\textsuperscript{m}40.3\textsuperscript{s} and Dec.(J2000) = $-$24$\degree$22$\arcmin$43$\arcsec$. The contour levels are 10\% to 100\% in steps of 10\% of the corresponding peak emissions. All maps are plotted using original beam sizes shown in the bottom left of each map.}
\label{fig:tex_n13co}
\end{figure*}

 \subsection{Ancillary data}
%\todo[inline]{FWY: Can you show lower contours of the ATLASGAL map in Fig.7?
%It seems that their is weak dust emission towards the CII in the west and
%also towards the molecular clump in the north-east. E.g. add a 5\% contour.} 
For a multi-wavelength view of M8 and in order to relate our observations to the dense and cold molecular cloud and to the hot ionized gas in M8, we compared our data with observations obtained at other wavelength ranges. The surveys chosen for this comparison are: firstly, we extracted data from the 870~$\mu$m APEX Telescope Large Area Survey of the Galaxy (ATLASGAL) \citep{2009A&A...504..415S} performed with the APEX-12 m telescope using the Large APEX BOlometer CAmera (LABOCA). The dust continuum emission probes dense and cold clumps in the interstellar medium of our Galaxy. Fig.~7 (a) shows our \cii\ velocity integrated intensity map overlaid with the ATLASGAL dust continuum image that peaks at Her 36 and also traces the cold molecular cloud in the north-west. The dust emission morphology is similar to the \textsuperscript{12}CO, \textsuperscript{13}CO and C\textsuperscript{18}O distribution. Secondly, we used data from the National Radio Astronomy Observatory (NRAO)/ Very Large Array (VLA) Archive Survey (NVAS\footnote{http://archive.nrao.edu/nvas/}). Fig.~7 (b) shows the \cii\ velocity integrated intensity map overlaid with the 1.3 cm radio continuum image\footnote{NRAO/VLA Archive Survey, (c) 2005-2007 AUI/NRAO} which peaks very close to Her 36 and traces free-free emission from the HII region NGC6523/33. This compact HII region, which is also traced by the H recombination lines with the IRAM 30m telescope, is also shown in Fig.~4 (right panel). Thirdly, the Wide-field Infrared Survey Explorer (WISE) imaged the sky at four mid-infrared wavelengths. Fig.~7 (c) and (d) compare the \cii\ velocity integrated intensity map with WISE 3.4 $\mu$m (band 1) and 4.6 $\mu$m (band 2) continuum images which peak closer to the \cii\ peak.\
Overall, the mid-infrared emission that originates from hot dust shows the best agreement with the morphology seen in the \cii\ image -- both probe hot material from HII regions and warm surfaces of PDRs.
%In addition to tracing hot gas similar to VLA i.e. quite good agreement with mid IR, these bands also show the clumpy structure at an offset of ($\Delta \alpha = 60\arcsec$, $\Delta \delta = 27\arcsec$) similar to what is seen in the \cii\ velocity channel maps. Once again, these surveys indicate the presence of bright HII region towards east of Her 36 and it is quite evident from the overlays that \cii\ peak matches well with those of the hot gas tracer surveys.\\

\subsection{Spectra of \textsuperscript{12}CO, \textsuperscript{13}CO, \ci\ and \cii\ emission lines at different offsets}

Fig.~8 shows a comparison between the spectra of \textsuperscript{12}CO, \textsuperscript{13}CO, \ci\ and \cii\ emission lines at different offsets relative to Her 36. Line parameters of Gaussian fits to profiles are reported in Table~2. In several cases the profiles show evidence of two velocity components, which were fit separately. The \textsuperscript{12}CO $J = 6\to 5$ and \textsuperscript{13}CO $J = 1\to 0$ transitions are representative of the general appearance of all \textsuperscript{12}CO and \textsuperscript{13}CO line profiles discussed in this paper. The different offsets were chosen along a curved line from the molecular cloud in the west to the east of Her 36 (see Fig.~\ref{fig:ancillary_data} (a)): the secondary C$^{+}$ peak that corresponds to the clump observed in the channel map of \cii\ at ($\Delta \alpha = 60.0\arcsec$, $\Delta \delta = 27.0\arcsec$), the C$^{+}$ peak which is the emission peak of the $^{2}P_{3/2}$ $\to$ $^{2}P_{1/2}$ transition of \cii\ at ($\Delta \alpha = 30.0\arcsec$, $\Delta \delta = -2.0\arcsec$), Her 36 is located at ($\Delta \alpha = 0.0\arcsec$, $\Delta \delta = 0.0\arcsec$), the mid-$J$ CO peak which is the mid-$J$ transition emission peak of \textsuperscript{12}CO and \textsuperscript{13}CO at ($\Delta \alpha = -13.0\arcsec$, $\Delta \delta = 8.0\arcsec$) and secondary ATLASGAL peak arises from deep into the molecular cloud to the west traced by ATLASGAL at ($\Delta \alpha = -53.0\arcsec$, $\Delta \delta = 23.0\arcsec$).\\

A lower velocity (2 km~s$^{-1}$ -- 6 km~s$^{-1}$) component is spectrally resolved at several positions. The higher velocity component emission lines have blue-shifted wings in the molecular cloud in the west while the emission is red-shifted towards the \cii\ peak toward the east compared to their emission peaking at 9 km s$^{-1}$ toward our reference position, Her 36. Furthermore, the peak of the lines shifts from the east to the west to lower velocities. The \textsuperscript{12}CO and \textsuperscript{13}CO line profiles are similar in being most intense with broadest line widths at the the mid-$J$ transition emission peak of \textsuperscript{12}CO and \textsuperscript{13}CO ($\Delta \alpha = -13.0\arcsec$, $\Delta \delta = 8.0\arcsec$) and at Her 36 itself while getting less intense with narrower line widths at the \cii\ peak. The \ci\ line profile gets most intense with broadest line width toward Her 36 itself with almost no emission from the clumpy structure in the HII region at an offset of ($\Delta \alpha = 60.0\arcsec$, $\Delta \delta = 27.0\arcsec$). As can also be seen from the comparison of different molecular transitions at Her 36 in  Fig.~9, CO and \ci\ are not associated with \cii\ at lower and higher velocities (see also Figs. 5 and 6). This is very similar to M17 SW as reported by \citet{2015A&A...575A...9P}, where \cii\ is not associated with other gas tracers at lower and higher velocities.\\ 

%\todo[inline]{FWY: above you say ALL lines show wings. Maybe you need a figure comparing e.g. at 0,0 the line profiles of all these transitions in a single figure.}

%\textsuperscript{12}CII profile peaks get broader in the HII region which is because the photons are energetic enough here to ionize carbon easily. While, \textsuperscript{12}CO emission becomes more intense in the cooler molecular cloud.\\
     
 %\todo[inline]{FWY: You have to multiply the 13CO and CI spectra by
% a factor, otherwise the profiles cannot be seen.}

\section{Analysis}

In this section we determine the temperature and density in the PDR of M8 with several complementary methods. We start by using the data for the $J$ = $6 \to 5$ transition of CO, which has the highest angular resolution, to estimate excitation temperatures and column densities throughout the PDR as probed by the mid-$J$ CO emission.

\subsection{Excitation temperature and column density estimates}
%The radiative transfer equation when scattering processes are ignored is given by \citep{2011piim.book.....D}:
%\todo[inline]{FWY: In the References section, give the full reference to the book. For a paper this is too detailed anyway. I would remove everything before "For a constant Tex..." }
%\begin{equation} \label{eq1}
 %   \frac{dI_\nu}{ds} = -\kappa_\nu I_\nu + j_\nu 
%\end{equation}\\
%where\\
%s $\equiv$ Path of propagation along the line of sight\\
%$I_\nu$  $\equiv$ Specific intensity\\
%$\kappa_\nu$ $\equiv$ Absorption coefficient\\
%$j_\nu$ $\equiv$ Emission coefficient\\

%In terms of optical depth, the equation changes to:
%\begin{equation} \label{eq2}
%\ dI_\nu = S_\nu d\tau_\nu - I_\nu d\tau_\nu
%\end{equation}\\
%where\\
%$d\tau_\nu$ $\equiv$ optical depth\\
%$S_\nu$ $\equiv$ Source function\\

%We assume that the medium through which the radiation travels is uniform at an excitation temperature T\textsubscript{ex} and is also infinitely large so the radiation field is defined as black body: I\textsubscript{$\nu$} = B\textsubscript{$\nu$}; the source function is equivalent to the Planck function at T\textsubscript{ex}: S\textsubscript{$\nu$} = B\textsubscript{$\nu$}(T\textsubscript{ex}), which is the Local Thermodynamical Equilibrium (LTE) \citep{2016PASP..128b9201M}. 
  A detailed description of spectral line radiative transfer relevant here can be found in \citet{2011piim.book.....D} and \citet{2016PASP..128b9201M}. For a constant excitation temperature $T_{\rm ex}$, we can integrate the radiative transfer equation to obtain the observable Rayleigh-Jeans equivalent temperature $T_{\rm R}^*$ \citep[Eq.~1 in][]{2012A&A...538A..12P}. The background radiation temperature comprises the cosmic background radiation of 2.73 K and the radiation from warm dust. The latter was calculated using the results from the Spectral and Photometric Imaging Receiver (SPIRE) of the European Space Agency's (ESA) Herschel Space observatory. Data obtained in the second band at 350~$\mu$m (close to the $^{12}$CO 7 $\to$ 6 line) wavelength was used. The maximum intensity in the analyzed region around Her 36 is about 1200~MJy/sr near the dense molecular cloud which corresponds to a Rayleigh-Jeans equivalent brightness temperature of about 0.06~K from dust. Thus, the total contribution from dust and background can be neglected as it contributes $\leq$ 1$\%$ to the resulting $T_{\rm R}^*$.

%\begin{equation} \label{eq3}
%\ T_B = 1.36 \frac{\lambda^2 S}{\theta^2}
%\end{equation}\\
%where\\
%$T_B$ $\equiv$ Brightness temperature\\
%$\lambda$ $\equiv$ wavelength in cm\\
%S $\equiv$ Intensity in mJy/beam\\
%$\theta$ $\equiv$ beam solid angle in arcsec\\

%\todo[inline]{FWY: You had already done a calculation for the dust radiation of the cloud itself as background. Here you could give an estimate of the impact of neglecting this term.}

%The $J = 6\to 5$ transition of \textsuperscript{12}CO and \textsuperscript{13}CO is used to calculate the excitation temperature and column densities because of their higher spatial resolution $\sim 9.58\arcsec$. 
%In order to determine the optical depths of the $J = 6\to 5$ transitions of \textsuperscript{12}CO and \textsuperscript{13}CO, we assume that the excitation temperature for \textsuperscript{12}CO and \textsuperscript{13}CO is the same. This leads to the following relation between their optical depths: 

%\begin{equation} \label{eq2}
%    \frac{T_{\rm R}^*(^{12}\mathrm{CO})}{T_{\rm R}^*(^{13} \mathrm{CO})} \cong \frac{1 - e^{-\tau(^{12}\mathrm{CO})}}{1 - e^{-\tau(^{13}\mathrm{CO})}}\,.
%\end{equation}\\

Assuming that the excitation temperature for \textsuperscript{12}CO and \textsuperscript{13}CO is the same and \textsuperscript{12}CO is optically thick:  

%\todo[inline]{FWY: since 12co is optically thick, you don't need the abundance ratio here.}

\begin{equation} \label{eq3}
    \frac{T^*_{\rm R}(^{12}\mathrm{CO})}{T^*_{\rm R}(^{13}\mathrm{CO})} = \frac{1}{1 - e^{-\tau(^{13}\mathrm{CO})}}\,.
\end{equation}\\

The excitation temperature of the \textsuperscript{12}CO $J = 6\to 5$ transition can be estimated by further assuming a beam filling factor of unity i.e. $T_{\rm R}^*$ = $T$\textsubscript{MB}:

\begin{equation}
   T_{\rm ex} = 33.2~\bigg[\rm ln\bigg(1 +\frac{33.2}{T_{\rm MB}(^{12}\mathrm{CO})}\bigg)\bigg]^{-1} K\,, 
\end{equation}\\
where $T$\textsubscript{MB} is the main beam brightness temperature in K, estimated from the peak temperature map of $^{12}$CO $J$ = 6 $\to$ 5 transition. The resulting $T_{\rm ex}$ distribution is shown in Fig.~\ref{fig:tex_n13co} (a). Formally, we are showing lower limits to the excitation temperature due to the assumption of a beam filling factor of unity. It is highest immediately in the north-west of Her 36 and decreases with distance from the star.\\

Using the computed $T$\textsubscript{ex} and the main beam brightness temperature $T$\textsubscript{MB} estimated from the peak temperature map of $^{13}$CO $J$ = 6 $\to$ 5, the total column density (Fig.~\ref{fig:tex_n13co} (b)) of \textsuperscript{13}CO can be calculated over the complete velocity range of the source from:
\begin{equation}
     N(^{13}\mathrm{CO}) = 1.06 \times 10^{12} (T_{\rm ex} + 0.88)~exp\Bigg(\frac{116.2}{T_{\rm ex}}\Bigg) \int T_{\rm MB}(^{13}\mathrm{CO})\rm dv~cm^{-2}\,,
 \end{equation}\\
 where $T_{\rm ex}$ is in K and $T_{\rm MB}$ dv is in K km$s^{-1}$. Fig.~\ref{fig:tex_n13co} (b) shows the resulting \textsuperscript{13}CO total column density with a peak value of  $\sim$ 5 $\times$ 10\textsuperscript{16} cm\textsuperscript{-2} north-west of Her 36. This results in an H\textsubscript{2} column density $N(\rm H_{\rm 2})$ of $\sim$ 3.7 $\times$ 10$^{22}$~cm$^{-2}$ adopting an isotopic abundance ratio [\textsuperscript{12}CO/\textsuperscript{13}CO] of $\sim$ 63 \citep{2005ApJ...634.1126M} and a CO abundance ratio [\textsuperscript{12}CO/H\textsubscript{2}] of $\sim$ 8.5 $\times$ 10$^{-5}$ \citep{2010pcim.book.....T}. The mass of the warm CO gas can be computed by integrating the column density over the whole clump in a region of 3.12 arcmin$^{2}$, which results in a mass of $\sim$ 467~M\textsubscript{\(\odot\)}. Complementary to this, the cold gas mass in a region of 5.1 arcmin$^{2}$ has an estimated value of 10$^{3}$~M\textsubscript{\(\odot\)}, calculated from a flux of $\sim$ 133 Jy at 870 $\mu$m measured with ATLASGAL \citep{2009A&A...504..415S} assuming an absorption coefficient of $k_{\rm \nu}$ = 1.85 g cm$^{2}$ and a temperature of 23 K \citep{2018MNRAS.473.1059U}, and not including potential uncertainties in the choice of these values. However, these mass estimations have an error of $\sim$ 26\% which accounts for errors of $\sim$ 16\% from the distance to the star \citep{2008hsf2.book..533T} and $\sim$~20\% from the calibration.  

 \begin{figure}[htp]
    \centering
    \subfigure{\includegraphics[width=90mm]{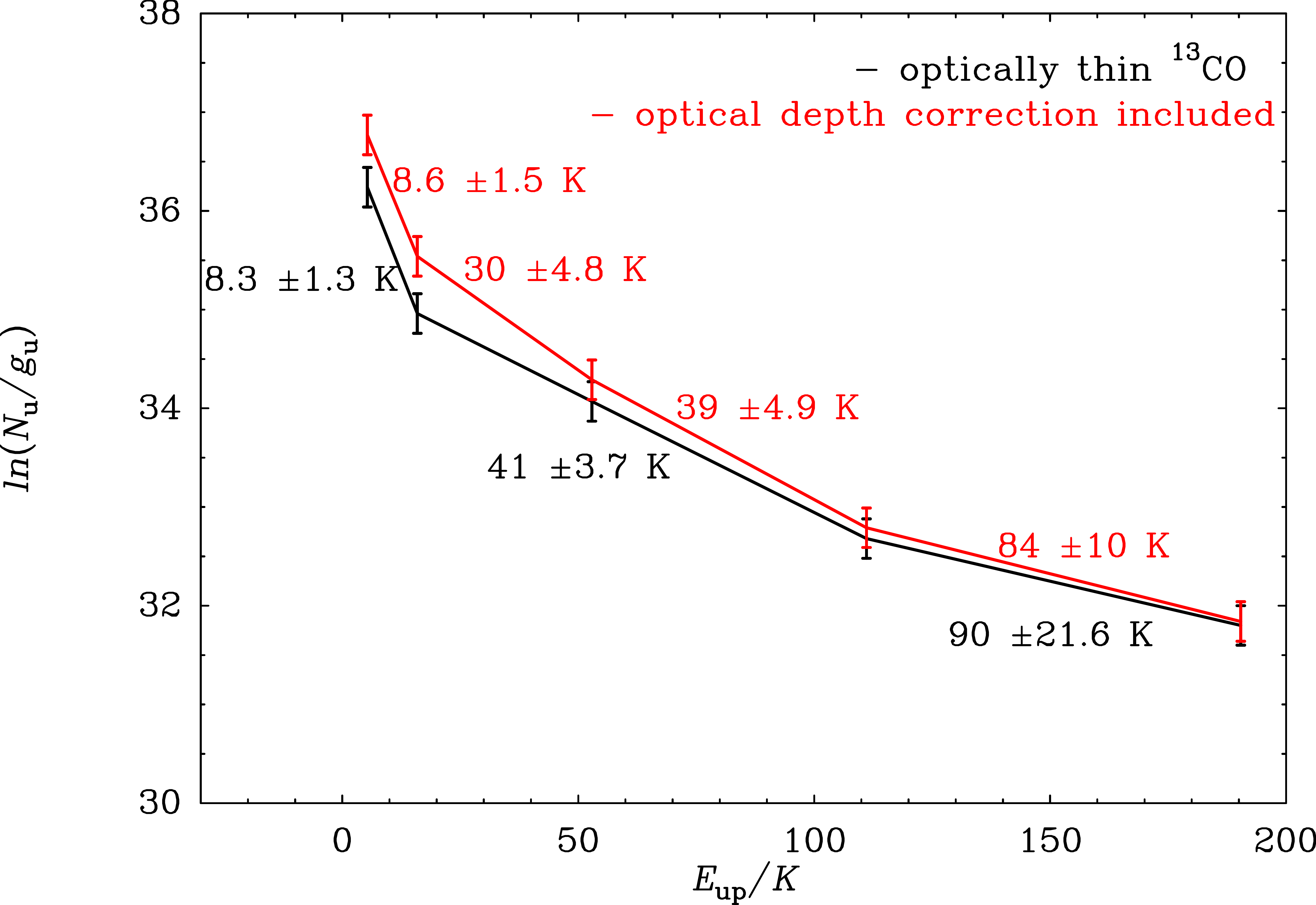}}
    \caption{The rotational diagrams of $J = 1\to 0$, $J = 2\to 1$, $J = 4\to 3$, $J = 6\to 5$ and $J = 8\to 7$ transitions of \textsuperscript{13}CO at Her 36. Shown in black is the rotational diagram when \textsuperscript{13}CO is assumed to be optically thin. In red the rotational diagram is shown including the optical depth correction factors. Rotation temperatures obtained from different slopes are indicated. Values of the velocity integrated intensities for different transitions were extracted from maps convolved to the same resolution of 31$\arcsec$. The error bars were calculated from the maximum noise level of the integrated intensities of individual transitions and from calibration uncertainties of 20\%}.
    \label{fig:rot_dia}
\end{figure}
 
  \begin{figure}[htp]
  \centering
  \subfigure{\includegraphics[width=90mm]{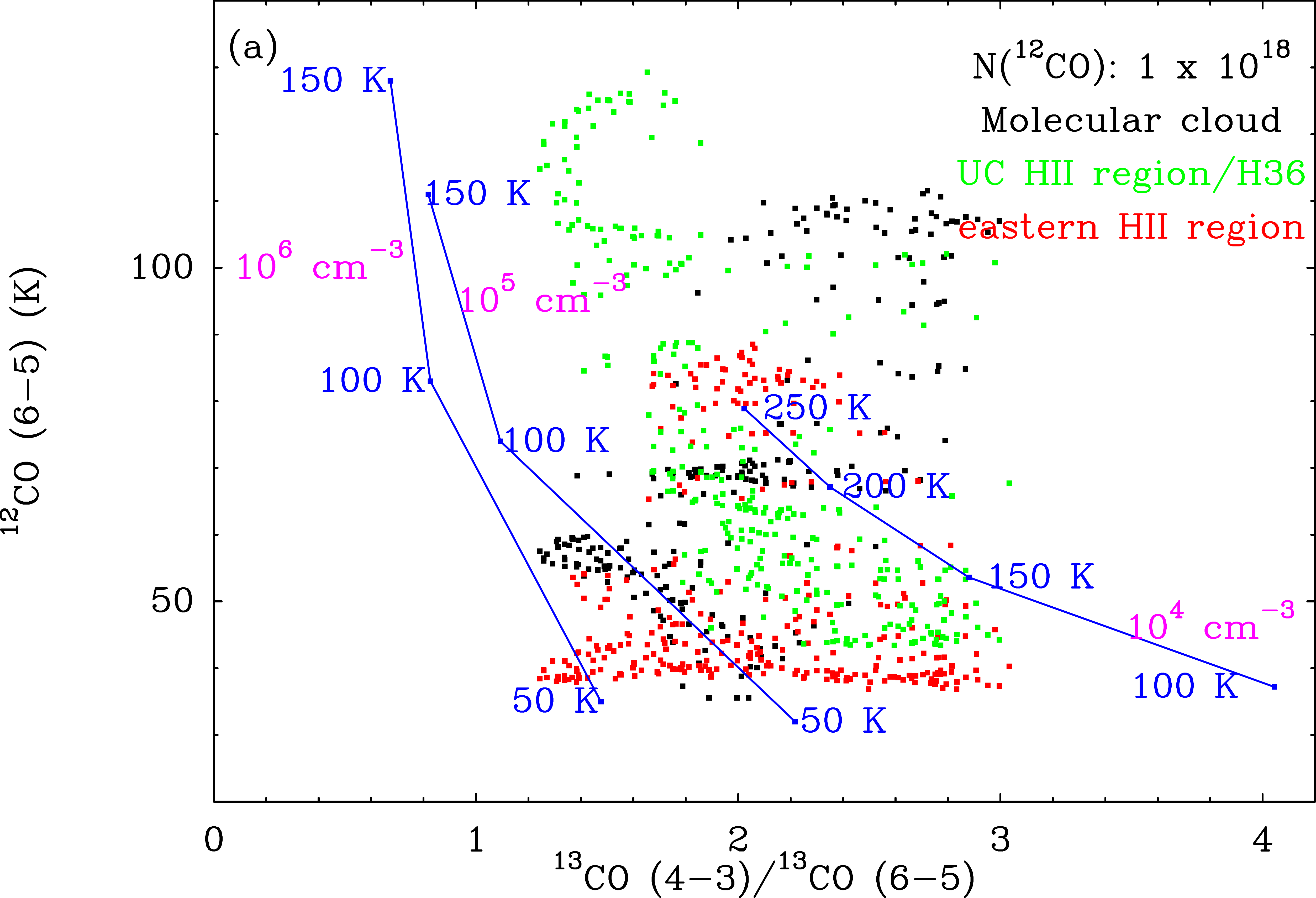}}
  \subfigure{\includegraphics[width=90mm]{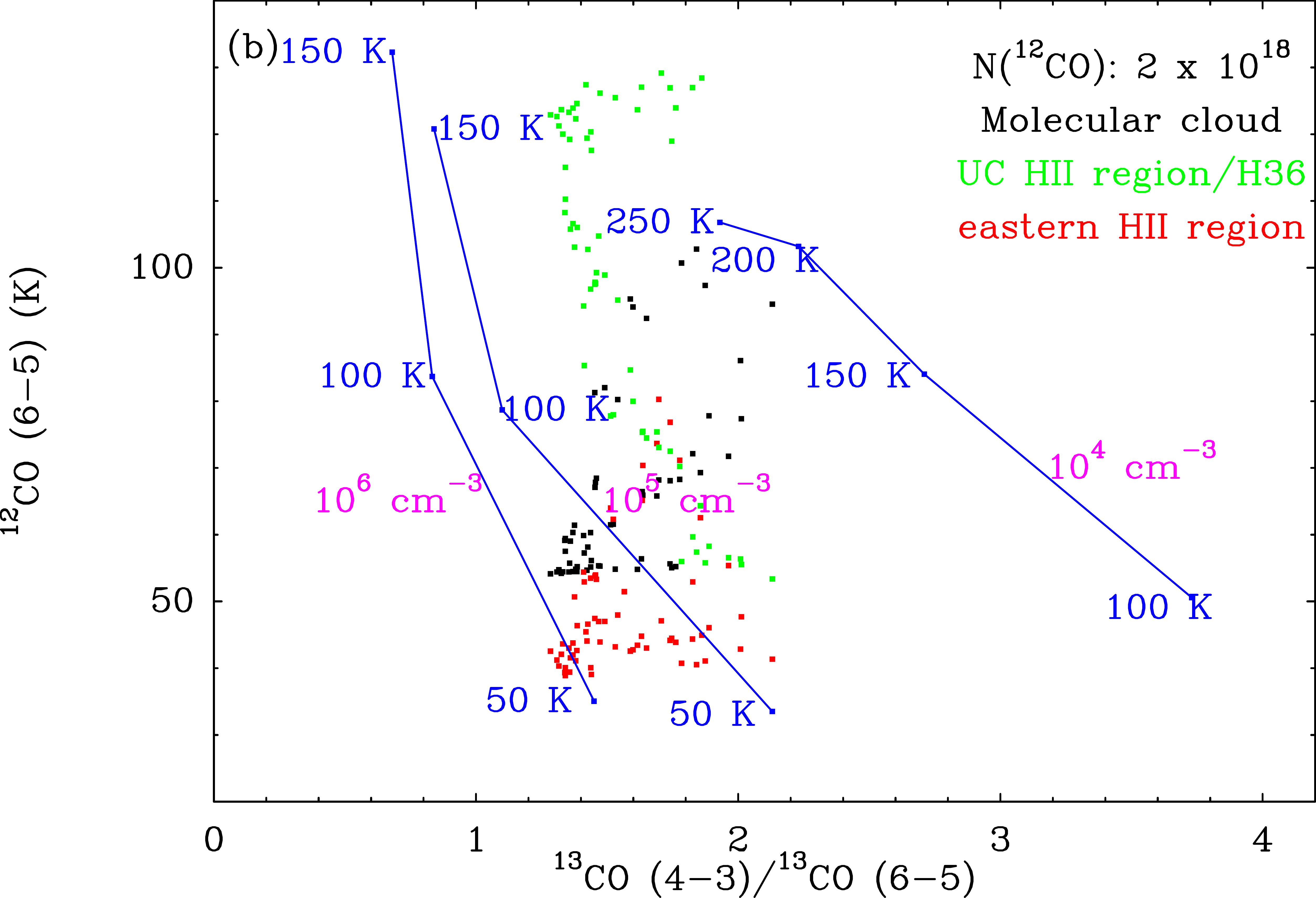}}
  \subfigure{\includegraphics[width=90mm]{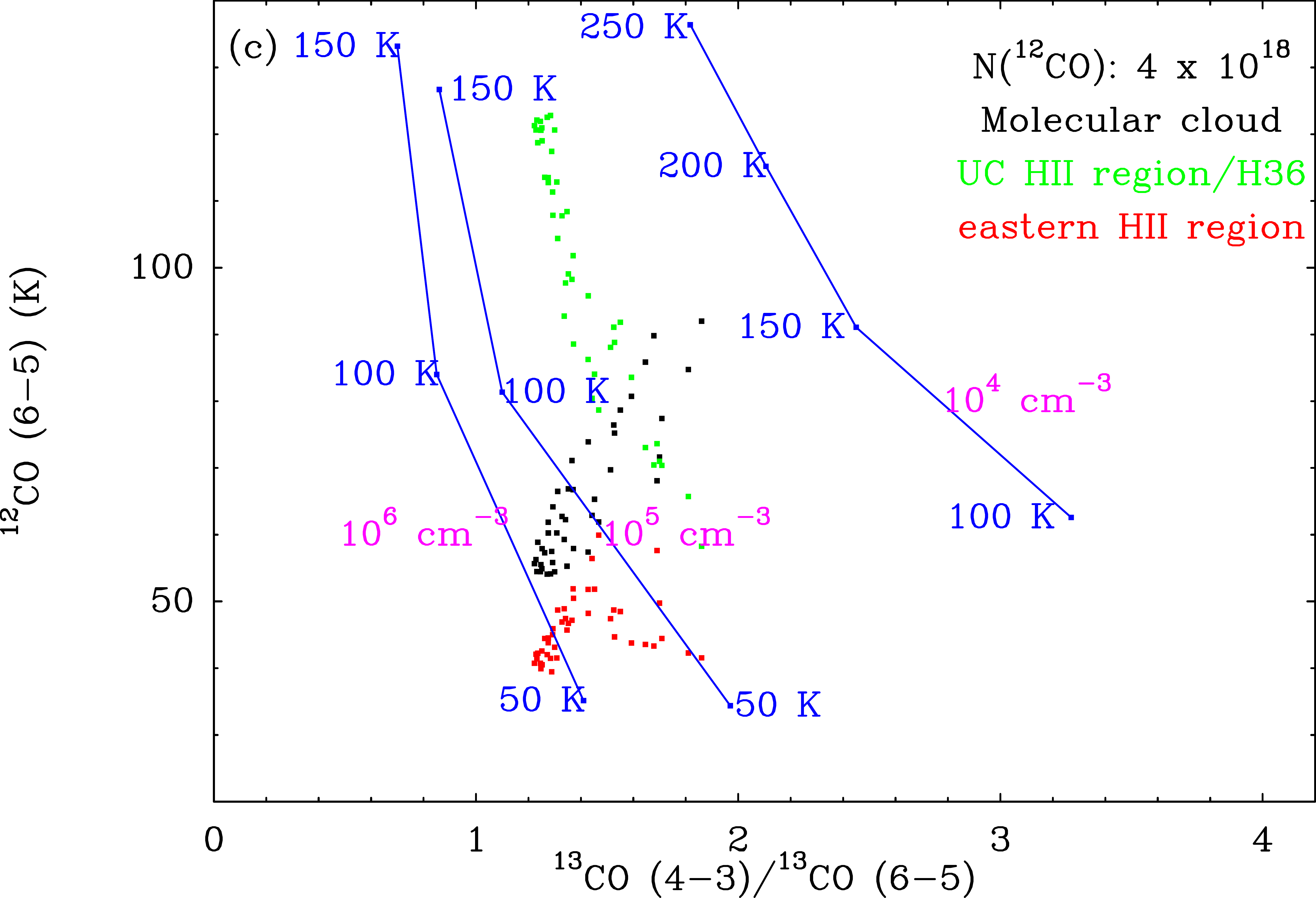}}

  \caption{For three different \textsuperscript{12}CO column densities, RADEX modeling results are shown for  \textsuperscript{12}CO $J$ = 6 $\to$ 5 peak main-beam brightness temperatures and corresponding \textsuperscript{13}CO $J$ = 4 $\to$ 3/\textsuperscript{13}CO $J$ = 6 $\to$ 5 ratios. The blue lines denote the RADEX modeling results for different kinetic temperatures at 3 different volume densities, and colour points represent data from near the HII region, PDR and molecular cloud (see Sect. 4.3). Each point denotes a \textsuperscript{12}CO $J$ = 6 $\to$ 5 temperature and \textsuperscript{13}CO $J$ = 4 $\to$ 3/\textsuperscript{13}CO $J$ = 6 $\to$ 5 temperature ratio obtained from a single pixel. Data points are extracted from peak temperature maps convolved to the same resolution of 16$\arcsec$. (a) The data points plotted are obtained for a \textsuperscript{12}CO column density range of 8 $\times$ 10$^{17}$ -- 1.8 $\times$ 10$^{18}$ cm$^{-2}$ and for modeling the chosen input \textsuperscript{12}CO column density is 1 $\times$ 10$^{18}$ cm$^{-2}$. (b) The data points plotted are obtained for a \textsuperscript{12}CO column density range of 1.8 $\times$ 10$^{18}$ -- 3.5 $\times$ 10$^{18}$ cm$^{-2}$ and for the modeling the input \textsuperscript{12}CO column density is 2 $\times$ 10$^{18}$ cm$^{-2}$. (c) The data points plotted are obtained for a \textsuperscript{12}CO column density range of 3.5 $\times$ 10$^{18}$ -- 5.1 $\times$ 10$^{18}$ cm$^{-2}$ and for the modeling input the \textsuperscript{12}CO column density is 4 $\times$ 10$^{18}$ cm$^{-2}$.}
  \end{figure}

 \begin{table}[ht]
  \tiny
 \caption{Physical parameters calculated from rotational diagrams.}
 \centering
 \begin{threeparttable}
 \begin{tabular}{c c c c }
 \hline\hline
 \noalign{\smallskip}
  \multicolumn{2}{c}{Optically thin $^{13}$CO} & \multicolumn{2}{c}{Correction factor included}\\
 \hline
 \noalign{\smallskip}
 $T_{\rm rot}$\tnote{a} (K) & $N(^{13}$CO) (cm\textsuperscript{-2})  & $T_{\rm rot}$\tnote{a} (K) & $N(^{13}$CO) (cm\textsuperscript{-2})\\
%\noalign{\smallskip}
\hline
 \noalign{\smallskip}
 8.3 $\pm$ 1.3 & 3.6 $\times$ $10^{16}$ & 8.6 $\pm$ 1.5 & 6.2 $\times$ $10^{16}$  \\
 41 $\pm$ 3.7 & 3.6 $\times$ $10^{16}$ & 30 $\pm$ 4.8 & 5.4 $\times$ $10^{16}$\\
 90 $\pm$ 21.6 & 1.9 $\times$ $10^{16}$ & 39 $\pm$ 4.9 & 4.6 $\times$ $10^{16}$\\
  & & 84 $\pm$ 10 & 2.1 $\times$ $10^{16}$\\
 \hline
 \noalign{\smallskip}
Transition & $\tau$ & Transition & $\tau$\\
 \hline
 \noalign{\smallskip}
 $J$ = 1 $\to 0$ & 1.68 &J = 1 $\to$ 0 & 1.92 \\
 $J$ = 2 $\to 1$ & 1.78 &J = 2 $\to$ 1 & 2.20 \\
 $J$ = 4 $\to 3$ & 1.25 &J = 4 $\to$ 3 & 0.74 \\
 $J$ = 6 $\to 5$ & 1.12 &J = 6 $\to$ 5 & 0.28 \\
 $J$ = 8 $\to 7$ & 1.04 &J = 8 $\to$ 7 & 0.10 \\
  \hline\hline
 \noalign{\smallskip}
 
  \end{tabular}
 \label{table:table}

\begin{tablenotes}
  \item[a] Calculated for different gradients with their errors in the plot as indicated in Fig.~11.
  \end{tablenotes}

\end{threeparttable}
\end{table}

\begin{figure*}[htp]
  \centering
 % \subfigure{\includegraphics[width=90mm]{radex_12co_vary_t_new.pdf}}\quad
 % \subfigure{\includegraphics[width=90mm]{radex_13co_vary_t_new.pdf}}
 %\subfigure{\includegraphics[width=90mm]{radex_12co_vary_d_new.pdf}}\quad
 %\subfigure{\includegraphics[width=90mm]{radex_13co_vary_d_new.pdf}} 
 
 \subfigure{\includegraphics[width=0.48\textwidth]{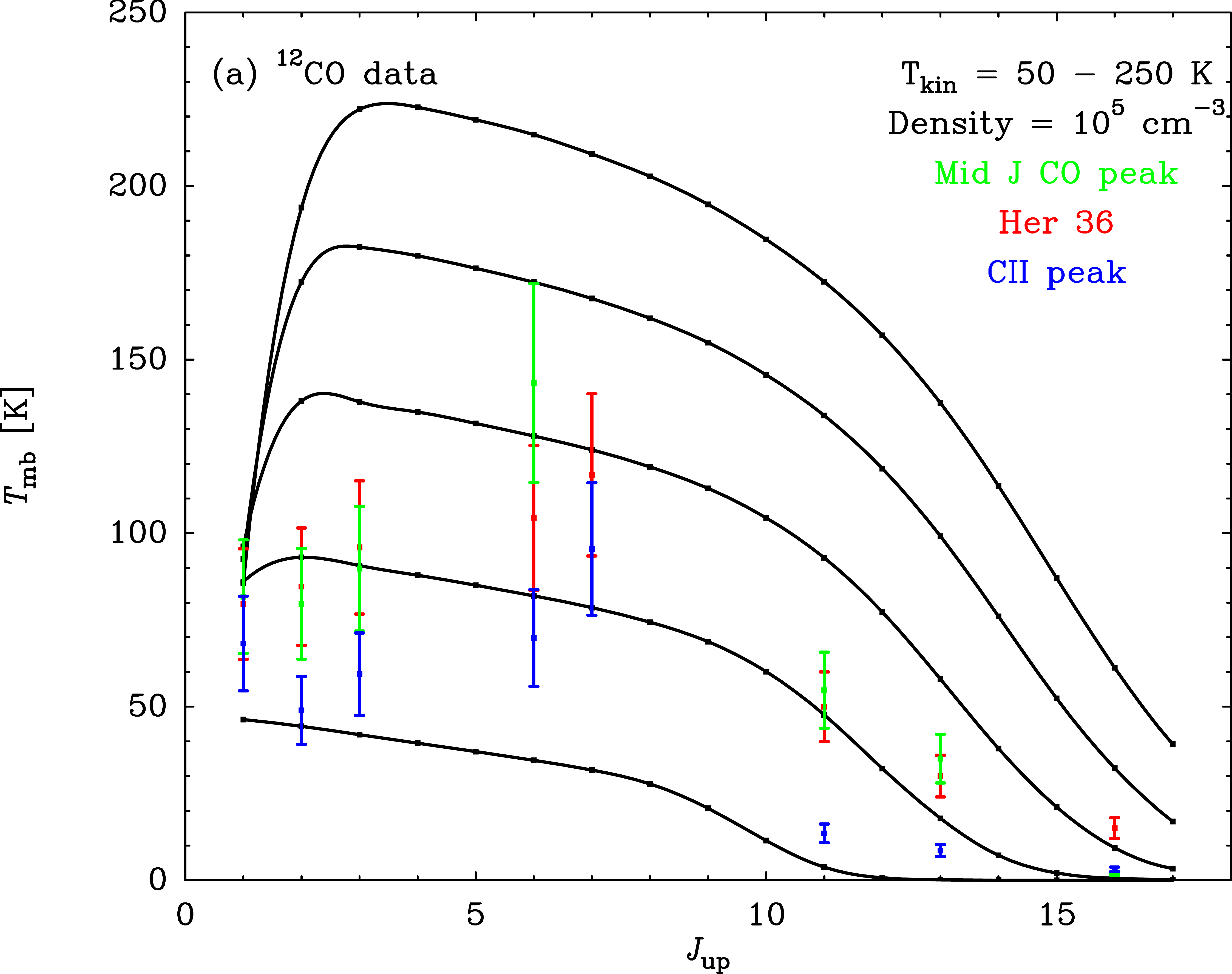}}\quad
  \subfigure{\includegraphics[width=0.48\textwidth]{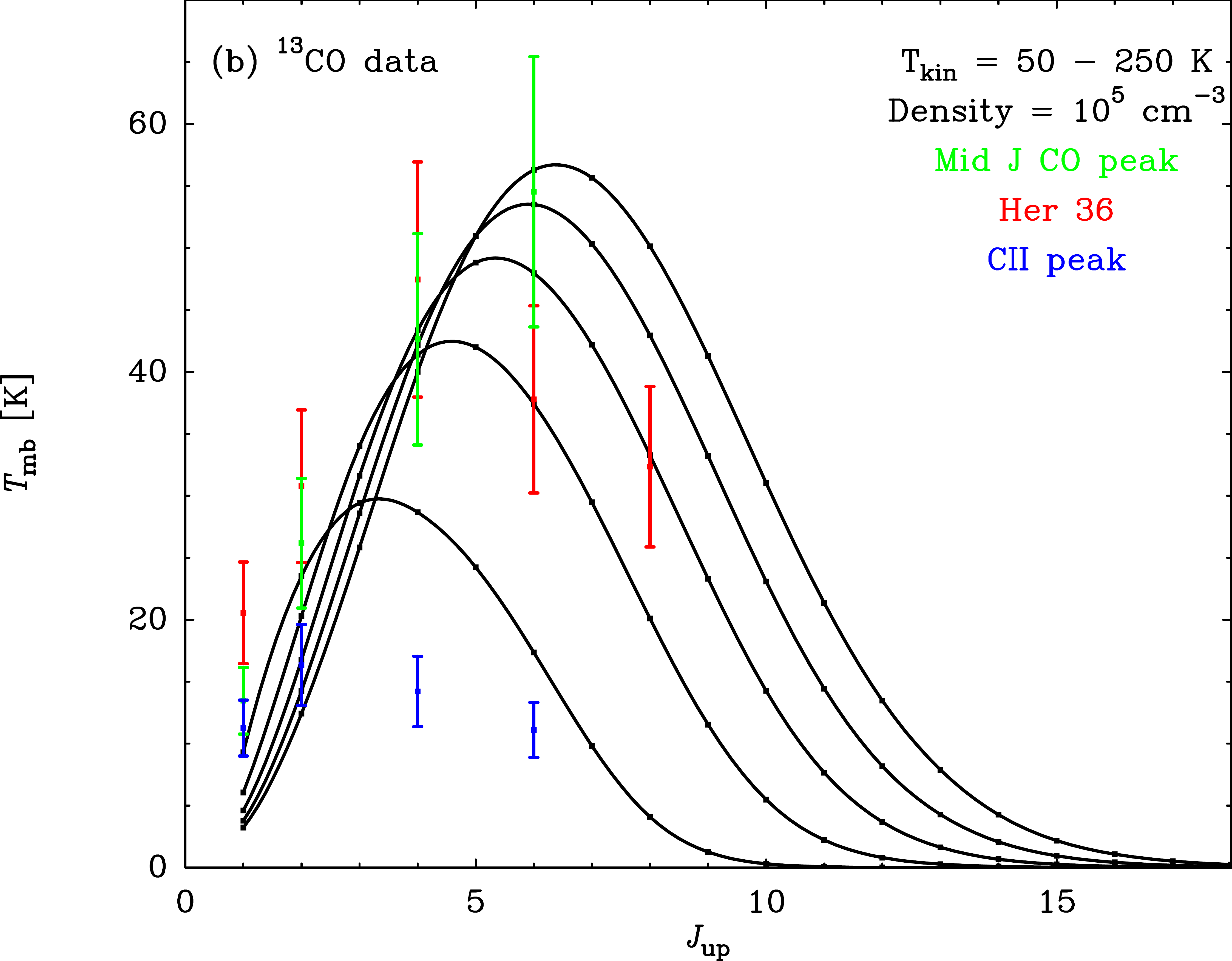}}
 \subfigure{\includegraphics[width=0.48\textwidth]{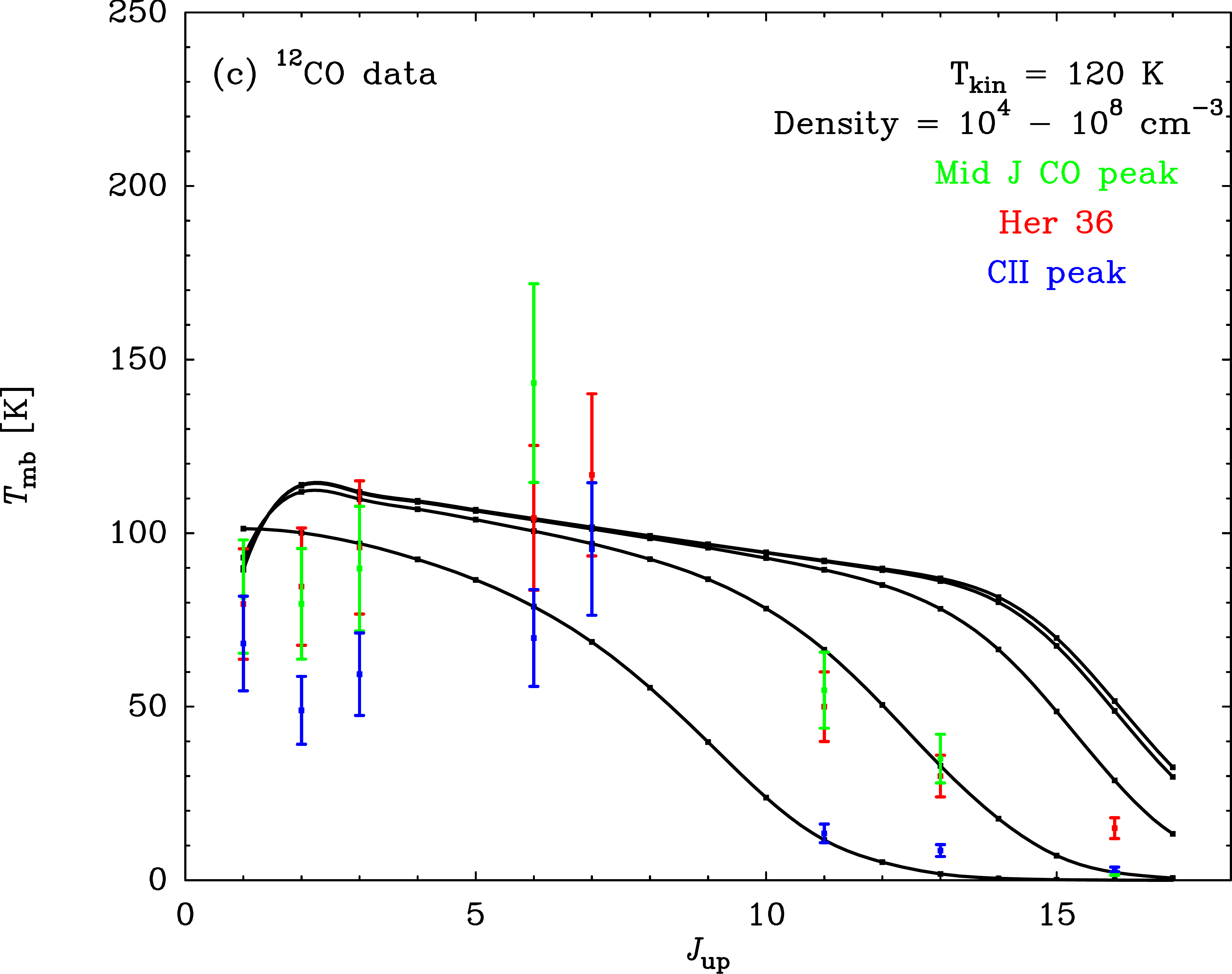}}\quad
 \subfigure{\includegraphics[width=0.48\textwidth]{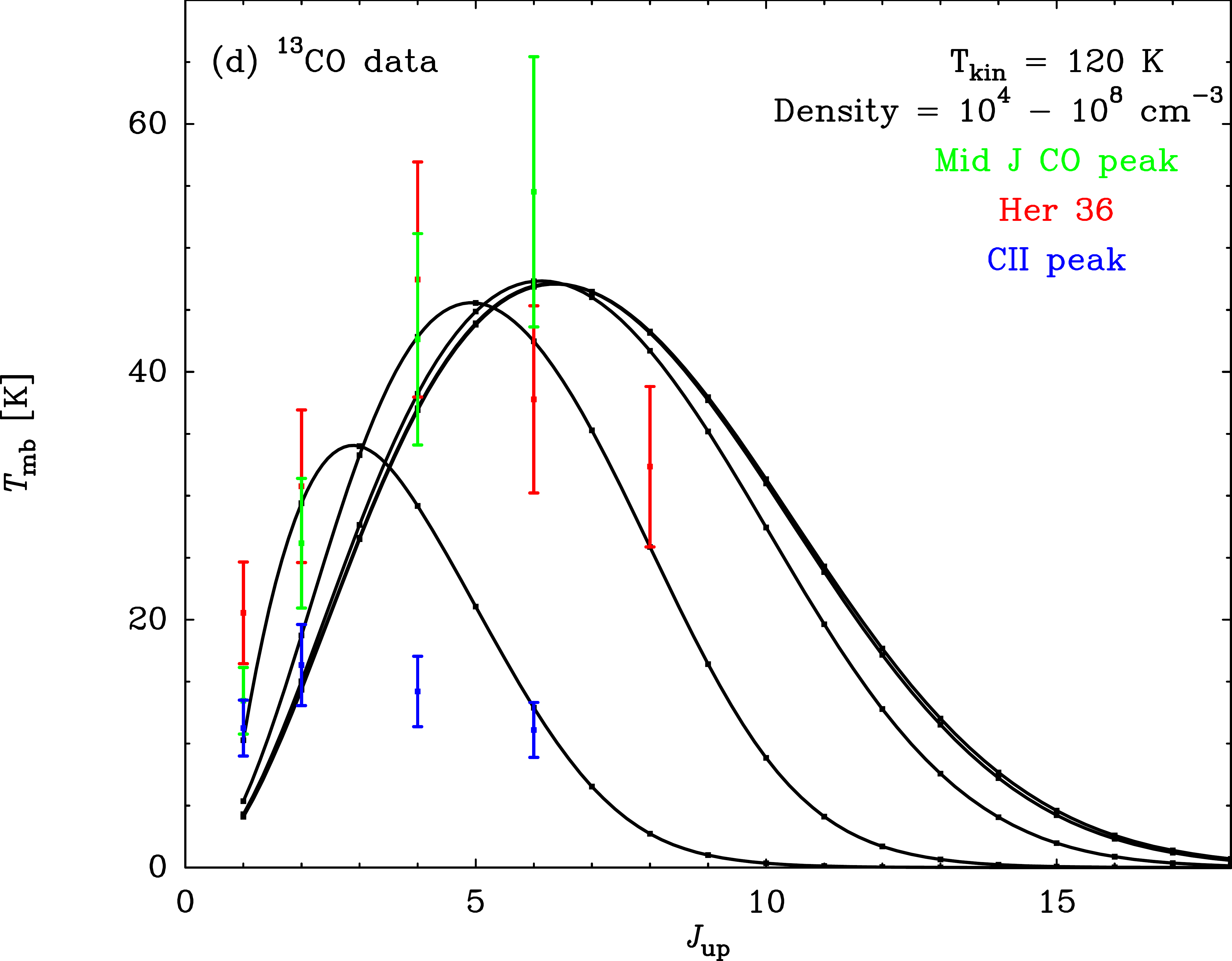}}

  \caption{Results obtained from RADEX modeling for \textsuperscript{12}CO and \textsuperscript{13}CO at the mid-$J$ CO peak ($\Delta\alpha$ = $-$13.0$\arcsec$, $\Delta\delta$ = 8.0$\arcsec$), at Her 36 ($\Delta\alpha$ = 0$\arcsec$, $\Delta\delta$ = 0$\arcsec$) and at the \cii\ peak ($\Delta\alpha$ = 30.0$\arcsec$, $\Delta\delta$ = $-$2.0$\arcsec$). Our observed data points are in green, red and blue and are extracted from peak temperature maps convolved to the same resolution of 31$\arcsec$. The column densities used in the modeling are for \textsuperscript{12}CO 4 $\times$ 10$^{18}$ cm$^{-2}$ and 8 $\times$ 10$^{16}$ cm$^{-2}$ for \textsuperscript{13}CO. (a) and (b) show results obtained by varying the kinetic temperature from 50 -- 250~K in steps of 50 K and keeping the density fixed at 10\textsuperscript{5} cm\textsuperscript{-3}; (c) and (d) show results obtained by varying the H\textsubscript{2} density from 10\textsuperscript{4} -- 10\textsuperscript{8} cm\textsuperscript{-3} in steps by a factor of 10 and keeping the kinetic temperature fixed at 120~K.}
  \label{radex2}
  \end{figure*}

\subsection{Rotational Diagrams of \textsuperscript{13}CO}

%The excitation temperature and column density determination for the $J = 6 \to 5$ transition of \textsuperscript{13}CO was straight forward. 
With observations of CO lines with different $J$, rotational diagrams can be used to study the excitation of the CO emitting gas. In a rotational diagram or "Boltzmann plot" the natural logarithm of the column density $N_{\rm u}/g_{\rm u}$ of different lines is plotted against their upper energies $E_{\rm up}/k$. Here $g_{\rm u}$ is the degeneracy of the upper energy level, ($\equiv 2J+1$), and $k$ is the Boltzmann constant. For a single temperature and optically thin emission these data points will fall onto a straight line. Deviations from a straight line indicate then either optical depth effects or temperature gradients in the clouds. A complete derivation of rotational diagrams for an LTE case can be found in \citet{1999ApJ...517..209G}.

\begin{figure*}[htp]
  \centering
  \subfigure{\includegraphics[width=58mm]{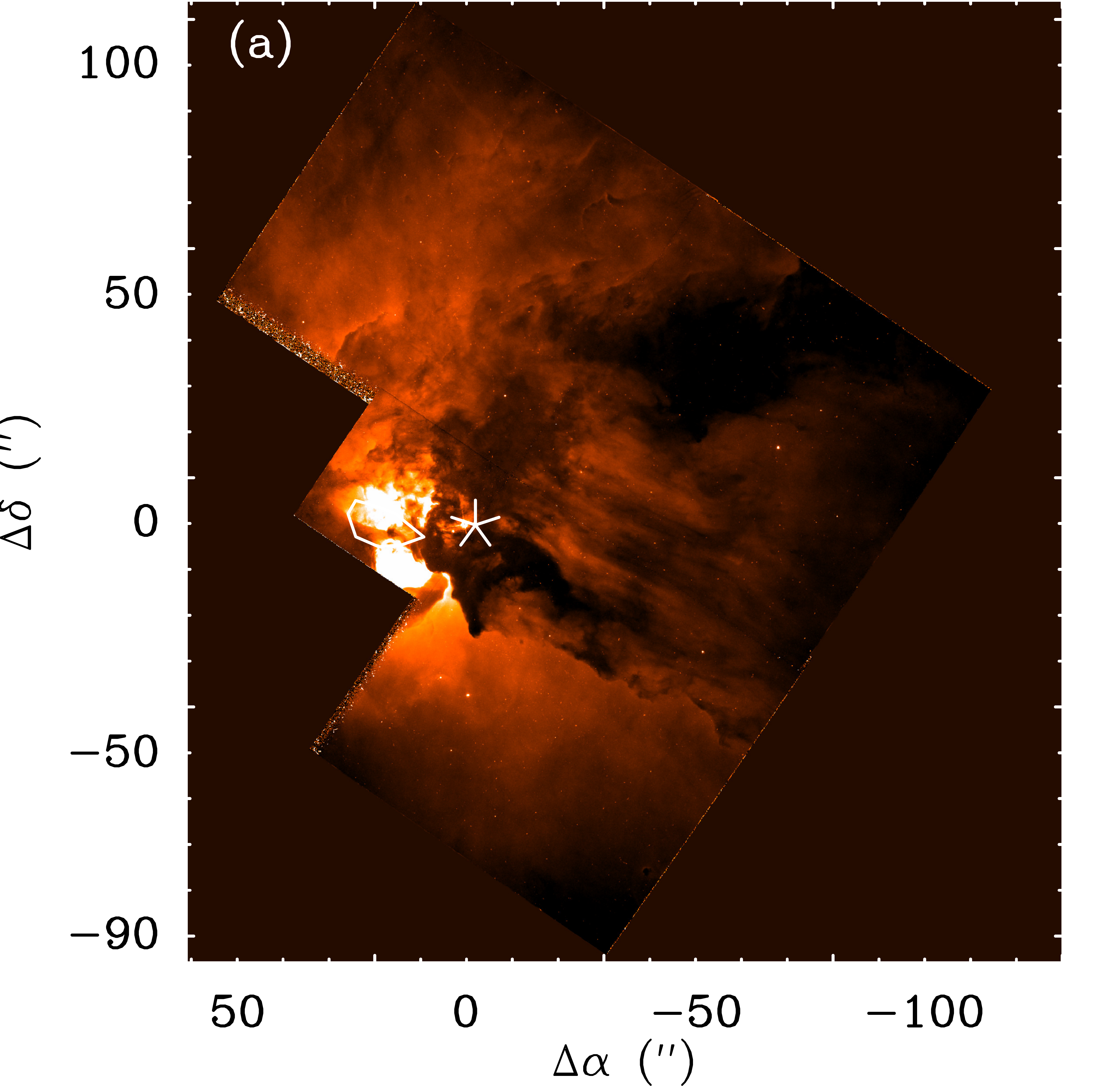}}\quad
  \subfigure{\includegraphics[width=58mm]{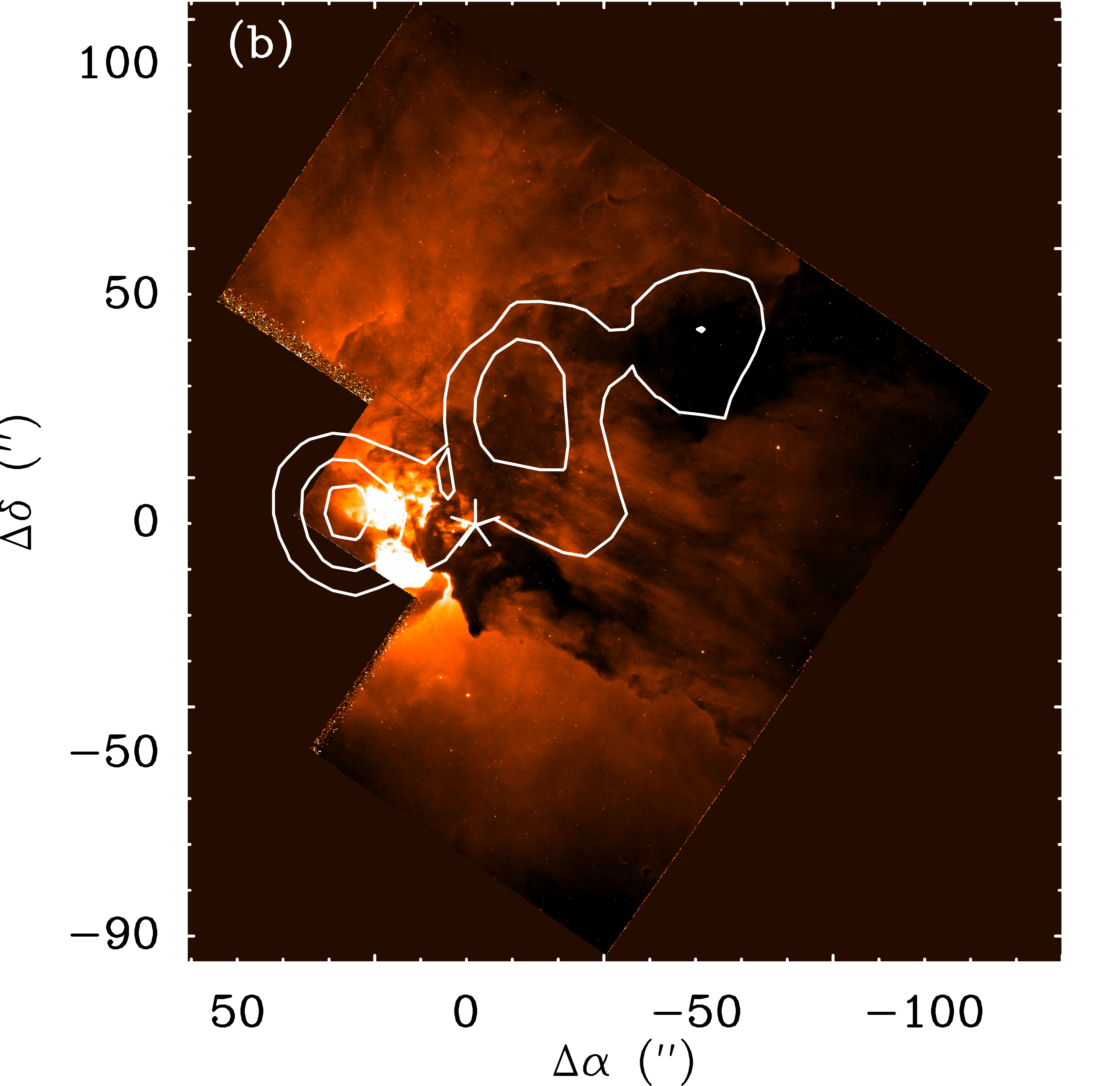}}\quad
 \subfigure{\includegraphics[width=58mm]{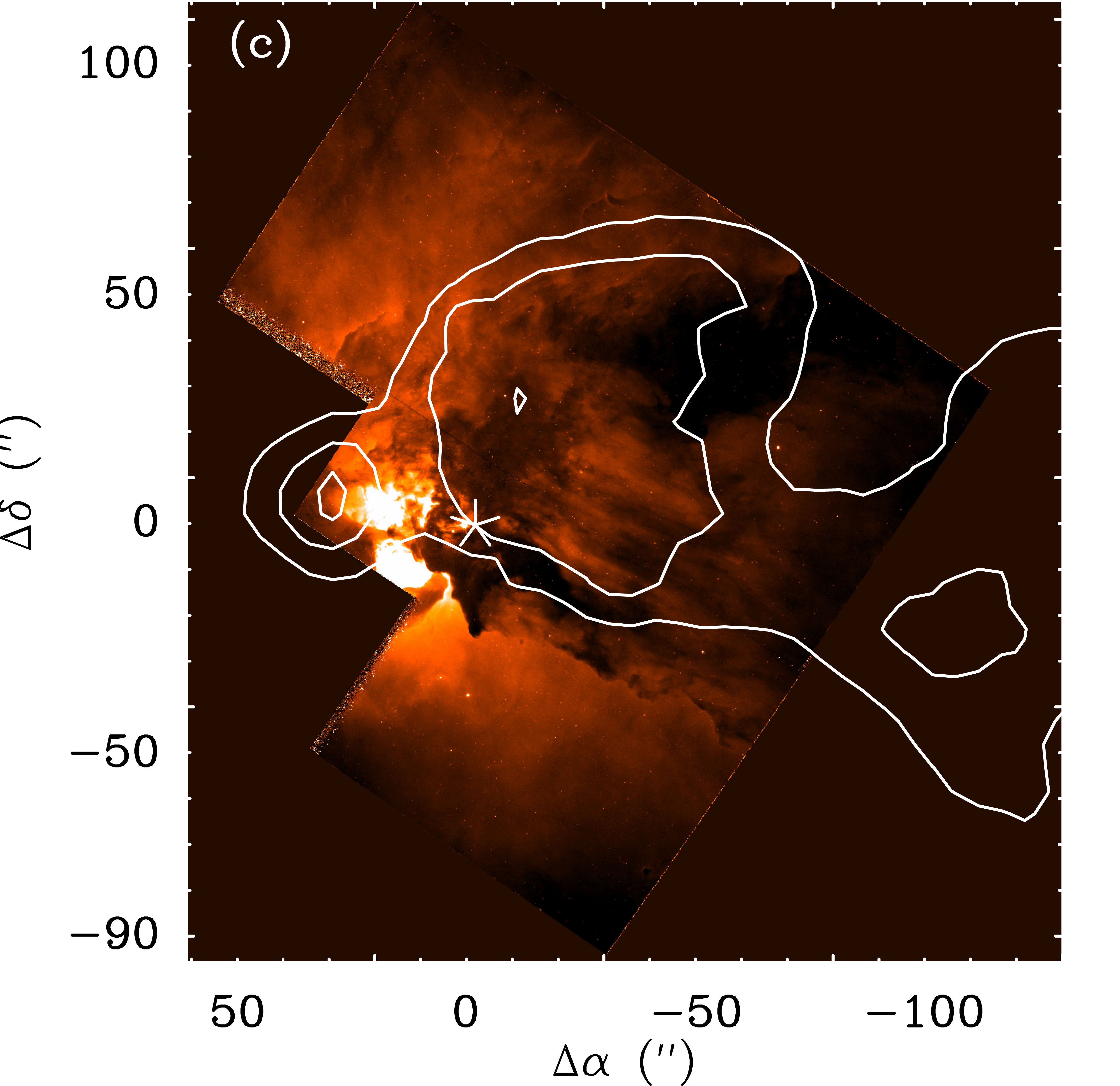}}
 
  \caption{HST F487N (4865 \AA) image of Her 36 and its surroundings overlaid with \cii\ channel map contours (in white) at: (a) 2 km s$^{-1}$; (b) 5 km\,s$^{-1}$ and (c) 7 km\,s$^{-1}$ to show the progression of the foreground material at lower velocities. Her 36 is the central position ($\Delta\alpha$ = 0, $\Delta\delta$ = 0) at R.A.(J2000) marked with an asterisk at R.A.(J2000) = 18\textsuperscript{h}03\textsuperscript{m}40.3\textsuperscript{s} and Dec.(J2000) = --24$\degree$22$\arcmin$43$\arcsec$, with the hourglass nebula representing the two bright hotspots about $\sim$ 15$\arcsec$ to the east. Contour levels are 10$\%$ to 100$\%$ in steps of 10\% of the peak emission observed at the channel with 10 km s$^{-1}$ velocity.}
\label{hst}  
\end{figure*}

%Population diagram helps us to determine the column densities of various levels and its dependence on the energies. It is also useful to confirm whether a transition is optically thin or not. 
%In LTE at gas temperature T, the column density of a given level can be related to the total column density $N$ of the molecule by:

%\begin{equation} \label{eq10}
%    N_{\rm u} = \frac{N}{Q_{\rm rot}}g_{\rm u} e^{-E_{\rm u}/kT}\,. 
%\end{equation}.

%Eq.~7 can then be written as:

%\begin{equation} \label{eq11}
%    \ln \frac{N_{\rm u}}{g_{\rm u}} = \ln N - \ln Q_{\rm rot} - \frac{E_{\rm u}}{kT}\,.
%\end{equation}

%If the optical depths of the lines are known, one can apply an optical depth correction factor $C_{\rm \tau} = \tau/(1 - e^{-\tau})$ \citep{1999ApJ...517..209G}
%so that Eq.~8 becomes
%\begin{equation} \label{eq12}
%    C_\tau = \frac{\tau}{1 - e^{-\tau}}
%\end{equation}\\

%Including this correction factor, eq.~11 becomes:

%\begin{equation} \label{eq12}
%    \ln \frac{N_{\rm u}^{\rm thin}}{g_{\rm u}} + \ln C_{\rm \tau} = \ln N - \ln Q_{\rm rot} - \frac{E_{\rm u}}{KT}\,.         
%\end{equation}

Firstly, by assuming \textsuperscript{13}CO to be optically thin i.e. the optical depth correction term, $C_{\rm \tau}$ is unity in Eq.~24 of \citet{1999ApJ...517..209G}, we plotted $\ln N_{\rm u}^{\rm thin}/g_{\rm u}$ vs $E_{\rm u}/K$ as shown by Fig.~\ref{fig:rot_dia} in black for the five \textsuperscript{13}CO lines observed towards Her 36. A curvature in a rotational diagram can be due to optical depth effects, therefore we estimate the expected optical depths for the computed column density from Eq.~25 of \citet{1999ApJ...517..209G} and applied the optical depth corrections $C_{\rm \tau}$ that lead to the corrected diagram as shown in red by Fig.~\ref{fig:rot_dia}. The new temperatures and column densities are then calculated as shown in Table~3. Further iterations would lead to corrections smaller than the error bars. After the optical depth correction the curvature in the rotational diagram remains and indicates temperature gradients in the gas, as expected in a PDR.\

The $J = 1\to 0$ and $J = 2\to 1$ transitions of \textsuperscript{13}CO seem to originate from colder gas, while the $J = 4\to 3$ and $J = 6\to 5$ transitions probe hotter gas, in agreement with the analysis done for $J = 6\to 5$ in Sect.~4.1. The $J = 8\to 7$ transition appears to originate from the hottest gas.

\subsection{RADEX modeling}
We used the non-LTE radiative transfer program RADEX \citep{2007A&A...468..627V} to verify the calculations done in Sects. 4.1 and 4.2, which assumed LTE to determine the temperatures and densities. RADEX uses the escape probability approximation for a homogeneous medium and takes into account optical depth effects. We chose a uniform sphere geometry. The Leiden Molecular and Atomic Database\footnote{http://www.strw.leidenuniv.nl/moldata/} (LAMBDA; \citet{2005A&A...432..369S}) provides rates coefficients for collisions of CO and H\textsubscript{2} used in the modeling. \textsuperscript{12}CO and \textsuperscript{13}CO transitions were modeled taking their line width from the average spectra of our data, i.e. 4 and 3 km s\textsuperscript{-1}, respectively. As input parameters we computed grids in temperature and volume density with a background temperature of 2.73~K,  
% 28 K as the background temperature as calculated from SPIRE of ESA's Herschel Space observatory in the 350 $\mu$m band; %\todo[inline]{how realistic is this background temperature? is there any data that may indicate some internal source of heating or measurements of dust temperatures in the background that should be used instead? - while the low-$J$ transitions are not sensitive to a higher background temperature (hence, it doesn't matter), the higher-$J$ lines are more affected by a hotter background temperature, depending on the column density...} 
kinetic temperatures in the range of 50 -- 250~K and H\textsubscript{2} densities in the range of 10\textsuperscript{4} -- 10\textsuperscript{8} cm$^{-3}$.
%and the column densities of \textsuperscript{12}CO in the range of 1 $\times$ 10\textsuperscript{18}, 2 $\times$ 10\textsuperscript{18} and 4 $\times$ 10\textsuperscript{18} cm\textsuperscript{-2} which were estimated from the LTE calculations done in Sects. 4.1 and 4.2.\\  
For a linear molecule line ratios from different $J$ depend on both temperature and density. To break this degeneracy, we compute with RADEX not only the line ratio of two \textsuperscript{13}CO transitions but also the temperature of the  \textsuperscript{12}CO line. The latter is optically thick, probes the excitation temperature (cf. Sect. 4.1) and can therefore be used to break the degeneracy between temperature and density.

In our first approach to determine the dominant kinetic temperatures and densities in M8 near Her 36, we selected from the CO maps data points for column densities of \textsuperscript{12}CO in three ranges: (a) 8 $\times$ 10\textsuperscript{17} -- 1.8 $\times$ 10\textsuperscript{18} cm\textsuperscript{-2}, (b) 1.8 $\times$ 10\textsuperscript{18} -- 3.5 $\times$ 10\textsuperscript{18} cm\textsuperscript{-2} and (c) 3.5 $\times$ 10\textsuperscript{18} -- 5.1 $\times$ 10\textsuperscript{18} cm\textsuperscript{-2} that were estimated from the LTE calculations done in Sects. 4.1 and 4.2. Another criterium for selection of these data points aims at selecting the most conspicuous regions inside the maps. The molecular cloud region consists of all data points in an area of 15$\arcsec \times 15\arcsec$ around the mid-$J$ CO peak, the UC HII region/ Her 36 encompasses all data points in an area of 20$\arcsec \times 20\arcsec$ around Her 36, and the eastern HII region has data points in an area of 20$\arcsec \times 20\arcsec$ around the \cii\ peak. 

%\todo[inline]{FWY: Are you sure about the regions, they seem to be small compared to the large number of points you have in the plot.}

Fig.~12 shows the $J$ = 4 $\rightarrow$ 3 and $J$ = 6 $\rightarrow$ 5 line ratios of \textsuperscript{13}CO vs. the $J$ = 6 $\rightarrow$ 5 \textsuperscript{12}CO line temperature for both the measured data points in the maps and the results of the RADEX computation of the density/temperature grid. The H$_2$ number densities in most of the regions are in the range of $\sim$ 10$^4$ -- 10$^6$ cm$^{-3}$. High densities are obtained for the molecular cloud and the eastern HII region, while the region consisting of the bright stellar system Her 36 and the ultra compact HII region has densities in the range of 10$^4$ -- 10$^5$ cm$^{-3}$. The kinetic temperatures from the modeling results are in a range of 100 -- 250~K for 10$^4$~cm$^{-3}$ and in a range of 50 -- 150~K for 10$^5$ -- 10$^6$ cm$^{-3}$.\ 

%However, these results are based on the $J$ = 4 $\rightarrow$ 3 and $J$ = 6 $\rightarrow$ 5 CO transitions only.\ 

%\todo[inline]{FWY: You need to explain how you selected data points from the different regions.}

In order to include all CO lines towards several positions in the RADEX analysis, their intensities as a function of $J$ are compared to RADEX results in Fig.~13. To fit the modeling results to the observed data points we varied the column densities of \textsuperscript{12}CO and \textsuperscript{13}CO. While for \textsuperscript{12}CO a column density of 4 $\times$ 10$^{18}$ cm$^{-2}$ was chosen, for \textsuperscript{13}CO a column density of 8 $\times$ 10$^{16}$ cm$^{-2}$ could make the modeling results fit the data points. This \textsuperscript{13}CO column density exceeds the value obtained by the LTE calculations and corresponds to a $^{12}$CO/$^{13}$CO ratio of 50. While this is lower than the assumed value of 63 in Sect. 4.1, it is still within the typical scatter of this ratio in the ISM \citep{2005ApJ...634.1126M}. The \textsuperscript{12}CO and \textsuperscript{13}CO observed data is compared to the RADEX model at the peak of the mid-$J$ CO transitions in the molecular cloud, at Her 36 and at the emission peak of \cii\ . Figs.~13 (a) and (b) show results with the kinetic temperature varied from 50 -- 250~K and keeping the H\textsubscript{2} density fixed at 10\textsuperscript{5} cm\textsuperscript{-3}, while (c) and (d) show results when the H\textsubscript{2} density is varied from 10\textsuperscript{4} -- 10\textsuperscript{8} cm$^{-3}$ while keeping the kinetic temperature fixed at 120~K.

  \begin{figure*}[htp]
  \centering
  \subfigure{\includegraphics[width=90mm,height=60mm]{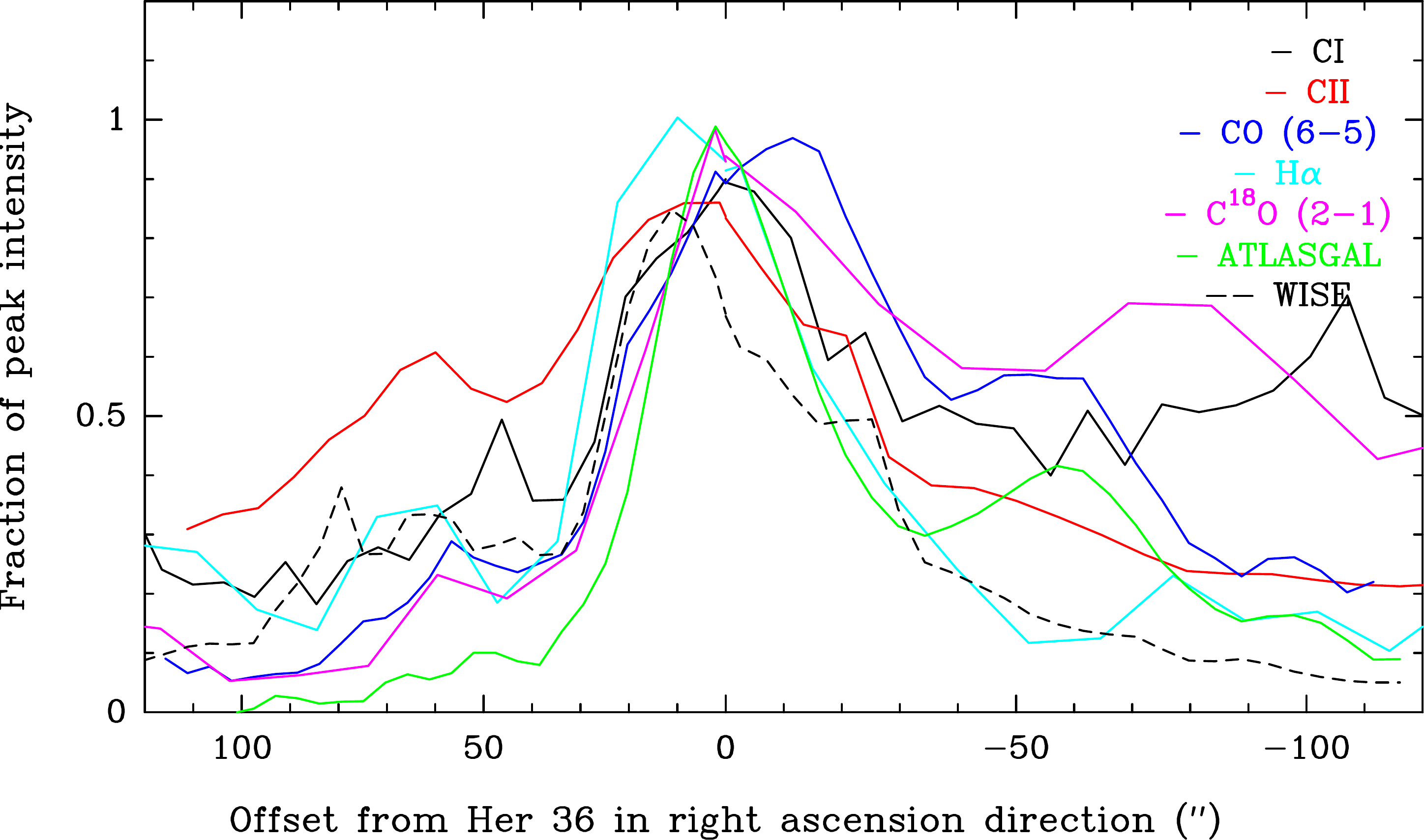}}\qquad
  \subfigure{\includegraphics[width=80mm]{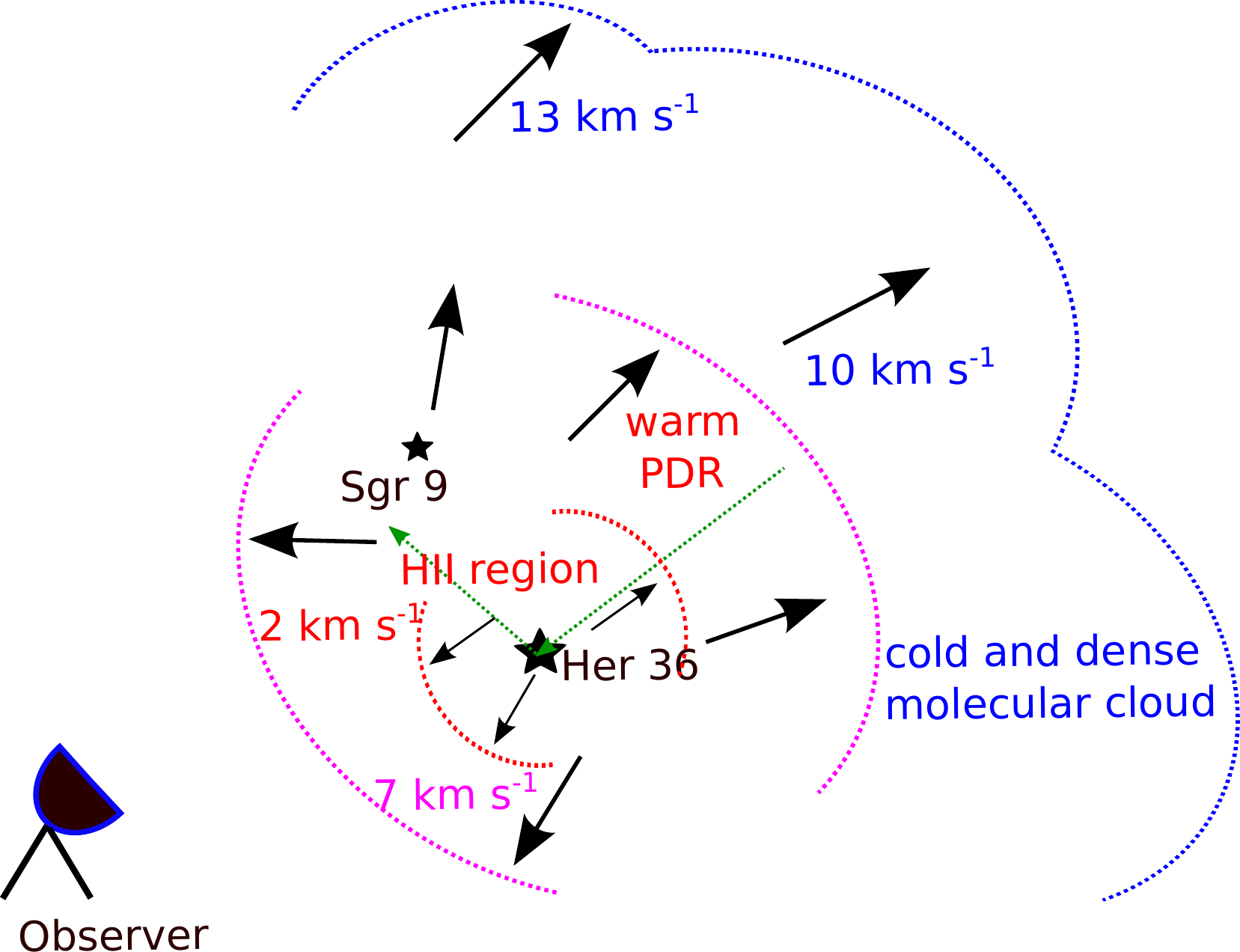}}
 
  \caption{Left: Velocity integrated intensities normalized to their value at peak vs offset ($\arcsec$) from Her 36 along the green arrows shown in the right. They follow a path from the second ATLASGAL peak at ($-$53$\arcsec$, 23$\arcsec$) to the second \cii\ peak at (60$\arcsec$, 27$\arcsec$) via Her 36 at (0$\arcsec$, 0$\arcsec$). Plots are shown for \cii\-, \ci\-, $J$ = $6\to 5$ $^{12}$CO, $J$ = $2\to 1$ C$^{18}$O, H$\alpha$, ATLASGAL 870 $\mu$m and WISE 3.4 $\mu$m emission; Right: Schematic diagram of the prominent optical features of M8 pertinent to the discussion in this paper. The cold and dense molecular cloud is in the background shown in blue. The foreground gas of the warm PDR veil is receding away from Her 36 ($\sim$ 9 km s$^{-1}$) with lower velocities (2 -- 7 km s$^{-1}$), while the HII region is powered by both stellar systems, Her 36 and 9 Sgr.}
 \label{morph}
\end{figure*}

It can be seen that no single kinetic temperature or H\textsubscript{2} density can fit all the observed data points. This suggests solutions with kinetic temperatures in the range of 100 -- 150~K and H\textsubscript{2} densities to be in the range of 10\textsuperscript{4} -- 10\textsuperscript{6} cm\textsuperscript{-3}. Such a spread in the ambient temperature was also implied by the rotational diagram analysis. Furthermore, these values are similar to the temperature and density ranges found in OMC~1 \citep{2012A&A...538A..12P}. 

\subsection{CO, \ci\ and \cii\ luminosities}

%\todo[inline]{fwy: it is strange to use here an extragalactic equation.
%can you at least change it to kpc?}
We obtained the total luminosities of the CO spectral line energy distributions (SLED), and of the \ci\ 609 $\mu$m and \cii\ lines, over the total observed region seen in the maps of Figs.~2, 3 and 4, as derived by \citet{1997ApJ...478..144S}; \citet{2013ARA&A..51..105C} and scaled it for galactic sources:
\begin{equation} \label{eq11}
    \ L = 1.04 \times 10^{-9} S \Delta V \nu D_{\rm L}^{2}\,,
\end{equation}  
where $L$ is the line luminosity in L\textsubscript{\(\odot\)}, $S$ $\Delta$$V$ is the velocity integrated flux in Jy km s$^{-1}$, $\nu$ is the transition frequency in GHz and $D_{\rm L}$ is the distance in kpc. A total CO luminosity of $L_{\rm CO}$ = 9.5~L\textsubscript{\(\odot\)} was calculated for the observed transitions and by accounting for the luminosities of the missing transitions. A \ci\ 609~$\mu$m line luminosity of $L_{\rm \ci\ }$ = 0.11 L\textsubscript{\(\odot\)} was obtained, which is a lower limit to the total \ci\ luminosity since the \ci\ 370~$\mu$m line was not observed. The estimated \cii\ luminosity is $L_{\rm \cii\ }$ = 95.8~L\textsubscript{\(\odot\)}. Similar to the mass estimations in Sect.~4.1, these luminosity estimations have an error of $\sim$ 26\%.  
%\todo[inline]{JPE: Are these luminosities computed for the total observed region? or for a particular position in the maps?. In either case it should be mentioned here.}

\section{Discussion}
\subsection{Overview of the PDR and HII region around Her 36}
%\todo[inline]{FWY: The figure you are talking about is Fig.14. Put labels into the figure environments and refer to them then with the ref command. I'm not sure whether 45degrees for the plot with the strips through the cloud is ideal. You seem to miss the second ATLASGAL peak, although I'm surprised by that since this should be the offset -40,35, hence close to the 45degrees. You could mark the location of the strip in the ATLASGAL figure with a dashed line. Then one sees also how the strip compares to the location of the positions selected for a more detailed analysis. We can also think about having the strip changing direction at Her 36 and from their to continue to the North-East through the " clumpy" CII position. This could be done by combining two strips with different angles. We also have to rethink the schematic. The molecular cloud needs to lie behind Her 35 AND 9 Sgr, with being closest to Her 36. I have also doubts about the current velocities in the figure. the most blue shifted velocity (the CII at 2-7km/s might be in the foregrund, being driven by the expansion of the nebula. the red-shifted stuff is in the background with the strongest redshifts towards the north-east in the direction of 9 Sgr. This is the opposite of what you are currently showing. What we could propose is that the cloud is in general in the background but that Her 36 is just being exposed from the cloud, hence it is still close to its dense natal environment, and some of the material seen in CII, the 2-7km/s compact peak, is still a remaining, now ionized dense core in the foreground.}

In Fig.~\ref{hst}, we present a F487N filtered 4865 {\AA} Hubble Space Telescope (HST) 
%\todo[inline]{FWY: this wavelengths must be wrong. Is there a reference to a published paper? What is the widths of the shown CII velocity channels? The contours are not the same as in the earlier channel maps. In 14c go down to a contour where the edge of the foreground veil is traced by CII. In 14a, what if you go to even lower velocities, will the contours move further towards the center of the hourglass and Her 36?} 
image\footnote{Based on observations made with NASA/ESA Hubble Space Telescope, and obtained from the Hubble Legacy Archive, which is a collaboration between the Space Telescope Science Institute (STScI/NASA), the Space Telescope European Coordinating Facility (ST-ECF/ESA) and the Canadian Astronomy Data Centre (CADC/NRC/CSA).} (observation ID number: 6227, observed in the year 1995) of Her 36 and its surroundings overlaid with contours from low velocity \cii\ channel maps. The \cii\ at the lowest velocity (2\,km s$^{-1}$) peaks at the Hourglass Nebula slightly to the east of Her 36. With increasing velocities, the \cii\ follows the dark patches in the HST images that form a foreground veil covering parts of the bright nebulosity excited by Her 36. The strong correlation of \cii\ and foreground absorption is particularly evident at the sharp southern edge of the veil seen at 7\,km s$^{-1}$. Therefore we suggest that the low velocity \cii\ probes  directly the gas of the veil that forms a foreground PDR illuminated by Her 36. On a fainter level, weak emission from this veil is also seen in the CO maps at low velocities (5 -- 7 km s$^{-1}$). %Assuming a temperatures of 200 - 500\,K, the measured low velocity \cii\ intensity corresponds to column densities of ??, or -- by assuming that most of the carbon is in \cii\ -- to a foreground visual extinction of ??. 
This foreground veil is receding away from Her 36 towards us and to the west with a change in the line-of-sight velocity. This is consistent with both high velocity red shifted and low velocity blue shifted emission of H$\alpha$, \nii\ and \sii\ doublets, \oiii\-, and absorption lines of the sodium D doublet as measured by \citet{2017A&A...604A.135D}. Assuming optically thin emission from \cii\ in this warm veil in the velocity range of 2 -- 7 km s$^{-1}$ and a kinetic temperature of 500~K \citep{1985ApJ...291..722T}, we calculated the \cii\ column density using Eq.~\ref{cii_colden} with a peak value of $\sim$ 9.6 $\times$ 10$^{17}$~cm$^{-2}$ at an offset of ($\Delta$$\alpha$ = 30$\arcsec$, $\Delta$$\delta$ = 10$\arcsec$) from Her 36. Taking C/H $\sim$ 1.2 $\times$ 10$^{-4}$, assuming all carbon being in ionized form and from the column density of H$_{\rm 2}$, $N({\rm H_{\rm 2}})$ = 9.4 $\times$ 10$^{20}$ cm$^{-2}$ ($A_{\rm v}$/mag) \citep{2008A&A...487..993K}, we derived the visual extinction $A_{\rm v}$. A maximum extinction of $A_{\rm v}$ $\sim$ 4.25 is obtained at the same position where the \cii\ column density peaks and the values get lower around it. This position is the same as position 13 in Fig. 5 of \citet{1986AJ.....91..870W}, where an $A_{\rm v}$ $\sim$ 3.9 was calculated.
%This is consistent with even more blue shifted radial velocities of stars in the larger cluster environment measured by \citet{2017A&A...604A.135D}.

In Fig.~\ref{morph} (left) we show how the intensity of various tracers evolves along a path from the direction of 9 Sgr to Her 36 and then continuing to the north-west along the molecular cloud. Towards the north-east \cii\ and the mid-IR WISE emission dominate, probing the extended HII regions towards 9 Sgr and the resulting PDR. All of the tracers peak on or close to Her 36, showing the tight spatial correlation between the ultra compact HII region (as seen in the recombination line), a dense clump in the larger scale molecular cloud ($870\, \mu$m dust and molecular emission), and the bright PDR illuminated by Her 36 (\cii\ and WISE). To the north-west, emission from the molecular cloud dominates with the second dense but colder clump at an offset of about 70$\arcsec$ from Her 36. \ci\ diffuse emission is very extended in this region.

Taking this into consideration along with our analysis of the morphology done in Sects.~3 and 4, we propose a geometry of the region illuminated by Her 36 and 9 Sgr as presented in Fig.~\ref{morph} (right). 
We propose that the cold dense molecular cloud is behind the bright stars Her 36 and 9 Sgr. Her 36 is still much closer to the dense part of the cloud in which it was born; in fact the foreground veil is part of the original cloud accelerated towards us by the strong radiation and wind of Her 36, showing the expansion of the nebula, blue shifted with respect to the molecular cloud (3 -- 7~km~s$^{-1}$) moving away from Her 36 ($\sim$ 9~km s$^{-1}$) towards the observer and the west, while the redshifted (11 -- 13 km s$^{-1}$) \cii\ probes the background material towards the north-east of Her 36. The ultra compact HII region is fueled by Her 36 around its vicinity and the more extended diffuse HII region by 9 Sgr towards the east \citep{2008hsf2.book..533T}. 
%The brightest emission emanates from a 4$\arcmin$ core around Her 36 and the hourglass nebula that is apparent in the optical regime \citep{1976ApJ...203..159L,2008hsf2.book..533T}.

%\todo[inline]{FWY: To me the hourglass in this picture just follows two openings within the veil which happen to have a shape like an hourglass. Please correct me if there a evidence for a different origin of the morphology of the nebula in the literature.}

\subsection{Comparison with the PDR of the Orion Bar}
The Orion Bar is a well known PDR and its properties are described in depth by \citet{1997ARA&A..35..179H} and \citet{2000A&A...364..301W}. It is part of the OMC-1 core at the edge of M42 and is illuminated by young massive stars which form a trapezium at the center of the Orion Nebular Cluster, mainly from $\theta$\textsuperscript{1} C, which is a O5 -- O7 star. Also, the PDR appears to be located at the edge of the HII blister tangential to the line of sight \citep{2012A&A...538A..12P}. A comparison of $L_{\rm {CO 2-1}}$, $L_{\rm \cii }$ and $L_{\rm FIR}$ between M8 and the Orion Bar is presented in Table 4. For M8, the $L_{\rm {CO 2-1}}$ = 0.128 $L$\textsubscript{\(\odot\)} (calculated in a similar way as done in Sect.~4.4), $L_{\rm \cii }$ is calculated in Sect.~4.4 and $L_{\rm FIR}$ $\sim$ 10$^{4}$ $L$\textsubscript{\(\odot\)} is obtained from \citet{1998ASPC..132..113W}. The Orion Bar's luminosities are taken from \citet{2015ApJ...812...75G}.
%\todo[inline]{FWY: in section 4.4 you did L(CO) not L(CO2-1).}

\begin{table}[ht]
\centering
\caption{CO $J$ = 2 $\to$ 1, \cii\ and FIR luminosity ratios of M8 and the Orion bar.}

\begin{tabular}{c c c  }
 \hline\hline
 \noalign{\smallskip}
 Luminosity Ratio & M8 & Orion bar \\
 \hline
   \noalign{\smallskip}

 $L_{\rm \cii\ }$/$L_{\rm FIR }$ & 10$^{-3}$ & 1.1 $\times$ 10$^{-3}$   \\
  $L_{\rm \cii\ }$/$L_{\rm CO 2-1 }$ & 7.5 $\times$ 10$^{2}$ & 1.6 $\times$ 10$^{3}$ \\
  $L_{\rm CO 2-1 }$/$L_{\rm FIR }$ & 1.3 $\times$ 10$^{-6}$ & 6.6 $\times$ 10$^{-7}$ \\
 
  \hline\hline
 \noalign{\smallskip}
 
  \end{tabular}
 \label{table:table}
\end{table}

%A comparison of L\textsubscript{\(\odot\)} and $L_{\rm FIR}$ between M8 and the Orion bar is presented in Table 4. For M8 the $L_{\rm {CO 2-1}}$, $L_{\rm \cii }$ are calculated for in Sect.~4.4 and $L_{\rm FIR}$ $\sim$ 10$^{4} L\textsubscript{\(\odot\)}. While, for that of Orion bar's are obtained from \citep{2015ApJ...812...75G}\ 

  We calculated the FUV radiation field, $G$\textsubscript{0} $\sim$ 0.6 -- 1.12 $\times$ 10$^{5}$ in Habing units and density, $n$ $\sim$ 0.97 -- 1.93 $\times$ 10$^{5}$~cm$^{-3}$ for the PDR of M8 by adopting an electron density $n$\textsubscript{e} of 2000 -- 4000~cm\textsuperscript{-3}, electron temperature $T$\textsubscript{e} of 7000 -- 9000~K
 % \todo[inline]{What do you mean by fluctuations? Is this important? Check the A\&A instructions and/or other papers for how to refer to books. What is the actual G0 value that you determined?} 
 \citep{1986AJ.....91..870W, 1999ApJS..120..113E}, using the values of stellar luminosity and number of ionizing photons for an O7 star from Sect.~7.2.1 of \citet{2010pcim.book.....T}. The densities match well with the RADEX calculations done in Sect.~4.4. Interestingly, these calculated values of $G$\textsubscript{0} and $n$ also match quite well with those calculated for the Orion Bar. This leads us to a direct comparison of the results of the PDR models of the Orion Bar \citep{1985ApJ...291..722T,1985ApJ...291..747T,1995A&A...303..541J,1995A&A...294..792H,1997ARA&A..35..179H,1999RvMP...71..173H,2017A&A...598A...2A} with the PDR of M8 since we expect similar chemical and thermal conditions in both PDRs. \citet{1985ApJ...291..747T} and \citet{1999RvMP...71..173H} calculated the structure of the Orion Bar as a function of visual extinction $A$\textsubscript{v}. They show a typical case where H\textsubscript{2} does not self-shield until dust attenuation of the far-ultraviolet photons creates an atomic surface layer. According to this, in the PDR of M8 the transition of atomic H to molecular H\textsubscript{2} will occur at $A$\textsubscript{v} = 2, the carbon balance will shift from C\textsuperscript{+} to C and CO at $A$\textsubscript{v} = 4 and except for the O in CO, all oxygen will be in atomic form until very deep into the molecular cloud at $A$\textsubscript{v} = 8. The gas in the surface layer will be much warmer at about $\sim$ 500~K than the dust, which will be at about $\sim$ 30 -- 75~K. Complementary to the H$_2$ column density calculation done in Sect.~4.1, assuming a dust temperature of 75~K and a maximum flux value of 5000 mJy/beam obtained from ATLASGAL data, allowed us to calculate the H$_2$ column density at Her 36, $N(\rm H_{\rm 2})$ $\sim$ 3.75 $\times$ 10$^{22}$~cm$^{-2}$ which is in reasonable agreement with that calculated in Sect.~4.1 within a factor $\sim$ 1.25.\

%\todo[inline]{FWY: what do you mean here with non LTE? Is there a dust temperature from an SED fit? Which LTE value do you refer to? There must be newer papers discussing the Orion Bar geometry. Check out whether it is discussed in recent Goicoechea ALMA/Herschel papers.}

 \citet{1995A&A...294..792H}; \citet{2000A&A...364..301W} and \citet{2017A&A...598A...2A} described the geometry of the Orion bar where the trapezium stars illuminate the PDR, which changes from a face-on to an edge-on orientation along the varying length of the line of sight. This geometry explains the \ci\ peak which is symmetric around the peak of the CO emission \citep{1994ApJ...422..136T}. The \cii\ emission peak is also distributed symmetrically around the ionization front \citep{1993ApJ...404..219S}. Contrary to this, in M8, we see that \cii\ peaks at the east of Her 36, \ci\ peaks at Her 36, while the CO transitions peak in the north-west of Her 36 which supports the proposition of a face on geometry.
%\todo[inline]{FWY: I need to continue here! Although the M17 section is better revised by JP. Concerning the conclusions, please update them based on the last changes of the paper. }
\subsection{Comparison with the PDR of M17 SW}
M17, the Omega nebula is also among the best nearby laboratories to study star formation. It has an edge-on geometry in contrast to the face-on geometry of M8. It has a bright HII region ionized by the rich cluster NGC 6618 \citep{2009ApJ...696.1278P} and beyond this HII region lies the bright PDR in the south west of M17 (M17 SW) which is responsible for the photoelectric heating of the warm gas \citep{2015EAS....75..205P}. M17 SW also contains a wide ranged clumpy molecular cloud studied widely by \citet{2010A&A...510A..87P}; \citet{2015EAS....75..205P}.\ 

In Sect.~3, the scatter plots related to M8 show only a weak correlation of \cii\ with \ci\ and \textsuperscript{12}CO. The channel maps and line profiles at different offsets also show different morphologies of \cii\ compared with those seen in \ci\ and CO except in a small range of intermediate velocities towards Her 36. This suggests that \cii\ and the molecular gas tracers on scales away from Her 36 don't originate from the same spatial region. This is similar to M17 SW as reported by \citet{2015EAS....75..205P}. In Sect.~4.3, the comparison between the observed \textsuperscript{12}CO and \textsuperscript{13}CO data with non-LTE RADEX modeling results show that the UC HII region/ molecular gas near the Her 36 region has the highest density and kinetic temperature, while the molecular gas near the eastern HII region has low density and lower kinetic temperature. In contrast to M17 SW \citep{2015A&A...583A.107P}, the \textsuperscript{12}CO SLED shapes we see in Fig.~\ref{radex2} towards Her 36 and mid-$J$ CO positions follow a similar trend. Thus, they do not indicate large fluctuations in gas temperatures of the molecular gas. However, similarly to the case of M17 SW, the higher-$J$ CO lines show significantly lower intensities at the \cii\ peak position. This is consistent with a PDR, where the \cii\ peak emission is expected to arise from less dense gas than at the Her 36 position.
%\todo[inline]{FWY: I toned down the differences between tracers, since towards Her 36 basically everything is bright. Furthermore, I don't understand that more diffuse moderately warm gas is needed for the high-j CO. I would have guessed that one needs much hotter gas given that the 16-15 transitions is already 750K above ground.}

\section{Conclusions}
In this paper, we presented for the first time velocity integrated intensity maps of $J$ = 11 $\to$ 10, $J$ = 13 $\to$ 12, $J$ = 16 $\to$ 15 \textsuperscript{12}CO and \cii\ 158 $\mu$m, observed towards Her 36 in M8 using the dual-colour Terahertz receiver GREAT on board of the SOFIA telescope; $J$ = 2 $\to 1$, $J$ = 3 $\to$ 2, $J$ = 6 $\to$ 5 and $J$ = 7 $\to$ 6 \textsuperscript{12}CO transitions; $J$ = 2 $\to$ 1, $J$ = 4 $\to$ 3 and $J$ = 6 $\to$ 5 \textsuperscript{13}CO transitions using the CHAMP\textsuperscript{+}, FLASH\textsuperscript{+} and PI230 receivers of the APEX telescope; and the $J$ = 1 $\to$ 0 transitions of \textsuperscript{12}CO and \textsuperscript{13}CO using the EMIR receiver of IRAM 30 m telescope.\

Combining the information obtained from Sects.~3 and 5.1, we put forward the geometry of the region surrounding Her 36. M8 has a face-on geometry where the cold dense molecular cloud lies in the background with Her 36 being still very close to the dense core of the cloud from which it was born. Her 36 is powering the HII region towards the east of it along with 9 Sgr, while the foreground veil of a warm PDR is receding away (at lower velocities) from Her 36 towards the observer.  

%Combining the information from Sect.~3, we found that \cii\ peaks towards the east of Her 36 where the HII region is illuminated from Her 36 and 9 Sgr; \ci\ and high - J transition lines of CO peak almost at Her 36 and; mid and low - J transition lines of \textsuperscript{12}CO and \textsuperscript{13}CO peak towards north west of Her 36 in the molecular cloud. Channel maps of \textsuperscript{12}C show a clumpy structure at an offset of ($\delta$$\alpha$ = 60.0$\arcsec$ $\times$ $\delta$$\rho$ = 27.0$\arcsec$) which is also traced by WISE 3.4 $\mu$m and 4.6 $\mu$m data in IR. The \textsuperscript{12}CO and \textsuperscript{13}CO line profiles are broad and most intense in the molecular cloud while \cii\ line profile is broad and most intense in the HII region.\
%Combining the information from section 3, we found about the morphology of M8. The HII region if found towards the east of Herschel 36 where the gas is hot enough to ionize \textsuperscript{12}C\textsuperscript{+}. This HII region is also powered by 9 Saggitarii (R.A. 18\textsuperscript{h}03\textsuperscript{m}52\textsuperscript{s}.4; Dec. -24$\degree$21'38".63). The region around Herschel 36, we have the PDR where the far-UV photons are dissociating CO isotopologues and ionizing \textsuperscript{12}C and further into the north-west of Herschel 36 is the cold molecular region.\

Using different techniques we studied the physical conditions in the molecular gas associated with M8. CO rotation diagrams indicate temperature gradients through the PDR. Low-$J$ $^{13}$CO transitions seem to originate from colder gas while the $J$ = 8 $\to$ 7 transition seems to originate from the hottest gas. Quantitative analysis including LTE approximation methods and the non-LTE RADEX program were used to calculate the temperatures and H\textsubscript{2} number density in the PDR around Her 36. Kinetic temperatures ranging from 100 -- 150~K and densities ranging from 10\textsuperscript{4} -- 10\textsuperscript{6} cm\textsuperscript{-3} were obtained. \\  

$Acknowledgements$. SOFIA is jointly operated by the Universities Space Research Association, Inc. (USRA), under NASA contract NAS2-97001, and the Deutsches SOFIA Institut (DSI) under DLR contract 50 OK 0901 and 50 OK 1301 to the University of Stuttgart. We are thankful to the SOFIA operations team for their help and support during and after the observations. M. Tiwari was supported for this research by the International Max-Planck-Research School (IMPRS) for Astronomy and Astrophysics at the Universities of Bonn and Cologne. We also thank Thushara Pillai for helpful discussions. The research reported here was motivated by a discussion between Don York and Karl Menten. 
 
%\todo[inline]{FWY: that's it from my side. please go again through the individual sections to see whether we missed anything that should be repeated in the conclusions.}

\bibliography{references1}
\bibliographystyle{aa}

\begin{appendix}

\section{\cii\ column density}

For optically thin, thermalised \cii\ emission and neglecting the effects from the background, the observed antenna temperature comes out to be: 

\begin{equation} \label{cii_colden}
    T_{\rm A}^{\rm *} = 3.43 \times 10^{-16} \big[1 + 0.5e^{91.25/T_{\rm kin}}]^{-1} \frac{N(\cii\ )}{\delta v}~\rm K,
\end{equation}\\
  where $T_{\rm kin}$ is the kinetic temperature, $N(\rm \cii\ )$ is \cii\ column density in cm$^{-2}$ and $\delta v$ is the line width in km~s$^{-1}$. 
\end{appendix}

\end{document}